\documentclass[twocolumn, twocolappendix]{aastex63}

\usepackage{graphicx}
\usepackage{apjfonts}
\usepackage{amsmath}
\usepackage[raggedright]{subfigure}
\usepackage{color}
\usepackage{hyperref}
\usepackage{lineno}
\usepackage{acronym}
\usepackage{mathtools}

\allowdisplaybreaks

\newcommand{\psr}{PSR\,J1311$-$3430}
\newcommand{\circsrc}{4FGL\,J1653.6$-$0158}
\newcommand{\ellisrc}{4FGL\,J0523.3$-$2527}
\def\Fermi{\textit{Fermi}}
\newcommand{\atlas}{ATLAS}
\newcommand{\EatH}{\textit{Einstein@Home}}
\newcommand{\pvec}{\boldsymbol{\lambda}}%\newcommand{\pvec}{\textbf{u}}
\newcommand{\tref}{t_\text{ref}}
\newcommand{\tobs}{T_\text{obs}}
\newcommand{\tcoh}{T_\text{coh}}

\newcommand{\tasc}{T_\text{asc}}
\newcommand{\forb}{f_{\text{orb}}}
\newcommand{\Porb}{P_\text{orb}}
\newcommand{\norb}{n_\text{orb}}
\newcommand{\nsky}{n_\text{sky}}
\newcommand{\hmf}{ \hat{m}_f }
\newcommand{\hmfd}{ \hat{m}_{\dot{f}} }
\newcommand{\hmo}{ \hat{m}_{\text{orb}}}
\newcommand{\hms}{ \hat{m}_{\text{sky}}}
\newcommand{\Omorb}{\Omega_\text{orb}}
\newcommand*\diff{\mathop{}\!\mathrm{d}}

\newcommand{\extraspace}{\vspace{0.0cm}}

\defcitealias{gaia2018}{Gaia}
\defcitealias{monet2003}{USNO-B1.0}

\shorttitle{Search Methods for Binary Gamma-Ray Pulsars} %%% < 44 characters
\shortauthors{\sc Nieder et al.}

\begin{document}
	
	%% Acronyms
	% no plural
	\newacro{LAT}[LAT]{Large Area Telescope}
	\newacro{SSB}[SSB]{solar system barycenter}
	\newacro{3FGL}[3FGL]{\Fermi~\acs{LAT} Third Source Catalog}
	\newacro{4FGL}[4FGL]{\Fermi~\acs{LAT} Fourth Source Catalog}
	\newacro{ATNF}[ATNF]{Australia Telescope National Facility}
	% normal plural (+s)
	\newacro{YP}[YP]{young pulsar}
	\newacro{MSP}[MSP]{millisecond pulsar}
	\newacro{SNR}[S/N]{signal-to-noise ratio}
	\newacro{FFT}[FFT]{fast Fourier transform}

	%% Title
	\title{Exploiting orbital constraints from optical data \\to detect binary gamma-ray pulsars}
	
	%% Authorlist
	\author[0000-0002-5775-8977]{L.~Nieder}
	\affiliation{Max-Planck-Institut f\"ur Gravitationsphysik (Albert-Einstein-Institut), 30167 Hannover, Germany}
	\affiliation{Leibniz Universit\"at Hannover, 30167 Hannover, Germany}

	\author[0000-0003-4285-6256]{B.~Allen}
	\affiliation{Max-Planck-Institut f\"ur Gravitationsphysik (Albert-Einstein-Institut), 30167 Hannover, Germany}
	\affiliation{Department of Physics, University of Wisconsin-Milwaukee, P.O. Box 413, Milwaukee, WI 53201, USA}
	\affiliation{Leibniz Universit\"at Hannover, 30167 Hannover, Germany}
	
	\author[0000-0003-4355-3572]{C.~J.~Clark}
	\affiliation{Jodrell Bank Centre for Astrophysics, Department of Physics and Astronomy, The University of Manchester, M13 9PL, UK}
	
	\author[0000-0002-1164-4755]{H.~J.~Pletsch}
	\affiliation{Max-Planck-Institut f\"ur Gravitationsphysik (Albert-Einstein-Institut), 30167 Hannover, Germany}
	
	% corresponding author
	\correspondingauthor{L.~Nieder}
	\email{lars.nieder@aei.mpg.de}
	
	% dates
	\received{2020 April 24}
	\revised{2020 July 30}
	\accepted{2020 August 12}
	\published{2020 October 05}

%%%%%%%%%%%%%%%%%%%%%%%%%%%%%%%%%%%%%%%%%%%
\begin{abstract} %%%<250words
	\noindent
	It is difficult to discover pulsars via their gamma-ray emission because current instruments typically detect fewer than one photon per million rotations.  This creates a significant computing challenge for isolated pulsars, where the typical parameter search space spans wide ranges in four dimensions.  It is even more demanding when the pulsar is in a binary system, where the orbital motion introduces several additional unknown parameters.  Building on earlier work by \citeauthor{pletsch2014}, we present optimal methods for such searches.  These can also incorporate external constraints on the parameter space to be searched, for example, from optical observations of a presumed binary companion.  The solution has two parts.  The first is the construction of optimal search grids in parameter space via a parameter-space metric, for initial semicoherent searches and subsequent fully coherent follow-ups.  The second is a method to demodulate and detect the periodic pulsations.  These methods have different sensitivity properties than traditional radio searches for binary pulsars and might unveil new populations of pulsars.
\end{abstract}
	

%%%%%%%%%%%%%%%%%%%%%%%%%%%%%%%%%%%%%%%%%%%
\section{Introduction} \label{s:intro}

	The \aclu{LAT} \citep[\acs{LAT};][]{atwood2009} on the \Fermi{} satellite has helped to increase the known Galactic population of gamma-ray pulsars to more than $250$ pulsars\footnote{\href{https://tinyurl.com/fermipulsars}{https://tinyurl.com/fermipulsars}} \citep[for a review see, e.g.,][]{caraveo2014}. However, in the recent \aclu{4FGL} \citep[\acs{4FGL};][]{4fgl} $1{,}525$ out of $5{,}098$ gamma-ray sources remain unassociated. Many of those are thought to be pulsars, perhaps in binary systems.

	Gamma-ray pulsars may be detected in three ways: (a) A known (radio or X-ray) pulsar position and ephemeris guides a follow-up gamma-ray pulsation search within a nearby \ac{LAT} source \citep[e.g.,][]{abdo2009a,abdo2009c,guillemot2012}. (b) A similar gamma-ray pulsation search is done for a known pulsar, but {\em without} an obvious gamma-ray source being present \citep{smith2017}. (c) A \replaced{``blind''}{``partially informed''} search\footnote{These searches have been called ``blind'' searches in previous literature.} hunts for gamma-ray pulsations around a \ac{LAT} source where no pulsar has yet been identified\added{, and hence several timing parameters, notably the spin period, are unknown in advance}.

	\replaced{Blind gamma-ray searches are}{Partially informed searches are} the focus of this paper.  Such searches have discovered more than $50$ \acp{YP} \citep[e.g.,][]{abdo2009d,parkinson2010,pletsch2012a,clark2017}, and three \acp{MSP} \citep{pletsch2012,clark2018}.  Many of these pulsars could not have been found via radio or X-ray emissions, which were not detected in extensive follow-up searches.  Such systems are of particular interest because they constrain models of pulsar emission and beaming. \replaced{Blind}{Partially informed} searches also have the potential to discover new populations of pulsar/neutron star objects.

	So far, most \replaced{blind}{partially informed} gamma-ray searches have targeted isolated pulsars.  The searches are a substantial computing effort, and have been carried out in campaigns or surveys that last several years.  More recent surveys find new systems because the ongoing \ac{LAT} operations provide additional data, which enables the detection of weaker pulsations \citep[e.g., ][]{clark2017}.  However, there is also a downside: the computing power required also increases quickly with longer observation time spans.

	Until now, \replaced{blind}{partially informed} gamma-ray searches have only found one binary \ac{MSP}, \psr{} \added{\citep{pletsch2012}}. This is tantalizing because three quarters of the known \acp{MSP} in the \ac{ATNF} Pulsar Catalogue\footnote{\label{n:atnf}\href{http://www.atnf.csiro.au/research/pulsar/psrcat}{http://www.atnf.csiro.au/research/pulsar/psrcat}} \citep{manchester2005} are in binaries.  So if search sensitivity were not limited by computing power, it might be possible to find many more.  But even for isolated pulsars it is expensive to search for high ($> 100$\,Hz) spin frequencies, and adding (at least three) additional orbital parameters makes it even more costly.  By improving the techniques, the methods presented here are a first step toward finding more of these systems.

	Much of our focus is on binary pulsars in so-called ``spider'' systems, in which the pulsar companion is being evaporated by an energetic pulsar wind.  A typical example is the first ``black widow'' pulsar to be discovered, PSR\,B1957$+$20 \citep{fruchter1988}.  This was found in radio, where pulsations are eclipsed for a large fraction of the orbit, presumably by material ablated from the companion.  Spider pulsars are categorized as black widows if the companion mass $M_\text{c}$ is very low ($M_\text{c} \ll 0.1\,M_\odot$) or as ``redbacks'' (another spider species) for larger companion masses (\replaced{$M_\text{c} \sim 0.15 - 0.4 \, M_\odot$}{$M_\text{c} \sim 0.15 - 0.7 \, M_\odot$}) \citep[e.g.,][]{roberts2013,strader2019}\added{, with one redback candidate likely having an even higher companion mass $\gtrsim 0.8\,M_{\odot}$ \citep{strader2014}}.

	For many of the known \acp{MSP} in spider systems, the companions are visible in the optical.  The light originates from nuclear burning, and/or from pulsar wind heating up the companion.  The orbital motion of the companion then leads to a detectable modulation of the orbital brightness.  The source of this modulation is not well understood.  It might be that the side of the companion facing the pulsar is hotter than the other side and is more visible at the companion's superior conjunction.  The companion might also be tidally elongated into an ellipsoid, whose projected cross section onto the line of sight varies over the orbit.

	The new \deleted{blind} search methods presented here are well suited to gamma-ray pulsars in spider systems, with nearly circular orbits (eccentricity $e < 0.05$) and for which optical observations of the pulsar's companion provide information about the orbital motion, and thus constrain the gamma-ray pulsation search space.

	For concreteness, we present the search designs for two promising gamma-ray sources: (a) \circsrc{}, a likely \ac{MSP} in a circular binary \citep{romani2014,kong2014}, and (b) \ellisrc{}, a probable \ac{MSP} in a slightly eccentric binary \citep{strader2014}.  These are ranked among the most likely pulsar candidates \citep{parkinson2016}.  We demonstrate the feasibility of a search using the computing resources of the distributed volunteer computing project \EatH{} \citep{allen2013}.

	The paper is organized as follows.  Section~\ref{s:background} reviews \replaced{blind }{partially informed }search methods for isolated gamma-ray pulsars and introduces the concepts required for such searches.  Section~\ref{s:methodcirc} extends the methods to gamma-ray pulsars in circular orbit binaries, and Section~\ref{s:methodecc} further extends these to eccentric orbit binaries.  In Section~\ref{s:alternatives} our methods are compared with alternatives used in radio and gravitational-wave astronomy.  Finally, in Section~\ref{s:conclusions} we discuss the feasibility of future \replaced{blind}{partially informed} searches for binary gamma-ray pulsars and also consider some specific sources.  This is followed by Appendices \ref{s:expectedvalues}, \ref{s:optimalmismatch}, and \ref{s:highorder} containing some technical details.

	In this paper, $c$ denotes the speed of light and $G$ denotes Newton's gravitational constant.

%%%%%%%%%%%%%%%%%%%%%%%%%%%%%%%%%%%%%%%%%%%
\extraspace
\section{Partially-informed gamma-ray searches for pulsars} \label{s:background}

	\replaced{Blind}{Partially informed} search methods for isolated gamma-ray pulsars have been studied in detail by \cite{pletsch2014}.  Here we summarize and extend their framework.  The following sections generalize the search methods to binary pulsars.

	The search for gamma-ray pulsations begins with a list of $N$ photons from a posited source, which we label with the index $j=1, \dots, N$.  The data available for these photons are their detector arrival time $t_j$, their direction of origin, and their energy, spanning an observation interval $\tobs$.

	We are dealing with many sums and products in this paper.  Sums and products over $j,k,\ell$ run from $1, \dots ,N$ unless otherwise specified.  Furthermore, we adopt the notation
	\begin{linenomath}
	\begin{equation}
		\sum_{j\ne k} \equiv	\sum_{j=1}^{N}\sum_{\substack{k=1 \\ j \ne k}}^N
	\end{equation}
	\end{linenomath}
	for simplicity reasons.

	Not all photons are equally significant.  Photons at low energies are less well localized than those at higher energies and cannot be so readily attributed to a target source.  Photons whose energy is more consistent with a distributed background are less likely to come from the pulsar.  Photons originating from a nearby point source might contaminate the data set.  For such reasons, searches may be improved by modeling the spatial and energy distribution of the sources.
	
	\replaced{This assigns a weight $w_j \in [0,1]$ to each photon, which is the only place where the energy and arrival direction of the photons enter our analysis.  The weight $w_j$ represents the probability that the $j$th photon originated at the nominal pulsar \citep{bickel2008, kerr2011}.}{To quantify the significance, we assign a weight $w_j \in [0,1]$ to
	each photon.  This weight $w_j$ represents the probability that the $j$th photon originated at the nominal pulsar \citep{bickel2008, kerr2011}.  The photon weights are determined from an assumed spectral and spatial model of gamma-ray sources in the region around the target pulsar, which is obtained using the standard methods for fitting gamma-ray sky maps\footnote{\href{https://fermi.gsfc.nasa.gov/ssc/data/analysis/scitools/}{https://fermi.gsfc.nasa.gov/ssc/data/analysis/scitools/}}.}

	\added{Each photon's weight is computed as the predicted fraction that the target pulsar contributes to the total photon flux at the photon's energy and arrival direction, after convolution with the \Fermi{}-\ac{LAT}'s energy-dependent point-spread function \citep{kerr2011,bruel2019}.  The weighting process, and hence the resulting $w_j$, is the only place where the energy and arrival direction of the photons enter our analysis.  In practice, the weights are computed using \texttt{gtsrcprob} from the \Fermi{} Science Tools\footnote{\href{https://fermi.gsfc.nasa.gov/ssc/data/analysis/software/}{https://fermi.gsfc.nasa.gov/ssc/data/analysis/software/}}, using, e.g., the 4FGL catalog \citep{4fgl} and associated Galactic and isotropic diffuse emission templates as the input model.}  These weights are used for noise suppression and to reduce computing cost by removing the lowest-weighted photons.

	In this paper, we assume that these weights have been determined in advance for each photon, so the only information available for the $j$th photon is its arrival time $t_j$ in the detector and the weight $w_j$.

	The question that we need to answer is, are the arrival times of these photons random, or is there an underlying periodicity?  To answer this question (in the statistical sense), we first need a model for the periodicity, which we assume is tied to the physical rotation of the pulsar.

\extraspace
\subsection{Pulse profile and photon arrival probability}

	For now, assume that ``in isolation'' the pulsar would have a linearly changing angular velocity.  Using $\Phi$ to denote the rotational phase in radians
	\begin{linenomath}
	\begin{equation} \label{eq:phasemodel}
		\Phi (t_{\rm psr}, \pvec) = 2 \pi f (t_{\rm psr} - \tref) +\pi \dot{f} (t_{\rm psr} - \tref)^2 \,,
	\end{equation}
	\end{linenomath}
	where $t_{\rm psr}$ is the time that would be measured by a fictitious observer freely falling with the center of mass of the pulsar, and $\tref$ is a reference time.  Note that detector time ticks at a different rate than $t_\text{psr}$, because the detector is moving around the Earth and the Sun, and because the pulsar might be orbiting a binary companion, or accelerating toward the Galaxy. Also note that without loss of generality we have set the phase at the reference time to zero.

	The parameters $\pvec$ describe the pulsar.  Here they are the spin frequency $f$ and its first time derivative $\dot{f}$ at reference time $\tref$.  This second-order Taylor approximation holds for many pulsars and most \acp{MSP}, but for very young and ``glitching'' pulsars, additional higher-order terms may be needed.

	The flux of photons can be broken into three parts.  The first does not come from the pulsar: it is a background that is uncorrelated with pulsar rotation.  We call these unpulsed photons ``background''.  The second part originates from the pulsar itself but is also uncorrelated with pulsar rotation.  We call these ``unpulsed source'' photons.  The last part is a periodically time-varying flux from the source, which we call ``pulsed''.  We use $p$ to denote the ratio of the number of pulsed photons to the total number of source photons (pulsed and unpulsed source).

	The pulsed photon flux may be described with a periodic function $F_\text{S}(\Phi)$ of the pulsar's phase around its rotational axis, $\Phi \in [0, 2\pi]$, and is time stable for most pulsars.  The normalized probability that a pulsed photon arrives in the phase interval $[\Phi, \Phi + \diff\Phi]$ is $F_\text{S}(\Phi) \diff\Phi$.  The function $F_\text{S}(\Phi)$ has minimum value zero and encloses unit area in the interval $[0, 2\pi]$.

	We can now give the probability density function for the rotation phase associated with a given photon.  This differs from one photon to the next because photons with small weight $w_j$ are more likely to have a phase-independent probability distribution.  The probability that the $j$th photon originates from a rotation phase interval $[\Phi_j, \Phi_j + \diff\Phi_j]$ is $F_j(\Phi_j) \diff\Phi_j$, where
	\begin{linenomath}
	\begin{equation} \label{eq:probdist1}
		F_j(\Phi_j) = \frac{1-w_j}{2 \pi} + w_j \biggl[ \frac{1-p}{2\pi} + p F_\text{S}(\Phi_j) \biggr] \,.
	\end{equation}
	\end{linenomath}
	The first term (with probability $1-w_j$) describes the background photons, and the second and third terms (with probability $w_j$) describe the unpulsed and pulsed source photons, respectively.

	The probability distribution of pulsed photons may be expressed as the Fourier series
	\begin{linenomath}
	\begin{equation} \label{eq:probdist2}
		F_\text{S}(\Phi) = \frac{1}{2\pi} + \frac{1}{2\pi} \sum_{n = 1}^\infty \left( \gamma_n {\rm e}^{i n \Phi} + \gamma_n^* {\rm e}^{-i n \Phi} \right) \,.
	\end{equation}
	\end{linenomath}
	The complex Fourier coefficients are
	\begin{linenomath}
	\begin{equation} \label{eq:gammadef}
		\gamma_n = \int_{0}^{2\pi} F_\text{S}(\Phi) {\rm e}^{-i n \Phi} \diff\Phi \,.
	\end{equation}
	\end{linenomath}
	Note that the Fourier coefficients $\gamma_n$ are constrained because $F_\text{S}$ has minimum value zero.  Note also that for known gamma-ray pulsars $|\gamma_n|^2$ decreases quickly with increasing index $n$ \citep{pletsch2014}.  In many cases the first five harmonics are sufficient to describe the pulse profile.

	In principle, to detect gamma-ray pulsations, we assume a rotational model $f, \dot f$ and then compute the rotational phase associated with each photon.  ``Binning'' these phases (mod $2\pi$) with weights $w_j$ provides an estimate of $F(\Phi) = \sum_j w_j F_j(\Phi)/\sum_j w_j$, from which we can estimate $F_\text{S}(\Phi)$ by shifting the minimum value to zero and rescaling to unit area.  If that function is compatible with zero (meaning: coefficients $\gamma_n$ are small), then no pulsations were detected.  Conversely, if the $\gamma_n$ are large for some values of $f$ and $\dot f$, we have found pulsations.

\extraspace
\subsection{Relationship of detector time $t$ to $t_{\rm psr}$} \label{ss:detectortime}

	The situation is slightly more complicated than described in the previous paragraph because computing $t_\text{psr}$ for each photon from its time of arrival at the \Fermi{} satellite also requires the pulsar's sky position (right ascension $\alpha$ and declination $\delta$).  The sky position allows for `` barycentric corrections'', e.g., to account for Doppler shifts due to the \ac{LAT}'s movement around the \ac{SSB}.  Thus, the photon's emission time $t_{\text{psr}}(t,\alpha,\delta)$ is a function of its arrival time $t$ at the \ac{LAT} and the putative pulsar's sky position.  The pulsar's putative phase is a function of $t$ and the four parameters $\pvec = \{f,\dot{f},\alpha,\delta\}$.

	In \replaced{blind}{partially informed} searches the spin parameters are unknown.  Although each photon is tagged with an arrival direction $\alpha$, $\delta$, these are not sufficiently precise to detect pulsations, so those location parameters must also be searched.  Hence, the parameter space search volume $\Lambda$ for isolated pulsars ($\pvec \in \Lambda$) is $4$-dimensional.  In Sections~\ref{s:methodcirc} and \ref{s:methodecc}, the higher-dimensional search spaces for binary pulsars in circular and elliptical orbits are discussed.

\extraspace
\vspace{0.8cm}
\subsection{Searching for pulsations} \label{s:multistage}

	For realistic searches the parameter space $\Lambda$ is too large to search by the straightforward computational process described above.  Instead, $\Lambda$ is explored with a multistage search based on several different test statistics \citep[e.g.,][]{meinshausen2009}.  This gives the greatest sensitivity at fixed computational cost \citep{pletsch2014}.  The approach is hierarchical. In the first stage, a coarse grid covering the parameter space $\Lambda$ is searched at low sensitivity using inexpensive test statistics.  These are relatively insensitive to mismatch between tested parameters and pulsar parameters.  In the following stages, smaller regions of $\Lambda$ around the most promising candidates are searched at higher sensitivity.  These use more expensive test statistics on finer, more closely spaced grids.  Thus, a search is defined by a test statistic/grid hierarchy.

	The spacing of the grids in parameter space is governed by the mismatch described above. For a given test statistic, we calculate a ``metric'', which is the fractional loss in the expected \ac{SNR}.  The details of this are found later in this section.

	The search described in this paper has four stages, which employ detection statistics $P_1$, $S_1$, and $H$. Here we briefly describe the overall structure. The test statistics are defined and characterized later in this section.

	The first three stages search for significant power in the first harmonic $|\gamma_1|^2$.  Each discards regions of parameter space that contain no signals; what remains is passed to the following stage.  The first stage uses the ``semicoherent'' test statistic $S_1$ with a low threshold.  The second stage tests $S_1$ on a finer grid, with a higher threshold.  The third stage uses the fully coherent test statistic $P_1$.  This searches coherently for power $|\gamma_1|^2$ over the full observation span $\tobs$ with much greater sensitivity and a finer grid than before.

	The fourth stage employs the expensive $H$ statistic, which combines $P_1$, \dots , $P_5$.  This coherently integrates over $\tobs$ to identify power in the first five harmonics $|\gamma_1|^2$, \dots ,$|\gamma_5|^2$.  By searching around the surviving candidate points in parameter space with a still finer grid, this completes the hierarchy.

\extraspace
\subsection{Coherent power test statistic $P$}

	The basis for all of our test statistics is the coherent Fourier power, evaluated over different periods of time.  For the $n$th harmonic, and including all of the photons, this is
	\begin{linenomath}
	\begin{equation} \label{eq:cohpower}
		P_n(\boldsymbol{\lambda}) = \frac{1}{\kappa^2} \Big| \sum\limits_{j} w_j {\rm e}^{-i n \Phi(t_j,\pvec)} \Big|^2.
	\end{equation}
	\end{linenomath}
	To simplify notation, from here on we use $\Phi(t_j, \pvec)$ to denote $\Phi(t_{\text{psr}}(t_j,\alpha, \delta),f,\dot{f})$, where $t_j$ is the photon arrival time measured at the \ac{LAT}. The normalization constant is
	\begin{linenomath}
	\begin{equation}
		\kappa^2 = \frac{1}{2} \sum\limits_{j} w_j^2 \,.
	\end{equation}
	\end{linenomath}
	How does $P_n$ behave in the absence of pulsations and in the presence of pulsations?

	To answer this question, we compute expectation values as shown in Appendix~\ref{s:expectedvalues}. The power $P_n$ has an expected value (Eq.~\ref{eq:finalexpec}) and variance (in the absence of a pulsed signal, $p=0$)
	\begin{linenomath}
	\begin{align} \label{eq:shortexpected}
		\operatorname{E}_p [P_n] &=  2 + \kappa^{-2} p^2  | \gamma_n |^2 \sum_{j \ne k} w^2_j w^2_k \\
		\operatorname{Var}_0[P_n] &=  \kappa^{-4} \sum_{j \ne k} w_j^2 w_k^2 \,.
	\end{align}
	\end{linenomath}

	The power $P_n$ is a {\it detection statistic} because it is sensitive to a nonvanishing pulse profile.  If $\gamma_n$ is nonzero, then $P_n$ should be larger than $2$.  It becomes larger as the fraction $p$ of pulsed to source photons increases (which we cannot control).  It also becomes larger as the number of photons (or equivalently, the observation time) grows.  But to understand what values of $P_n$ correspond to statistically significant detections, we need to know about its statistical fluctuations, meaning the variance in $P_n$.

	Note that the diagonal-free double sum in these expressions can be reexpressed as $(\sum_j w_j^2)^2 - \sum_j w_j^4$.  Thus, the variance can be written as
	\begin{linenomath}
	\begin{equation} \label{eq:var_pn}
		\operatorname{Var}_0[P_n] = 4 - 4\frac{\sum_{j} w_j^4}{( \sum_{j} w_j^2)^2} \,.
	\end{equation}
	\end{linenomath}
	If there are many photons from the source and the weights are relatively uniformly distributed, then it follows that the numerator in Eq.~\eqref{eq:var_pn} is $\mathcal{O}(N)$ and the denominator is $\mathcal{O}(N^2)$.  Hence, the variance $\operatorname{Var}_0[P_n] \rightarrow 4 - \mathcal{O}(1/N)$ approaches $4$.  In this limit, and with the statistical assumptions of Appendix \ref{s:expectedvalues}, $P_n$ has a noncentral $\chi^2$-distribution with two degrees of freedom \citep{pletsch2014}.  The noncentrality parameter is the second term appearing in Eq.~(\ref{eq:shortexpected}).

	The expected \ac{SNR} associated with $P_n$ is
	\begin{linenomath}
	\begin{equation} \label{eq:coh_snr_p}
		\begin{aligned}
			\theta_{P_n}^2 = \frac{\operatorname{E}_p[P_n] - \operatorname{E}_0[P_n]}{\sqrt{\operatorname{Var}_0[P_n]}} &=  p^2 |\gamma_n|^2 \sqrt{ \sum_{j \ne k}  w_j^2 w_k^2 } \\
			&= p^2 |\gamma_n|^2  \mu \, \tobs \,.
		\end{aligned}
	\end{equation}
	\end{linenomath}
	In the many-photon limit the quantity $\mu \rightarrow \sum_j w_j^2 /\tobs$ is proportional to the mean weighted photon arrival rate.

\extraspace
\subsubsection{Loss of $P$ from parameter mismatch} \label{s:par_mismatch}

	In a real search, we compute detection statistics at a grid of discrete values of the signal parameters $\pvec$.  If there is a signal present, its actual (true) parameters might be close to one of these discrete values but will not match it exactly.  There will always be {\it some} offset between the tested parameters and the true parameters.  Here we quantify how much \ac{SNR} is expected to be lost because of this mismatch.

	Assume that the tested parameters $\pvec$ are close to the true pulsar parameters $\pvec_\text{psr} \,,$ and introduce the notation
	\begin{linenomath}
	\begin{equation}
		\diff \lambda^a = \lambda^a - \lambda^a_\text{psr}
	\end{equation}
	\end{linenomath}
	for the small parameter offsets.  Here and elsewhere in the paper we index the parameter space dimension with lowercase Latin letters ``$a$'' and ``$b$''.  These offsets change the pulsar rotation phase by
	\begin{linenomath}
	\begin{equation}
		\Delta \Phi(t) = \Phi(t,\pvec) - \Phi(t,\pvec_\text{psr}) \approx \partial_a \Phi \diff\lambda^a \,,
	\end{equation}
	\end{linenomath}
	where the notation
	\begin{linenomath}
	\begin{equation}
		\partial_a \Phi = \left.\frac{\partial \Phi}{\partial \lambda^a}\right|_{\pvec = \pvec_\text{psr}}
	\end{equation}
	\end{linenomath}
	is introduced and we neglect higher powers in $\diff \lambda$.  We also adopt the {\it Einstein summation convention} that repeated parameter space indices are summed over all the dimensions of the parameter space.

	We now compute the fractional loss in expected S/N associated with this parameter mismatch.  For the offset parameters the coherent power is 
	\begin{linenomath}
	\begin{equation}
		P_n(\pvec) = 2 + \kappa^{-2}  \sum_{j \ne k} w_j w_k  {\rm e}^{i n (\Phi_j - \Phi_k)}  {\rm e}^{i n (\Delta \Phi_j - \Delta \Phi_k )} \,,
	\end{equation}
	\end{linenomath}
	where $\Phi_j = \Phi(t_j,\pvec_\text{psr})$ and $\Delta \Phi_j = \Delta \Phi(t_j)$.  Following Appendix \ref{s:expectedvalues}, the expectation value of this is
	\begin{linenomath}
	\begin{equation}
		\operatorname{E}_p[P_n(\pvec)] = 2 + \kappa^{-2}  p^2 |\gamma_n|^2 \sum_{j \ne k} w^2_j w^2_k {\rm e}^{i n (\Delta \Phi_j - \Delta \Phi_k)} \,.
	\end{equation}
	\end{linenomath}
	It follows that for the mismatched signal the expected S/N is
	\begin{linenomath}
	\begin{equation} \label{eq:SNR_Pn}
		\theta_{P_n(\pvec)}^2 =  p^2 |\gamma_n|^2  \left[ \sum_{j \ne k}  w_j^2 w_k^2 \right]^{-1/2} \!\!\!\!\!\!\!\!\!\!\!\sum_{j \ne k}  w_j^2 w_k^2  {\rm e}^{i n \bigl(\Delta \Phi_j - \Delta \Phi_k \bigr)} \,.
	\end{equation}
	\end{linenomath}
	The fractional loss in S/N (often called the ``mismatch'') is
	\begin{linenomath}
	\begin{align}
		& m(\pvec, \pvec_\text{psr}) = \frac{ \theta_{P_n}^2(\pvec_\text{psr}) - \theta_{P_n}^2(\pvec)}{\theta_{P_n}^2(\pvec_\text{psr})} \nonumber\\
		&= \sum_{j \ne k}  w_j^2 w_k^2  \left[ 1 - {\rm e}^{i n (\Delta \Phi_j - \Delta \Phi_k )} \right] / \sum_{j \ne k}  w_j^2 w_k^2 \\
		&=\!\!\! \left. \left[ \!\!\! \left(\sum_{j}  w_j^2\right)^2\!\!\!\! - \left| \sum_{j}  w_j^2 {\rm e}^{i n \Delta \Phi_j} \right|^2 \right] \middle/ \left[ \!\!\! \left(\sum_{j}  w_j^2\right)^2\!\!\!\! - \sum_{j}  w_j^4\right] \right. \,. \nonumber
	\end{align}
	\end{linenomath}
	We need the mismatch to help set the spacings of the parameter space search grids, but for that purpose, approximations suffice.

	Assume that there are many photons and the weights are uniformly distributed in time (or at least slowly varying in a way that is not correlated with the pulsar rotation phase).  The sums over the weights may then be replaced with simple integrals over time, giving
	\begin{linenomath}
	\begin{equation} \label{eq:mismatch_coh}
		m(\pvec, \pvec_\text{psr})  \approx 1 - \left| \prescript{}{\tobs \!\!\!\!}{\Big\langle} {\rm e}^{-in \Delta\Phi(t)} \Big\rangle (t_0)  \right|^2 \,.
	\end{equation}
	\end{linenomath}
	Here we introduce the ``angle bracket'' notation for an average over a time interval of length $T$ centered around an arbitrary time $t_0$.  This takes an input function $Q(t')$ and outputs a new function of time $t$ defined by
	\begin{linenomath}
	\begin{equation}
		\prescript{}{T \! \!}{\langle} Q(t')\rangle(t) \equiv \frac{1}{T} \int_{t - T/2}^{t + T/2} Q(t') \diff t' \,,
	\end{equation}
	\end{linenomath}
	which is the average of $Q$ around the time $t$.
	
\extraspace
	\subsubsection{Parameter space metric $g_{ab}$} \label{s:cohmetric}

	Since the sensitivity of these searches is limited by available computing power, we need to construct a grid that covers the relevant parameter space with the smallest number of grid points.  This means that the parameters  $\pvec_\text{psr}$ of any possible pulsar should be close enough to a grid point that we do not lose too much S/N from the mismatch, but the grid should have as few points as possible.

	The distance metric on the search space is a useful tool for such constructions \citep{balasubramanian1996, owen1996}. It provides an analytical approximation to the mismatch. For example, the coherent mismatch in Equation~\eqref{eq:mismatch_coh} can be approximated by the ``coherent metric'' $g_{ab}$
	\begin{linenomath}
	\begin{equation} \label{eq:def_cohmetric}
		m(\pvec, \pvec_\text{psr})  =  n^2 g_{ab}(\pvec)  \diff\lambda^a \diff\lambda^b + \mathcal{O}(\diff \lambda^3)
	\end{equation}
	\end{linenomath}
	for small coordinate offsets $\diff \lambda^a$ from the true pulsar parameters.

	Expanding the exponential that appears in Eq.~(\ref{eq:mismatch_coh}) to first order, one finds
	\begin{linenomath}
	\begin{equation} \label{eq:cohmetrcomp}
		g_{ab} = \prescript{}{\tobs \! \!}{\langle} \partial_a \Phi \partial_b \Phi \rangle(t_0) -  \prescript{}{\tobs \! \!}{\langle} \partial_a \Phi \rangle (t_0) \prescript{}{\tobs \! \!}{\langle} \partial_b \Phi \rangle(t_0) \,.
	\end{equation}
	\end{linenomath}
	To evaluate the metrics, we need to account for the way in which the detected pulsar rotation phase depends on the different pulsar parameters.
	
\extraspace
\subsubsection{Evaluation of $g_{ab}$ for isolated pulsars}

	As seen by an observer freely falling at the center of mass of the pulsar, the rotation phase just depends on the intrinsic frequency $f$ and its derivative $\dot{f}$ as given in Eq.~\eqref{eq:phasemodel}.  But as explained in Sec.~\ref{ss:detectortime}, these must be converted to detector time.

	For computing the metric, we do not need a conversion that is accurate to microseconds, but only one that takes into account the largest shifts between detector and pulsar time, of order $\approx 500$\,s, arising from the motion of the Earth around the Sun \citep{pletsch2014}.  We denote the orbital angular frequency by $\Omega_\text{E} = 2\pi/\text{yr}$, the orbital light-crossing time by $r_\text{E} = 1\,\text{AU}/c$, and the obliquity of the ecliptic by $\epsilon = 23.4^\circ$.

	If we choose a coordinate axis $z$ along the line of sight to the pulsar, then the projected motion is
	\begin{linenomath}
	\begin{equation} \label{eq:r_sky}
		r_{z, \text{sky}}(t) = r_\text{E} \left[n_\text{x} \cos \left(\Omega_\text{E} t + \varphi_\text{ref}\right) + n_\text{y} \sin \left(\Omega_\text{E} t + \varphi_\text{ref}\right) \right] \,,
	\end{equation}
	\end{linenomath}
	where
	\begin{linenomath}
	\begin{align}
		n_\text{x} &= \cos \alpha  \cos \delta \,, \\
		n_\text{y} &= \cos \epsilon  \sin \alpha  \cos \delta  + \sin \epsilon  \sin \delta \,,
	\end{align}
	\end{linenomath}
	and the sky location is given by the right ascension $\alpha$, and the declination $\delta$.  The (arbitrary) choice for the origin of the time coordinate determines the constant $\varphi_\text{ref}$, which is the Earth's orbital phase at that moment.

	Note that this simplified version of the R\o mer delay does not account for the motion of the \Fermi{} satellite around the Earth.  It is not accurate enough to use in a search for pulsations and is only used in the metric calculation.

	For the purpose of computing the metric we can model the detected pulsar rotation phase as the sum of Eq.~\eqref{eq:phasemodel} and the additional phase cycles introduced by the R\o mer delay \eqref{eq:r_sky}:
	\begin{linenomath}
	\begin{align}
		\Phi (t, \pvec) =&\,\, 2 \pi f (t - \tref) + \pi \dot{f} (t - \tref)^2 \\
		& + 2 \pi f r_\text{E} \left[n_\text{x} \cos \left(\Omega_\text{E} t + \varphi_\text{ref}\right) + n_\text{y} \sin \left(\Omega_\text{E} t + \varphi_\text{ref}\right) \right] \,.\nonumber
	\end{align}
	\end{linenomath}
	Here the search parameters are $\pvec = \{f, \dot{f}, n_x, n_y\}$, and the terms correcting the arrival times $t$ have been neglected for the $\dot{f}$ summand.

	The metric for the coherent power $P_1$ follows from Eq.~\eqref{eq:cohmetrcomp}.  The formulae are complicated, but if we keep only the most significant terms, then they simplify.  To determine these, consider the relative size of the different quantities:
	\begin{linenomath}
	\begin{equation} \label{eq:assumptions}
       	\begin{aligned}
			\tobs   &\approx 10~\text{yr} \approx 3 \times 10^{8} \,\text{s} \,,\\
			|t_0  -  t_\text{ref}| &\lesssim \tobs \,,\\
			\Omega_\text{E}  &\approx 2 \pi/\text{yr} \approx 2 \times 10^{-7}\,\text{s}^{-1} \,,\\
			r_\text{E} &\approx 5 \times 10^2 \, \text{s} \,,\\
			f &\approx (100 - 700) \, \text{s}^{-1}  \,,\\
			\dot{f} &\approx (10^{-16} - 10^{-14}) \, \text{s}^{-2} \,.
		\end{aligned}
	\end{equation}
	\end{linenomath}
	Most \acp{MSP} have parameters $f$ and $\dot{f}$ in the given range.  With these in mind, one finds diagonal metric components
	\begin{linenomath}
	\begin{equation} \label{eq:g_ab_fs_d}
		\begin{aligned}
			g_{f f} &= \frac{1}{3} \pi^2 \tobs^2 \left[ 1 + \mathcal{O}( r_\text{E} / \tobs ) \right] \,,\\
			g_{\dot{f} \dot{f}} &= \frac{1}{180} \pi^2 \tobs^4  \left[ 1 + 60 \frac{(t_0 - \tref)^2 }{\tobs^2} \right] \,,\\ 
			g_{n_x n_x} &= 2 \pi^2 f^2 r_\text{E}^2\left[ 1   + \mathcal{O}(1/\Omega_\text{E} \tobs) \right] \,,\\
			g_{n_y n_y} &= 2 \pi^2 f^2 r_\text{E}^2 \left[ 1   + \mathcal{O}(1/\Omega_\text{E} \tobs) \right] \,.
		\end{aligned}
	\end{equation}
	\end{linenomath}
	Most of the off-diagonal metric components are negligible.

	Determining whether off-diagonal metric components are significant requires some care because they need to be compared to the corresponding diagonal components.  This arises here and in several other places in the paper.  Here we show in detail how this significance is determined.  The same reasoning is used for the other cases that arise later but is not elaborated.

	Since the fundamental quantity of interest is the mismatch $m$, for fixed $a$ and $b$ (no Einstein summation convention), consider $m = g_{a a} (\diff \lambda^a)^2 + g_{b b} (\diff \lambda^b)^2 + 2 g_{a b} \diff \lambda^a \diff \lambda^b$.  Rescale the coordinates $\{\lambda^a,\lambda^b\}$ to new coordinates $\{\lambda^{a'} = u \lambda^a,\lambda^{b'} = w \lambda^b\}$ such that the two diagonal components of the metric in the new coordinates are both unity.  (Here $u$ and $w$ denote the rescaling factors.)  This implies that $g_{a a} (\partial \lambda^a / \partial \lambda^{a'})^2 = g_{a a} u^{-2} = 1$ and $g_{b b} (\partial \lambda^b / \partial \lambda{b'})^2 = g_{b b} w^{-2} = 1$.  Then, all off-diagonal metric components are of $\mathcal{O}(1/\Omega_\text{E} \tobs)$, apart from
	\begin{linenomath}
	\begin{equation} \label{eq:g_ab_fs_o}
		g_{f \dot{f}} = \frac{1}{3} \pi^2 \tobs^3 \left[ \frac{(t_0-\tref)}{\tobs} + \mathcal{O}(r_\text{E} / \tobs) \right]\,.
	\end{equation}
	\end{linenomath}
	Note that all the off-diagonal terms may be neglected in the case that the integration time $\tobs \gg 1\,\text{yr}$ and the reference time $\tref = t_0$.

	For this case the  diagonal ``coherent metric'' terms reduce to
	\begin{linenomath}
	\begin{equation}
		\begin{aligned}\label{eq:g_ab_fs_diag}
			g_{f f} &= \frac{1}{3} \pi^2 \tobs^2 \,,\\
			g_{\dot{f} \dot{f}} &= \frac{1}{180} \pi^2 \tobs^4 \,,\\
			g_{n_x n_x} &= 2 \pi^2 f^2 r_\text{E}^2 \,,\\
			g_{n_y n_y} &= 2 \pi^2 f^2 r_\text{E}^2 \,.
			\end{aligned}
		\end{equation}
	\end{linenomath}

\extraspace
\subsection{Semicoherent power test statistic $S$}

	The coherent power $P_n$ in Eq.~\eqref{eq:cohpower} provides a good statistical basis to find pulsations (meaning $\gamma_n$ nonzero) but is inefficient to compute.  Hence, the first two stages of our searches use the ``semicoherent'' Fourier power $S_n$. Its definition is similar to $P_n$ except that photons are only combined if their arrival time difference is smaller than a coherence time, $\tcoh \ll \tobs$. This makes it less expensive to compute (but also less sensitive).  The coherence time in a typical search in the first stage is $\tcoh = 2^{21} \, \text{s} \approx 24 \, \text{d}$, in the second stage it is $\tcoh = 2^{22} \, \text{s} \approx 48 \, \text{d}$, and the observation span $\tobs$ (i.e. the operation time of the \ac{LAT}) is more than $10$ years.

	For convenience the statistic $S_n$ differs from $P_n$ in one other way: we omit the diagonal $j = k$ terms in the sum.  This ensures that in the no-signal ($p=0$) case the expected value of $S_n$ vanishes, with
	\begin{linenomath}
	\begin{equation} \label{eq:semipower}
		S_n(\boldsymbol{\lambda}) = \frac{1}{\bar{\kappa}} \sum_{\ j \ne k} w_j w_k {\rm e}^{-in[\Phi(t_j,\pvec) - \Phi(t_k,\pvec)]} \hat{W}_{\tcoh}(\tau_{jk}) \,.
	\end{equation}
	\end{linenomath}
	The rectangular window function restricts the sum to photons in which the arrival time difference $\tau_{jk} = t_j - t_k$ (or ``lag'') is not larger than $\tcoh$:
	\begin{linenomath}
	\begin{equation}
		\hat{W}_{\tcoh}(\tau) = 
		\begin{cases}
		1 &\mbox{for } |\tau| \le \tcoh / 2 \,,\\
		0 &\mbox{otherwise} \,.
		\end{cases}
	\end{equation}
	\end{linenomath}
	The semicoherent normalization constant is chosen to be
	\begin{linenomath}
	\begin{equation} \label{eq:snvar}
		\bar{\kappa} = \sqrt{ \sum_{j \ne k} w_j^2 w_k^2 \hat{W}_{\tcoh}^2 (\tau_{j k}) } \,,
	\end{equation}
	\end{linenomath}
	which ensures that in the no-signal ($p=0$) case $S_n$ has unit variance \citep{clark2017}.

	To characterize this detection statistic, we calculate the expectation value and variance with the calculational framework of Appendix~\ref{s:expectedvalues}, obtaining
	\begin{linenomath}
	\begin{align}
		\operatorname{E}_p[S_n] &=  \frac{1}{\bar{\kappa}} p^2 |\gamma_n|^2 \sum_{j \ne k} w^2_j w^2_k \hat{W}_{\tcoh}(\tau_{jk}) \,, \\
		\operatorname{Var}_0[S_n] &= 1 \,.
	\end{align}
	\end{linenomath}
	The expectation value is the same as the second term of $P_n$ in Eq.~(\ref{eq:shortexpected}), except that the sum is restricted to the lag window. In fact, the formulae above hold for any choice of window function.

	The \ac{SNR} for the semicoherent Fourier power $S_n$ is simplified by assuming a rectangular window function (which equals its square). This gives
	\begin{linenomath}
	\begin{equation} \label{eq:scoh_snr_p}
		\begin{aligned}
			\theta_{S_n}^2 = \frac{\operatorname{E}_p[S_n] - \operatorname{E}_0[S_n]}{\sqrt{\operatorname{Var}_0[S_n]}}
			&= p^2 |\gamma_n|^2 \sqrt{\sum_{j \ne k} w_j^2 w_k^2 \hat{W}_{\tcoh} (\tau_{j k}) } \\
			&= p^2 |\gamma_n|^2 \mu \sqrt{\tcoh \tobs} \,.
		\end{aligned}
	\end{equation}
	\end{linenomath}
	The second line adopts the definition of $\mu$ given after Eq.~(\ref{eq:coh_snr_p}) and makes the same assumptions of steady photon flux and large photon number.

	In practice, how large are these detection statistics?  A typical gamma-ray pulsar might have a pulsed flux for which $|\gamma_1|^2 \approx 0.2$ and a $70\%$ fraction of pulsed photons for which $p^2 \approx 0.5$.  The weighted flux of source photons detected might be $\sum_j w_j^2\approx 500$ over $T_{\text{obs}}=10$\,yr, implying a rate $\mu \approx 50$\,yr$^{-1}$.  With $T_{\text{coh}} = 24$\,d, this leads to coherent and incoherent \acp{SNR} of order $\theta_{P_1}^2 \approx 50$ and $\theta_{S_1}^2 \approx 4$, significant at the $50\sigma$ and $4\sigma$ levels, respectively.

\extraspace
\subsubsection{Loss of $S$ from parameter mismatch}

	We now turn to the metric for the semicoherent statistic.  To compute the mismatch for the semicoherent detection statistic $S_n$, with the same assumptions as above, we can replace the sums with integrals, obtaining
	\begin{linenomath}
	\begin{equation} \label{eq:mismatch_scoh}
		\begin{aligned}
			\bar{m}(\pvec, \pvec_\text{psr}) &= 1 - \frac{\theta_{S_n}^2(\pvec)}{\theta_{S_n}^2(\pvec_\text{psr})} \\
			&= 1 - \prescript{}{\tobs \!\!\!\!}{\biggl\langle} {\rm e}^{-in \Delta\Phi(t')}  \prescript{}{\tcoh \!\!\!}{\Bigl\langle} {\rm e}^{in\Delta\Phi(t'')} \Bigr\rangle (t') \biggr\rangle(t_0) \,.
		\end{aligned}
	\end{equation}
	\end{linenomath}
	Note that the inner integral in the second line can include times outside the observation span $t'' \in [t_0-\tobs/2, t_0+\tobs/2]$, going down to $t''=t_0 - \tobs/2 - \tcoh/2$ or up to $t''=t_0 + \tobs/2 + \tcoh/2$.  In such cases the integrand should be set to zero and normalized so that $\langle 1 \rangle = 1$.

\extraspace
\subsubsection{Parameter space metric $\bar{g}_{ab}$}

	We now evaluate these mismatches to lowest order, obtaining a \textit{distance metric} on the parameter space.  We evaluate the integrals in Eq.~\eqref{eq:mismatch_scoh} naively, without setting the integrands to zero outside of the ``valid data range''.  This gives rise to terms (complex or linear in $\diff\lambda^a$) that are not present in the exact expression.  We assume that $\tcoh \ll \tobs$ (typically $\tcoh = 24\,{\rm d}$ and $\tobs > 10\,{\rm yr}$).  In that case, these terms are small, and we discard them.

	The partial derivatives with respect to $\lambda^a \in \{f,\dot{f},n_x,n_y\}$, under the assumption that $\tcoh \ll 1\,{\rm yr} \ll \tobs$, can be approximated as
	\begin{linenomath}
	\begin{subequations} \label{eq:assum_iso}
		\begin{align}
			\partial_a \Phi &\approx \prescript{}{\tcoh \!\!}\langle \partial_a \Phi \rangle(t) \,,\\
			\partial_a \partial_b \Phi &\approx \prescript{}{\tcoh \!\!}\langle \partial_a \partial_b \Phi \rangle(t) \,,
		\end{align}
	\end{subequations}
	\end{linenomath}
	as \cite{pletsch2014} did.  (Here and in what follows, for readability, the time dependence of phase derivatives such as $\partial_a \Phi$ is not shown explicitly.)

	With these assumptions the semicoherent mismatch Eq.~\eqref{eq:mismatch_scoh} can be approximated by the semicoherent metric
	\begin{linenomath}
	\begin{equation} \label{eq:def_scohmetric}
		\bar{m}(\pvec, \pvec_\text{psr}) \approx n^2 \bar{g}_{ab} \diff\lambda^a \diff\lambda^b + \mathcal{O}(\diff\lambda^3) \,,
	\end{equation}
	\end{linenomath}
	where $\diff \lambda^a = \lambda^a - \lambda^a_\text{psr}$ as earlier.  Note that Eq.~\eqref{eq:def_scohmetric} has the same form as the coherent mismatch in Eq.~\eqref{eq:def_cohmetric}.

	The metric components are
	\begin{linenomath}
	\begin{align}
		\bar{g}_{ab} &= \prescript{}{\tobs \!\!\!\!}{\Bigl\langle} \prescript{}{\tcoh \!\!}\langle \partial_a \Phi \partial_b \Phi \rangle(t') - \prescript{}{\tcoh \!\!}\langle \partial_a \Phi \rangle(t') \prescript{}{\tcoh \!\!}\langle \partial_b \Phi \rangle(t') \Bigr\rangle(t_0) \,, \nonumber \\
		&= \frac{1}{2} \prescript{}{\tobs \!\!\!\!}{\Bigl\langle} \tilde{g}_{ab}(t') \Bigr\rangle(t_0) \,,
	\end{align}
	\end{linenomath}
	where we have introduced
	\begin{linenomath}
	\begin{equation} \label{eq:gtilde}
		\tilde{g}_{ab}(t') = \prescript{}{\tcoh \!\!}\langle \partial_a \Phi \partial_b \Phi \rangle(t') - \prescript{}{\tcoh \!\!}\langle \partial_a \Phi \rangle(t') \prescript{}{\tcoh \!\!}\langle \partial_b \Phi \rangle(t') \,,
	\end{equation}
	\end{linenomath}
	which is exactly the coherent metric given in Eq.~\eqref{eq:cohmetrcomp}, but with $\tobs$ replaced by $\tcoh \ll 1\,{\rm yr}$ and $t_0$ replaced by $t'$.  Thus, terms of $\mathcal{O}(1/\Omega_\text{E}\tcoh)$, similar to those appearing in Eq.~\eqref{eq:g_ab_fs_d}, cannot be neglected.

\extraspace
\subsubsection{Evaluation of $\bar{g}_{ab}$ for isolated pulsars} \label{ss:semicoherentisolated}

	The nonvanishing semicoherent metric components are
	\begin{linenomath}
	\begin{align}
		\bar{g}_{f f} =& \frac{1}{6} \pi^2 \tcoh^2 \left[1 + \mathcal{O}(r_{\rm E}/\tobs)\right] \,,\nonumber\\
		\bar{g}_{f \dot{f}} =& \frac{1}{6} \pi^2 \tcoh^2 \tobs \left[ \frac{(t_0 - \tref)}{\tobs} + \mathcal{O}(r_{\rm E}/\tobs) \right]\,,\nonumber\\
		\bar{g}_{\dot{f} \dot{f}} =& \frac{1}{72} \pi^2 \tcoh^2 \tobs^2 \left(1 + 12 \frac{(t_0 - \tref)^2}{\tobs^2} + \frac{\tcoh^2}{5 \tobs^2} \right) \,, \label{eq:semicoherentisolated}\\
		\bar{g}_{n_x n_x} =& \pi^2 f^2 r_\text{E}^2 \left[1 - \frac{4}{\Omega_\text{E}^2 \tcoh^2} \sin^2\!\!\left(\frac{\Omega_\text{E}\tcoh}{2}\right) + \mathcal{O}(1/\Omega_{\rm E}\tobs)\right] \,, \nonumber\\
		\bar{g}_{n_y n_y} =& \pi^2 f^2 r_\text{E}^2 \left[1 - \frac{4}{\Omega_\text{E}^2 \tcoh^2} \sin^2\!\!\left(\frac{\Omega_\text{E}\tcoh}{2}\right) + \mathcal{O}(1/\Omega_{\rm E}\tobs)\right] \,.\nonumber
	\end{align}
	\end{linenomath}
	The semicoherent metric $\bar{g}$ is diagonal for $t_0 = \tref$, as was the case for the coherent metric $g$ in Eq.~\eqref{eq:g_ab_fs_diag}. Note that $g_{n_x n_x}$ and $g_{n_y n_y}$ are not equal because the neglected terms of $\mathcal{O}(1/\Omega_{\rm E}\tobs)$ have opposite signs.

	The metric component $\bar{g}_{\dot{f} \dot{f}}$ differs from that given by \cite{pletsch2014}, but our results are identical in the limit of a large number of photons $N$ homogeneously distributed over the observation span.  This is the case, since we assumed it in deriving Eqs.~\eqref{eq:mismatch_coh} and \eqref{eq:mismatch_scoh}.

	Comparison of Eqs.~\eqref{eq:g_ab_fs_d} and \eqref{eq:semicoherentisolated} illustrates the benefits of the multistage search process described in Section~\ref{s:multistage}.  For grids with the same mismatch $m = \bar{m}$, the ratio between the density of the coherent grid and the semicoherent grid would be
	\begin{linenomath}
	\begin{equation} \label{eq:density_ratio}
		\frac{\text{coherent grid density}}{\text{semicoherent grid density}} = \sqrt{\frac{\det g}{\det \bar{g}}} = \frac{48 \tobs^2}{\sqrt{5}\Omega_{\rm E}^2\tcoh^4} \,.
	\end{equation}
	\end{linenomath}
	For the timescales $\tcoh$ and $\tobs$ given above, the ratio is $\sim 10^6$.  This is why the semicoherent search stage is beneficial.

\extraspace
\subsection{Multiple harmonic test statistic H}

	In the last and most sensitive stage of the multistage search, we adopt the widely used statistic
	\begin{linenomath}
	\begin{equation}
		H(\boldsymbol{\lambda}) = \max_{1 \le M \le M_\text{max}} \left( 4 - 4 M + \sum\limits_{n = 1}^{M} P_n(\boldsymbol{\lambda}) \right) \,,
	\end{equation}
	\end{linenomath}
	which incoherently sums the coherent power from up to the first $M_\text{max}$ harmonics in the pulse profile.  The $H$ statistic provides a sensitive test for unknown (generic) pulse profiles.  The original simulations by \cite{dejager1989} recommended $M_\text{max} = 20$, and to assess the false-alarm probability, they carried out a numerical study of the distribution of $H$ in pure noise.

	Later results by \cite{kerr2011} show that the single-trial probability $\rho$ of exceeding a value $H_\text{threshold}$ in pure noise is well modeled by $\rho \approx \exp(-0.398 \, H_\text{threshold})$ if the number of harmonics $M_\text{max}$ is very large.  Obviously, if $M_\text{max}$ is reduced, then the single-trial probabilities are smaller than this, so $\exp(-0.4 \, H_\text{threshold})$ is a reliable upper bound.

	To avoid overfitting, we generally use smaller limits $M_\text{max} = 3,4, \text{ or } 5$ on the number of harmonics.  Typical \replaced{blind}{partially informed} search gamma-ray pulsar detections have $H$ values in the hundreds, corresponding to single-trial $\rho$ values that must lie below $10^{-30}$.

	Normally, the last search stage is not computationally limited.  Hence, we use a grid fine enough to secure power in the higher harmonics, while overcovering the search space for power in the lower harmonics. In practice, the grid is built using the coherent metric presented in Section \ref{s:cohmetric} with $n = M_{\rm max}$.

\extraspace
\subsection{Searches for isolated pulsars}

	\replaced{Blind}{Partially informed} searches for isolated pulsars within gamma-ray data recorded by the \ac{LAT} have been very successful \citep[see, e.g.,][]{clark2017}.  The key ingredients are the utilization of the powerful volunteer-distributed computing system \EatH{} \citep{allen2013} and searches that use these computing resources as efficiently as possible.

	Most of the tools for constructing efficient searches have been presented in the earlier sections.  To discard unpromising regions in parameter space, the multistage approach is used as described in Section~\ref{s:multistage}.  For the first and computationally most crucial search stage, efficient grids covering the parameters $\dot{f}$, $n_x$, $n_y$ are built based on the distance metric, and $f$ is searched using \ac{FFT} algorithms \citep{frigo2005}.  In later search stages, $f$ is also gridded with the metric, but it is not efficient to use \acp{FFT} on the small ranges in $f$ around the few most significant candidates from the semicoherent search stage.

%%%%%%%%%%%%%%%%%%%%%%%%%%%%%%%%%%%%%%%%%%%
\extraspace
\section{Search Method: Circular Binary Orbits} \label{s:methodcirc}

	The main problem in \replaced{blind}{partially informed gamma-ray} searches for pulsars is that the phase model from Eq.~\eqref{eq:phasemodel} depends on the (photon emission) time at the pulsar, while a gamma-ray detector records the time of arrival at the telescope.  For binary pulsars the largest corrections to shift between these two times arise from the line-of-sight motion of the \Fermi{} satellite around Earth and Sun $r_{z, \text{sky}}(t)$ and of the pulsar around its companion $r_{z, \text{cir}}(t_\text{psr})$.

	The line-of-sight motion of a binary pulsar in a circular orbit can be described via three parameters, which are usually taken to be the orbital frequency $\Omorb$, the projected semimajor axis $x$ in seconds, and the epoch of ascending node $\tasc$.  With these, the two times are related by
	\begin{linenomath}
	\begin{equation}
		t_\text{psr} + r_{z, \text{cir}}(t_\text{psr}) = t + r_{z, \text{sky}}(t) \,,
	\end{equation}
	\end{linenomath}
	where the corrections, also called R\o mer delays, are expressed in seconds.

	The simplest expression of the pulsar's orbital line-of-sight motion $r_{z,\text{cir}}$ depends on the time measured at the pulsar $t_\text{psr}$.  In many cases, this time may be replaced with the detector time because
	\begin{linenomath}
	\begin{equation}
		r_{z, \text{cir}}(t_\text{psr}) = r_{z, \text{cir}}(t) \left[1 + \mathcal{O} \left( x \Omorb \right) \right] \,,
	\end{equation}
	\end{linenomath}
	and the quantity $x \Omorb \ll 1$.  In such cases
	\begin{linenomath}
	\begin{equation} \label{eq:timeapproximation}
		t_\text{psr} \approx t + r_{z, \text{sky}}(t) - r_{z, \text{cir}}(t) \,.
	\end{equation}
	\end{linenomath}
	This holds for most black widow and some redback systems with projected semimajor axes on the order of a few light-seconds \citep[see, e.g., the \ac{ATNF} Pulsar Catalogue\footnote{\href{http://www.atnf.csiro.au/research/pulsar/psrcat}{http://www.atnf.csiro.au/research/pulsar/psrcat}} by][]{manchester2005}.  In all cases, it is accurate enough to compute the metric, and in many cases accurate enough for maintaining phase coherence in a search.

	The R\o mer delay can be expressed in terms of the three orbital parameters as
	\begin{linenomath}
	\begin{equation} \label{eq:r_cir}
		r_{z, \text{cir}}(t) = x \sin [\Omorb (t - \tasc)] \,.
	\end{equation}
	\end{linenomath}
	Here the orbital frequency $\Omorb$ is connected to the orbital period $\Porb$ via $\Porb = 2 \pi / \Omorb $.

	In gamma-ray searches, in addition to the R\o mer delay, we also have to correct for other effects like the Shapiro and Einstein delays.  In contrast to radio observations, we do {\it not} have to account for the frequency-dependent dispersion caused by the interstellar medium (ISM) because gamma rays are well above the plasma frequency of the ISM.

	All of these effects are described by \cite{lorimer2004} and \cite{edwards2006}. While these corrections must be included in gamma-ray searches, only the largest effects need to be included in the phase model for the derivation of a distance metric approximation.

%%%%%%%%%%%%%%%%%%%%%%%%%%%%%%%%%%%%%%%%%%%
\extraspace
\subsection{Parameter space metrics} \label{s:metric-circ}

	In order to compute the metric, a simplified phase model can be used that accounts for the corrections \eqref{eq:r_sky} and \eqref{eq:r_cir}:
	\begin{linenomath}
	\begin{align} \label{eq:pm_circ}
		\Phi (t, \pvec) =& 2 \pi f (t - \tref) + \pi \dot{f} (t - \tref)^2 \nonumber\\
		&+ 2 \pi f r_\text{E} \left[n_\text{x} \cos \left(\Omega_\text{E} t + \varphi_\text{ref}\right) + n_\text{y} \sin \left(\Omega_\text{E} t + \varphi_\text{ref}\right) \right] \nonumber\\
		&- 2 \pi f x \sin [\Omorb (t - \tasc)] \,.
	\end{align}
	\end{linenomath}
	Here the search parameters are $\pvec = \{f, \dot{f}, n_x, n_y, \Omorb, x, \tasc\}$ and the terms correcting the arrival times $t$ have been neglected for the $\dot{f}$ summand.  This phase model is not sufficient for searches because it would not maintain phase coherence with a true pulsar signal.  However, it is sufficient to describe how varying the signal parameters leads to loss of \ac{SNR}.

	The dominant components of the coherent metric for the orbital parameters are
	\begin{linenomath}
	\begin{align}
		g_{x x} &= 2 \pi^2 f^2 [1 + \mathcal{O}(1/\Omorb\tobs)] \,, \nonumber\\
		g_{\Omorb \Omorb} &= \frac{1}{6} \pi^2 f^2 x^2 \tobs^2 \left[ 1 + 12 \frac{(t_o - \tasc)^2}{\tobs^2} + \mathcal{O}(1/\Omorb\tobs) \right] \,, \nonumber\\
		g_{\tasc \tasc} &= 2 \pi^2 f^2 x^2 \Omorb^2 [1  + \mathcal{O}(1/\Omorb\tobs)] \,, \label{eq:g_ab_orb} \\
		g_{\Omorb \tasc} &= -2 \pi^2 f^2 x^2 \Omorb \tobs \left[ \frac{(t_o - \tasc)}{\tobs} + \mathcal{O}(1/\Omorb\tobs) \right] \,,\nonumber
	\end{align}
	\end{linenomath}
	where we have assumed that the integration time span $\tobs$ is much larger than the orbital period $\Porb$.  Compared to the diagonal terms, as done in the text below Eq.~\eqref{eq:g_ab_fs_d}, all other components are of $\mathcal{O}(1/\Omorb\tobs)$.

	The off-diagonal component $g_{\Omorb \tasc}$ is vanishingly small if the epoch of the ascending node is close to the middle of the gamma-ray data set, $\tasc \approx t_0$.  In principle, $\tasc$ can be shifted forward or backward by an integer number $\cal{N}$ of orbital periods $\Porb$ to achieve this.  However, when $\tasc$ is constrained, for example, by optical observations, this is undesirable because it introduces uncertainties in the shifted value of $\tasc$ that grow linearly with $\cal{N}$.

	Even if $\tasc \not \approx t_0$, our current searches ignore the off-diagonal term in the metric.  The only negative consequence is that the grids are more closely spaced than needed, which reduces the efficiency of the search.

	If we include the additional orbital parameters, the semicoherent mismatch \eqref{eq:mismatch_scoh} can still be written in metric form,
	\begin{linenomath}
	\begin{equation}
		\bar{m}(\pvec, \pvec_\text{psr}) \approx n^2 \bar{g}_{ab} \diff \lambda^a \diff \lambda^b + \mathcal{O}(\diff \lambda^3) \,.
	\end{equation}
	\end{linenomath}
	However, the assumptions made previously in Eq.~\eqref{eq:assum_iso} to calculate this only hold for the ``isolated pulsar'' parameter space coordinates  $\pvec_\text{iso}=\{f, \dot{f}, n_x, n_y\}$.  They do not hold for the additional orbital parameters $\pvec_\text{orb} = \{\Omorb, x, \tasc\}$.

	If $\lambda^a \in \pvec_\text{orb}$ is an orbital parameter and $\Porb \ll \tcoh$ (typical coherence time $\tcoh \approx 24\,\text{d}$), the approximations
	\begin{linenomath}
	\begin{subequations} \label{eq:vanishing}
		\begin{align}
			\prescript{}{\tcoh \!\!}\langle \partial_a \Phi \rangle(t) &\approx 0 \,,\\
			\prescript{}{\tcoh \!\!}\langle \partial_a \partial_b \Phi \rangle(t) &\approx 0 \,,\\
			\prescript{}{\tobs \!\!\!\!}{\Big\langle} \prescript{}{\tcoh \!\!}\langle \partial_a \Phi \partial_b \Phi \rangle(t') \Big\rangle(t_0) &\approx \prescript{}{\tobs \!\!}\langle \partial_a \Phi \partial_b \Phi \rangle(t_0)
		\end{align}
	\end{subequations}
	\end{linenomath}
	are valid.  By this, we mean that the ratio of the resulting metric to the correct metric is $1+ \mathcal{O}(\Porb/\tcoh)$.

	With these assumptions the semicoherent metric $\bar{g}_{ab}$ is composed of three types of components.  For the first type, the parameters $\lambda^a,\lambda^b \in \pvec_\text{orb}$ are orbital.  For these components,
	\begin{linenomath}
	\begin{equation} \label{eq:con_g_gbar_orb}
		\bar{g}_{ab} = \prescript{}{\tobs \!\!}\langle \partial_a \Phi \partial_b \Phi \rangle(t_0) = g_{ab} \,,
	\end{equation}
	\end{linenomath}
	giving the coherent result from Eq.~\eqref{eq:g_ab_orb}.

	For the second type, the parameters $\lambda^a,\lambda^b \in \pvec_\text{iso}$ are isolated.  For these components
	\begin{linenomath}
	\begin{align}
		\bar{g}_{ab} & =  \frac{1}{2} \prescript{}{\tobs \!\!\!\!\!}{\Big\langle} \prescript{}{\tcoh \!\!}\langle \partial_a \Phi \partial_b \Phi \rangle(t') - \prescript{}{\tcoh \!\!}\langle \partial_a \Phi \rangle(t') \prescript{}{\tcoh \!\!}\langle \partial_b \Phi \rangle(t') \Big\rangle(t_0) \nonumber \\
		& = \frac{1}{2} \prescript{}{\tobs \!\!}\langle \tilde{g}_{ab}(t') \rangle(t_0) \,,
	\end{align}
	\end{linenomath}
	which is the semicoherent result found in Eq.~(\ref{eq:semicoherentisolated}).

	For the third type, one of $a$ or $b$ is in $\pvec_\text{orb}$ and the other is in $\pvec_\text{iso}$.  One obtains the same equation as for the second type.  This vanishes by virtue of Eq.~(\ref{eq:vanishing}) and because $\prescript{}{\tcoh \!\!}\langle \partial_a \Phi \partial_b \Phi \rangle(t')$ is of order $\mathcal{O}(\Porb /\tcoh)$.

	In short, the nonvanishing semicoherent metric components reduce to earlier results.  For the orbital parameters, they are the same as the coherent metric components.  For the isolated (spin and celestial) parameters, they are the same as the semicoherent metric components for an isolated pulsar.  To reiterate, the nonvanishing semicoherent orbital metric components are
	\begin{linenomath}
	\begin{equation} \label{eq:gbar_ab_orb}
		\begin{aligned}
			\bar{g}_{x x} &=  2 \pi^2 f^2 \,,\\
			\bar{g}_{\Omorb \Omorb} &=  \frac{1}{6} \pi^2 f^2 x^2 \tobs^2 \left(1 + 12\frac{(t_0 - \tasc)^2}{\tobs^2} \right) \,,\\
			\bar{g}_{\tasc \tasc} &=  2 \pi^2 f^2 x^2 \Omorb^2 \,,\\
			\bar{g}_{\Omorb \tasc} &= -2 \pi^2 f^2 x^2 \Omorb (t_0 - \tasc) \,.
		\end{aligned}
	\end{equation}
	\end{linenomath}
	As before, for epoch of ascending node close to the middle of the dataset, i.e. $\tasc \approx t_0$, the semicoherent metric is diagonal.

	At the end of Section~\ref{ss:semicoherentisolated}, we discussed the relative densities of the coherent and semicoherent grids for isolated sources.  Now we have added three additional (orbital) dimensions to the parameter space.  Because the metric factors into a product of a metric on the orbital parameters and a metric on the isolated parameters, the grid may also be constructed as a product of the grids on the corresponding subspaces.  For the isolated parameters, the ratio between the density of the coherent grid and the semicoherent grid is the same as for the search for isolated pulsars.  For the orbital parameters, the number of grid points needed is the same as in the coherent case.  Hence, the ratio of grid densities is the same as in Eq.~\eqref{eq:density_ratio}.

	\begin{figure*}
		\centering
		\includegraphics[width=\textwidth]{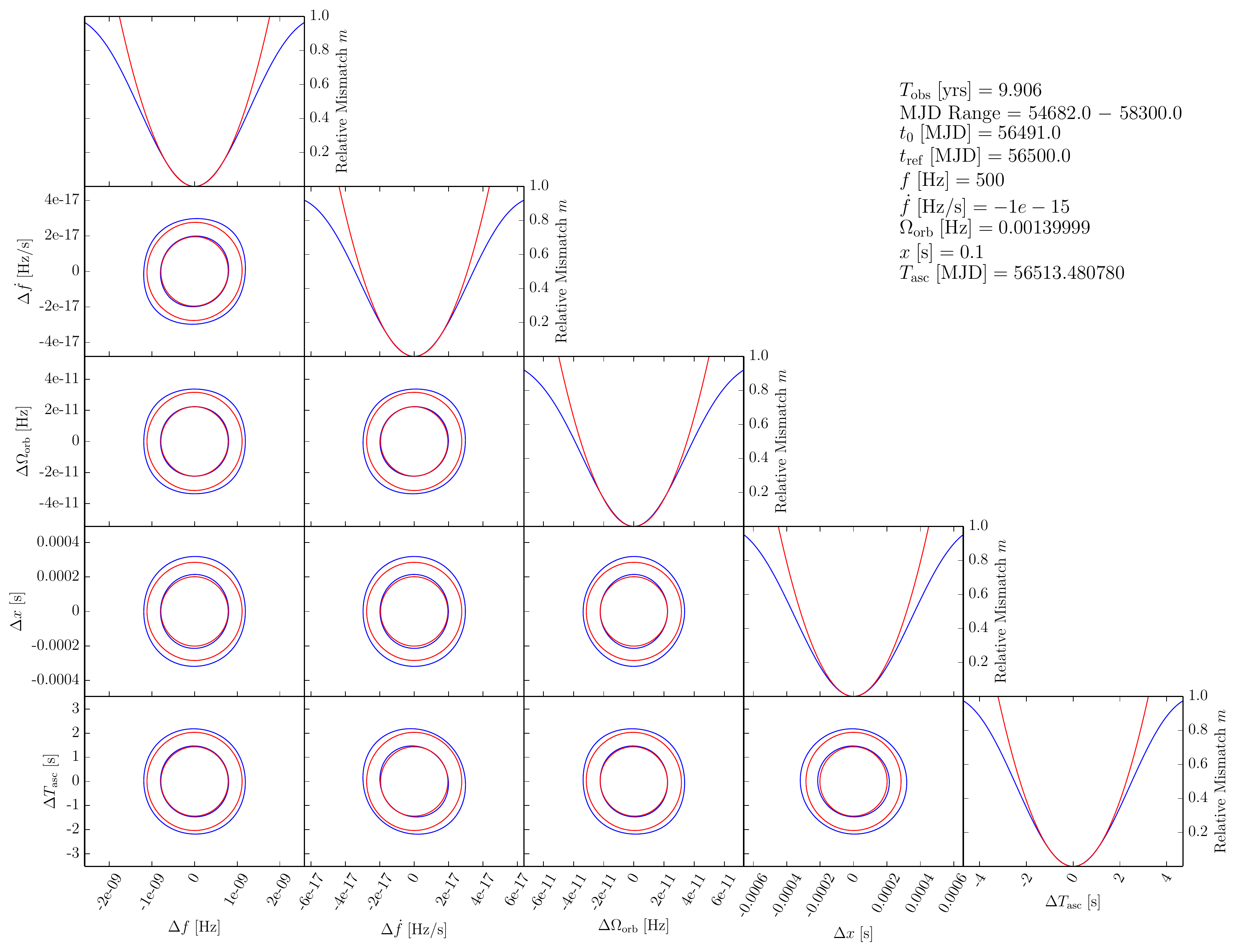}
		\caption{\label{f:triplot_circ} Comparison of the coherent metric approximation to the actual mismatch, for parameters of a simulated circular orbit binary pulsar in \circsrc{}.  Blue contours show the actual mismatch and red contours the metric approximation, at $m = 0.2$ and $0.4$.  As is generally the case \citep{allen2019}, the metric contours  are conservative and lie inside the actual mismatch contours.}
	\end{figure*}

	In Figure~\ref{f:triplot_circ} the mismatch and its coherent metric approximation are compared for small parameter offsets, for a realistic simulated pulsar.  The corresponding plot for the semicoherent mismatch looks very similar but has different $f$- and $\dot{f}$-scales.  The mismatch and its metric approximation agree well for mismatch $m \le 0.4$.  This is a typical value for a search: in Appendix~\ref{s:optimalmismatch}, we show that maximum sensitivity for a given computing resource is obtained for an average mismatch $\hat{m} = 0.383$ (see Table~\ref{t:optimalmismatch}).

	The celestial parameters are not shown in Figure~\ref{f:triplot_circ}; for spider pulsars they are usually known to high precision from optical observations \citep[e.g., from the Gaia DR2 Catalog; ][]{gaia2018}, so no grid is required.  For other pulsars where the sky position is less constrained, a grid may be needed.

	The search ranges for the orbital parameters are very large, and without further knowledge a \replaced{full blind}{partially informed} search is not possible.  On the other hand, some searches are possible if the pulsar's companion is visible in optical/X-ray observations, which constrains the search parameters.  In the next section, we discuss a gamma-ray pulsar search design for \circsrc{}, which is thought to be an \ac{MSP} in a circular orbit binary \citep{romani2014,parkinson2016}.

%%%%%%%%%%%%%%%%%%%%%%%%%%%%%%%%%%%%%%%%%%%
\extraspace
\subsection{Search design for circular binary} \label{s:J1653}

	This section shows how to reduce the binary pulsar search parameter space by exploiting orbital constraints from the companions.

	We use the gamma-ray source \circsrc{}, which is predicted to be a spider pulsar \citep{romani2014,kong2014}, as an example.  \added{In previous \ac{LAT} source catalogs the gamma-ray source is named 3FGL\,J1653.6$-$0158 and 2FGL\,J1653.6$-$0159.}  It was ranked second in \citeauthor{parkinson2016}'s list (published $2016$) of the most significant \ac{3FGL} unassociated sources predicted to be pulsars.  The list also classifies it as a likely \ac{MSP}.  The gamma-ray source \circsrc{} shows typical pulsar properties: a time-stable photon flux and a spectrum described by an exponential cutoff power law.

	The search ranges in spin frequency $f$ and spin-down parameter $\dot{f}$ are guided by the known pulsar population and computational constraints. The search range is divided into \acp{YP}, with lower frequencies ($f < 44 \, \text{Hz}$), and \acp{MSP}, with higher frequencies ($44 \, \text{Hz} < f < 1500 \, \text{Hz}$)\footnote{The high-frequency limit is around the second harmonic of the fastest known pulsar.}.  Correspondingly, the spin-down lies between $0$ and $-10^{-10} \, \text{Hz} \, \text{s}^{-1} \,,$ for \acp{YP} and between $0$ and $-10^{-13} \, \text{Hz} \, \text{s}^{-1}$ for \acp{MSP}.

	The constraints for $f$ and $\dot{f}$ define a region in parameter space that has to be searched.  The frequency dimension can be efficiently scanned using the \ac{FFT} algorithm \citep{frigo2005} as described by \cite{pletsch2014} and \cite{clark2016,clark2017} for isolated pulsars.  The $\dot{f}$-dimension can be covered by an uniformly spaced lattice.  Special treatment for these parameters is possible: since their metric components are independent of the other parameters, so is the spacing.

	In practice, the \acp{FFT} are computed in frequency intervals of bandwidth $f_\text{BW} = 8$\,Hz.  These have $f_\text{BW} \tcoh$ frequency grid points, with frequency spacing $1/\tcoh$.  In the semicoherent stage, for two points separated by half the grid spacing, this gives a worst-case metric mismatch $m=\pi^2/24 \approx 0.411$.  (As discussed in Appendix~\ref{s:optimalmismatch} following Eq.~\eqref{eq:mbar}, this can be reduced by interpolation to a worst-case value of $m=0.14$, at no significant cost.) Thus, for one $f_\text{BW}$ interval, the computing cost is the product of the cost of a single \ac{FFT} multiplied by the number of parameter space grid points in the other dimensions.

	The sky position is tightly constrained because a likely optical and X-ray counterpart with significant light-curve modulation was found \citep{romani2014,kong2014,hui2015} and proposed to be an irradiated pulsar companion.  At the time, the best estimate for the position of the likely optical counterpart was from the \citetalias{monet2003} Catalog \citep{monet2003}.  Using this instead of the \ac{3FGL} position makes it possible to search $3\sigma$ ranges of the sky parameters with only one semicoherent sky grid point.  At high frequencies extra sky grid points are needed only in the follow-up stages.  The computing costs of these are negligible compared to the semicoherent stage.  The same optical source can now be identified in the Gaia DR2 Catalog \citep{gaia2018}; see Table~\ref{t:const1}.  For this, the uncertainty in sky position is small enough that even at $f = 1.5 \, \text{kHz}$ no extra sky points are needed.

	The orbital parameters $\Omega_\text{orb}$ and $\tasc$ are directly constrained by \cite{romani2014} using optical observations of the companion.  As shown in Table~\ref{t:const1}, they found a significant modulation at a period of $\Porb = 0.05194469 \pm 1.0 \times 10^{-7} \, \text{d}$, with epoch of ascending node $\tasc = 56513.48078 \pm 5.2 \times 10^{-4} \, \text{MJD}$.

	Additional observations allow the third orbital parameter, the projected semimajor axis of the pulsar $x = a_1 \sin i/c$ (in units of light travel time), to be constrained.  Here we denote the neutron star with subscript ``1'' and the companion with subscript ``2''.  Measurements of the companion's velocity amplitude $K_2 = 666.9 \pm 7.5 \, \text{km} \, \text{s}^{-1}$, together with the orbital period, imply that the pulsar mass function has the value
	\begin{linenomath}
	\begin{equation} \label{eq:massfunc_j1653}
		f(M_1, M_2) = \frac{\Porb K_2^3}{2 \pi G} = \frac{M_1 \sin^3 i}{(1 + q)^2} = 1.60 \pm 0.05 \, M_\odot \,,
	\end{equation}
	\end{linenomath}
	where the mass ratio is $q = M_2 / M_1$.  This implies that the neutron star has mass $M_1 > 1.60 \pm 0.05 \, M_\odot$.  Since redback companions \added{often} have masses $M_2 \lesssim 0.4 \, M_\odot$ \citep{roberts2013,strader2019}, \added{and black widow companions are even lighter,} \replaced{this in turn implies}{we assume} $q<0.25$.  \added{The extremely short orbital period supports this, since evolutionary models would suggest a black widow companion \citep{chen2013}.}  From Eq.~\eqref{eq:massfunc_j1653}, a mass ratio of $q=0.25$ allows neutron star masses up to $2.5 \, M_\odot$ for $i = 90\arcdeg$.  (This is reassuringly conservative, since the most massive known neutron star \citep{cromartie2020} has mass $2.14 \, M_\odot$.)  Combining the mass function with Kepler's third law $(a_1 + a_2)^3 = G(M_1 + M_2) (\Porb/2\pi)^2$ and the center-of-mass definition $a_1 M_1 = a_2 M_2$ gives
	\begin{linenomath}
	\begin{equation} \label{eq:x}
		x = \frac{q K_2 \Porb}{2 \pi c} \,.
	\end{equation}
	\end{linenomath}
	The upper limit for $q$ then implies an upper limit $x \lesssim 0.2 \, \text{s}$.

%------------------------------------------------------------------------------
	\begin{deluxetable}{ll}
		\tablewidth{0.99\columnwidth}
		\tablecaption{\label{t:const1} Parameters and constraints for \circsrc{} }
		\tablecolumns{2}
		\tablehead{
			\colhead{Parameter} &
			\colhead{Value}
		}
		\startdata
		Range of observational data (MJD) \dotfill  & $54682$ -- $58300$ \\[0.15em]
		Reference epoch (MJD)\dotfill  & $56500.0$   \\[-0.2em]
		\cutinhead{Initial companion location from \citetalias{monet2003} catalog}
		R.A., $\alpha$ (J2000.0)\dotfill & $16^{\rm h}53^{\rm m}38\fs07(10)$ \\[0.15em]
		Decl., $\delta$ (J2000.0)\dotfill    & $-01\arcdeg58\arcmin36\farcs7(2)$ \\[-0.2em]
		\cutinhead{Precise companion location from \citetalias{gaia2018} catalog}
		R.A., $\alpha$ (J2000.0)\dotfill & $16^{\rm h}53^{\rm m}38\fs05381(5)$ \\[0.15em]
		Decl., $\delta$ (J2000.0)\dotfill    & $-01\arcdeg58\arcmin36\farcs8930(5)$ \\[-0.2em]
		\cutinhead{Constraints from probable counterpart \citep{romani2014}}
		Ascending node epoch, $\tasc$ (MJD) \dotfill & $56513.48078 \pm 5.2\times10^{-4}$\\[0.15em]
		Companion velocity, $K_2$ ($\text{km} \,\, \text{s}^{-1}$) \dotfill & $666.9\pm 7.5$\\[0.15em]
		Orbital period, $\Porb$ (d) \dotfill & $0.05194469 \pm 1.0\times 10^{-7}$\\[0.15em]
		{\hfill equivalent to \hfill}\\
		Orbital frequency, $\Omorb$ ($10^{-3}$ Hz) \dotfill & $1.3999901 \pm 2.7\times10^{-6}$ \\[-0.2em]
		\cutinhead{Derived search range}
		Projected semimajor axis\tablenotemark{a}, $x$ (s) \dotfill &  $0$ -- $0.2$ \\[0.5em]
		\enddata
		\tablecomments{
			The JPL DE405 solar system ephemeris has been used, and times refer to TDB.}
		\tablenotetext{a}{Assuming a mass ratio of $q < 0.25$; see text following Eq.~\eqref{eq:massfunc_j1653}.}
	\end{deluxetable}
%------------------------------------------------------------------------------

	It is challenging to build a search grid that covers the three-dimensional orbital parameter space with as few points as possible.  This is because (as can be seen from the metric) the orbital parameter space is not flat, so a constant-spacing lattice is not optimal.  A solution to this is presented by \cite{fehrmann2014}, starting with ``stochastic search grids'' \citep{babak2008,harry2009}.  A stochastic grid is built by placing grid points with a random distribution that follows the expected distribution of metric distances, while ensuring a preset minimum distance between them.  The resulting grid is then optimized by nudging grid points toward regions where neighboring grid points have higher-than-average separation.  The resulting search grid is efficient and has a \textit{well-behaved} mismatch distribution, which simplifies the \ac{SNR} distribution in the absence of signals.

	The minimum number of grid points needed to cover the orbital parameter search space at mismatch $m$ can be estimated from the proper $3$-volume
	\begin{linenomath}
	\begin{equation} \label{eq:norb}
		N_\text{orb} \approx m^{-3/2} \int \sqrt{\det \bar{g}} \diff{\pvec_\text{orb}} \,.
	\end{equation}
	\end{linenomath}
	Here the integral is over the relevant range of orbital parameter space, $g$ denotes the orbital metric from Eq.~\eqref{eq:g_ab_orb}, and numerical factors of order unity related to the efficiency (technically ``thickness''; see Appendix~\ref{s:optimalmismatch}) of the grid lattice have been dropped.  To understand how this depends on parameters, note that the integral is proportional to
	\begin{linenomath}
	\begin{equation} \label{eq:norb_dep}
		N_\text{orb} \propto f^3 \tobs \left( x_\text{max}^3 - x_\text{min}^3 \right) \Omega_\text{orb}\Delta\Omega_\text{orb} \Delta\tasc \,,
	\end{equation}
	\end{linenomath}
	where the search range for $x$ is $[x_{\rm min},x_{\rm max}]$.  $\Delta\Omorb$ and $\Delta\tasc$ are the search ranges around the values of $\Omorb$ and $\tasc$ estimated from the optical modeling. Furthermore, we make the assumption that $\Delta\Omorb \ll \Omorb$. The strong dependency of $N_{\text{orb}}$ on $x_{\rm max}$ and $f$ means that searches for \acp{YP} (smaller $f$) in tight binary orbits (smaller $x_{\rm max}$) are computationally much cheaper than searches for \acp{MSP} in wide orbits.  The latter are only possible if the orbital constraints are very narrow.

	If the parameter space is small in a particular direction, this reduces the effective dimension of the parameter space and changes the formulae above. For example, denote the range of $x$ by $[x_{\rm min},x_{\rm max}]$. Now consider the case where $\Delta x = x_{\rm max} - x_{\rm min}$ is small enough that $g_{x x} \Delta x^2 \ll m$. Then, only a single grid point is needed in the $x$-direction, and Eq.~\eqref{eq:norb} must be replaced with a two-dimensional integral, and the exponent on $m$ must be replaced with $-1$. Since the orbital metric components in Eq.~\eqref{eq:gbar_ab_orb} depend on the parameters, for example, $g_{x x} = 2 \pi^2 f^2$, this reduction in dimension can take place for certain ranges of parameters (here small frequency $f$) and not for others.

	We can estimate the computing cost of a search for \circsrc{} by computing the number of grid points in parameter space.  We take $f \in [0,44]$\,Hz and $\dot{f} \in [-10^{-10},0]$\,Hz\,s$^{-1}$ for the \ac{YP} search and $f \in [44,1500]$\,Hz and $\dot{f} \in [-10^{-13},0]$\,Hz\,s$^{-1}$ for the \ac{MSP} search from early in this section.  The remaining parameter space search ranges are taken from Table~\ref{t:const1} (no grid is needed over sky location).  The frequency range is gridded in intervals of bandwidth $f_\text{BW} = 8 \, \text{Hz}$ as discussed earlier in this section.  The total computing cost is obtained by multiplying the cost of one FFT, the number of $\dot{f}$ grid points, and the number of orbital grid points (which depends on the $f$ interval) and then summing over the $f$ intervals.  Since the orbital grid depends on frequency, a new search grid is constructed for each frequency interval, using the metric at the maximum frequency of that interval.

\begin{figure}
		\centering
		\includegraphics[width=\columnwidth]{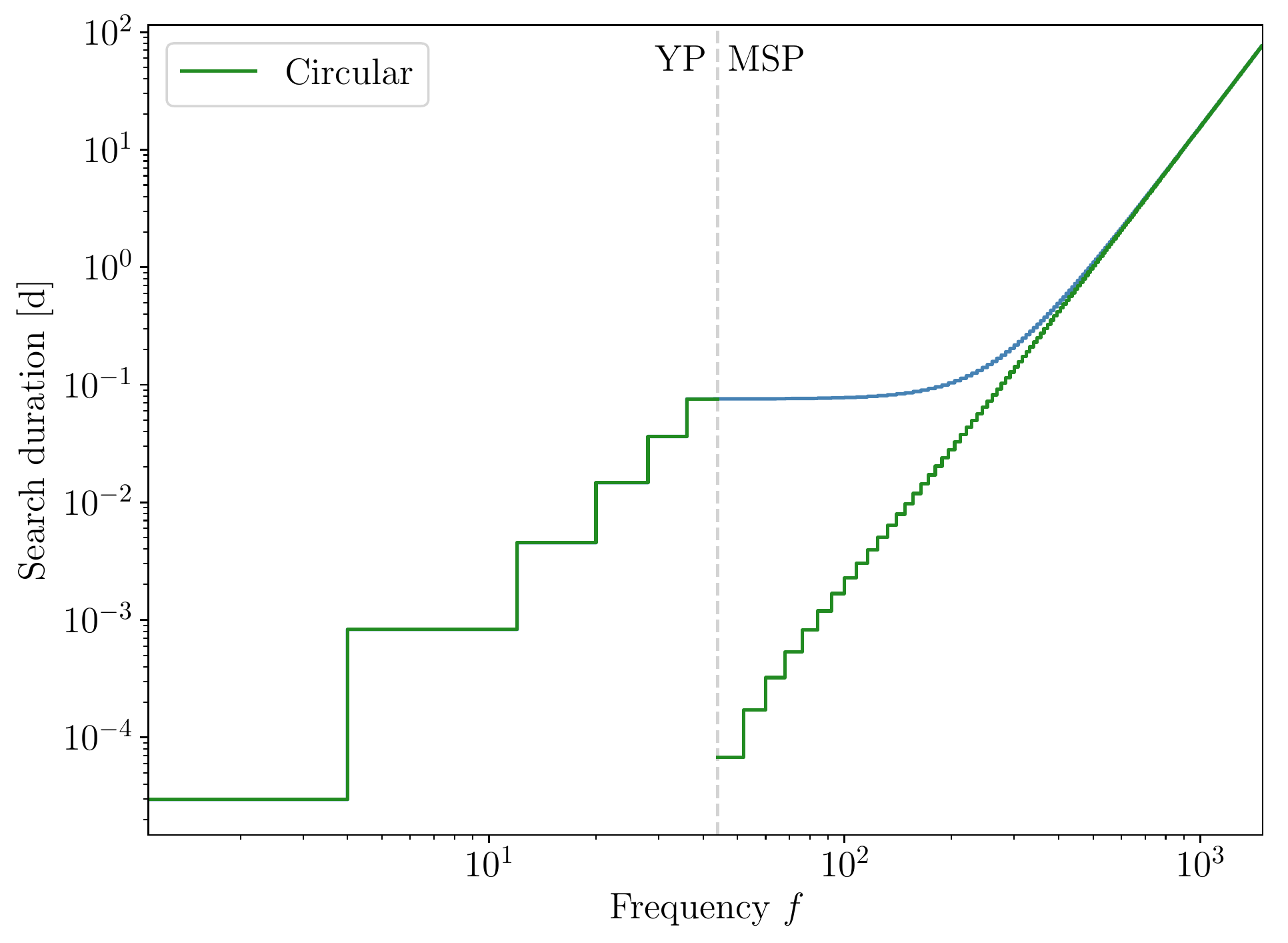}
		\caption{\label{f:fft_circ} Predicted days on \EatH{} needed to search \circsrc{}, assuming a circular orbit. The left green curve shows the cumulative duration of a \ac{YP} search from $0$\,Hz up to maximum frequency $f$. The right green curve shows the cumulative duration of an \ac{MSP} search from $44$\,Hz up to maximum frequency $f$. Their slopes are $\propto f^4$ because they are an integral over the number of orbital templates in Eq.~\eqref{eq:norb_dep}. The blue curve shows the sum: the cumulative duration of a combined YP and \ac{MSP} search.}
	\end{figure}

	A convenient way to express the computing cost is in terms of search duration on \EatH{}, where we assume that the project provides $25{,}000$ GPU-hr/week.  This is shown in Figure~\ref{f:fft_circ} as a function of the maximum frequency searched.  Searching up to $f = 1500$\,Hz requires less than $80\,\text{d}$.  Note that the search cost in one frequency step is proportional to the number of orbital grid points.  To search $3\sigma$ ranges in $\tasc$ and $\Omorb$ within a reasonable amount of time, either the maximum $f$ or $x$ needs to be reduced.

	We can also give a general estimate for the \ac{MSP} search duration.  Since the semimajor axis is typically not well constrained, we assume $x_{\rm min} = 0$.  We evaluate Eq.~\eqref{eq:norb_dep}, using Kep\-ler's third law to replace $x_{\rm max}$ with the corresponding maximum searched mass ratio $q_{\rm max}$, obtaining
	\begin{linenomath}
	\begin{equation}
		N_{\rm orb} \propto \frac{\tobs G M_1}{4 \pi^2 c^3} f^3 \frac{q_{\rm max}^3}{(1 + q_{\rm max})^2} \left(\frac{\Delta\Omorb}{\Omorb}\right) \Delta\tasc \,,
	\end{equation}
	\end{linenomath}
	where $M_1$ is the neutron star mass.  As before, we assume $\dot{f} \in [-10^{-13},0]$\,Hz\,s$^{-1}$ for an \ac{MSP} search.  The search duration up to a maximum frequency $f_{\rm max}$ is then
	\begin{linenomath}
	\begin{equation}
		A \biggl(\frac{B(q_{\rm max})}{0.01}\biggr) \biggl(\frac{f_{\rm max}}{1\,\text{kHz}}\biggr)^4 \biggl(\frac{\Delta\Porb/\Porb}{10^{-6}}\biggr) \biggl(\frac{\Delta\tasc}{1\,\text{min}}\biggr) \,,
	\end{equation}
	\end{linenomath}
	where the dimensionless parenthetical factors are of order unity for typical systems of interest, and
	\begin{linenomath}
	\begin{equation}
		B(q_{\rm max}) = \frac{q_{\rm max}^3}{(1 + q_{\rm max})^2} \,.
	\end{equation}
	\end{linenomath}
	For redbacks (\added{typically:} $q < 0.3$) one has $B(q) < 0.02$, whereas for black widows ($q < 0.08$) one has $B(q) < 4 \times 10^{-4}$.  The time $A$ depends on the details of the search and the available computing resources.  A typical \EatH{} search as described in this section has $A \sim 10 \,\text{d}$.

	In summary, this section has shown how the circular orbit binary pulsar search for \circsrc{} can be carried out.  It is computationally expensive, but by exploiting the orbital constraints, it is feasible, even for high \ac{MSP} frequencies.  In practice, a search would start at low frequencies, gradually working up to $1.5 \, \text{kHz}$.  To further reduce cost, the search should be stopped if a pulsar is found.

	While here we have considered one specific example, these methods are more broadly applicable.  With them, circular orbit binary pulsar searches are practical if there are good orbital constraints from optically visible companion stars and if the pulsar's projected semimajor axis is not too large.

%%%%%%%%%%%%%%%%%%%%%%%%%%%%%%%%%%%%%%%%%%%
\extraspace
\section{Search Method: Eccentric Binary Orbits} \label{s:methodecc}

	For pulsars in eccentric binary orbits, the photon arrival times have to be corrected for the line-of-sight motion $r_{z, \text{ell}}(t)$, which is the projection of the eccentric orbit in the line-of-sight direction.  In analogy with Eq.~\eqref{eq:timeapproximation}, we can approximate the photon emission time at the pulsar as
	\begin{linenomath}
	\begin{equation} \label{eq:tpsr}
		t_\text{psr} \approx t + r_{z, \text{sky}}(t) - r_{z, \text{ell}}(t)
	\end{equation}
	\end{linenomath}
	up to $\mathcal{O}(x\Omega_\text{orb})$.  Compared with the circular case, two extra parameters are needed to describe the projected line-of-sight motion, $r_{z, \text{ell}}(t)$.  For now, we take these to be the orbital eccentricity $e$ and the angle $\omega$ between the ascending node and the pericenter.

	We note that the approximation to $\mathcal{O}(x\Omega_\text{orb})$ is sufficient for the elliptical example source considered in this paper.  If the value of $x$ were larger, a higher-order approximation in $x$ would also be required \citep{edwards2006}.

	\acp{YP} with main-sequence stars as companions can have very eccentric orbits.  For small orbits the pulsars tidally deform the companion, which dissipates energy.  This tidally locks the companion, so that the same side of the companion faces the pulsar and over time circularizes the orbit \citep{phinney1992}.  This explains why old, spun-up \acp{MSP} are usually found in binaries with small or unobservable eccentricity.  Only a few exceptions are known \citep{knispel2015}.

	If the energy loss in a spider system is small for each orbit, the pulsar moves around a smaller ellipse and the companion around a larger ellipse.  The fixed center of mass is a focus of both ellipses, and the separation vector between pulsar and companion also traces an ellipse.

	The line-of-sight variation due to the elliptical motion, $r_{z, \text{ell}}(t)$, was derived by \citet{blandford1976} and can be written as
	\begin{linenomath}
	\begin{equation} \label{ecc_los_motion}
		r_{z, \text{BT}}(t) = x \left[ \sin \omega \, (\cos E - e) + \sqrt{1 - e^2} \,\cos\omega \, \sin E \right] \,.
	\end{equation}
	\end{linenomath}
	In this formula the label ``ell'' is replaced by ``BT'' to denote that this is the \citeauthor{blandford1976} model.

	The eccentric anomaly $E$ is a parameter along the pulsar path that increases with time.  If $\psi$ is the angular position of the pulsar measured from the center of the ellipse, then $\tan \psi = (1-e^2)^{1/2} \tan E$.  Equivalently, project the pulsar's position parallel to the semiminor axis, onto a circle whose radius is the semimajor axis, and whose center is the center of the ellipse.  Then, $E$ is the angular position of that projected point on the circle. $E$ obeys Kepler's equation
	\begin{linenomath}
	\begin{equation} \label{eq:kepler}
		M = E - e \sin E \,,
	\end{equation}
	\end{linenomath}
	where $M$ is the mean anomaly.  This is a linear function
	\begin{linenomath}
	\begin{equation} \label{eq:meananomaly}
		M = \Omorb (t - T_0) \,,
	\end{equation}
	\end{linenomath}
	where $T_0 = \tasc + \omega/\Omorb$ is the epoch of pericenter passage.

	Unfortunately, there are some problems with the BT model and this parameterization.  Kepler's equation \eqref{eq:kepler} cannot be solved in closed form to find $E$ as a function of $t$.  Furthermore, in small-eccentricity orbits, the pericenter is not well defined and the mismatch arising from offsets in $T_0$ and $\omega$ does not take the simplest possible form.  For these reasons, we shift to an uncorrelated set of parameters and Taylor-expand $r_{z, \text{BT}}$ as function of $e$.

	A new set of parameters was suggested by \cite{lange2001}.  These are the time of ascending node $\tasc$ and two Laplace-Lagrangian parameters $\epsilon_1$ and $\epsilon_2$ defined via
	\begin{linenomath}
	\begin{align}
		\tasc &= T_0 - \omega / \Omorb\,,\\
		\epsilon_1 &= e \sin\omega \,, \label{eq:eps1}\\
		\epsilon_2 &= e \cos\omega \,. \label{eq:eps2}
	\end{align}
	\end{linenomath}
	The parameters $\{T_0, e, \omega\}$ are given by
	\begin{linenomath}
	\begin{align}
		T_0 &= \tasc + \Omorb^{-1}\arctan(\epsilon_1/\epsilon_2) \,,\\
		e &= \left(\epsilon_1^2 + \epsilon_2^2\right)^{1/2} \,,\\
		\omega &= \arctan(\epsilon_1/\epsilon_2) \,.
	\end{align}
	\end{linenomath}
	With the old parameters, the region of constant mismatch around a grid point is an ellipsoid whose principal directions are not parallel to the $\{T_0, e, \omega\}$ axes.  In the next section, we show that with the new parameters the region of constant mismatch is a sphere.  This simplifies the code used to optimize grid point locations.

	The R\o mer delay $r_{z, \text{BT}}$ for the pulsar's motion can be expanded to first order in $e$.  Following convention, we use the label ``ELL1'' for this linear-in-$e$ model: $r_{z,\text{BT}} = r_{z, \text{ELL1}} + \mathcal{O}(e^2)$.  This can be described using the parameters $\{T_0,e,\omega\}$ or the parameters $\{\tasc,\epsilon_1,\epsilon_2\}$ as
	\begin{linenomath}
	\begin{align}
		r_{z, \text{ELL1}}(t) &= x \left[ \sin(M + \omega) + \frac{e}{2} \sin(2M + \omega) - \frac{3 e}{2} \sin \omega \right] \label{eq:ell1} \\
		&= x \left[ \sin \phi + \frac{\epsilon_2}{2} \sin 2 \phi - \frac{\epsilon_1}{2} \cos 2 \phi - \frac{3}{2} \epsilon_1 \right] \,. \label{eq:ELL1}
	\end{align}
	\end{linenomath}
	We have introduced
	\begin{linenomath}
	\begin{equation}
		\phi = \Omorb (t - \tasc) \,,
	\end{equation}
	\end{linenomath}
	which is similar to $M$ in Eq.~\eqref{eq:meananomaly} but shifted from pericenter to ascending node.  (Note that the term $-3e\sin\omega/2 = -3\epsilon_1/2$ is typically dropped, as it is time independent.)

	The ELL1 approximation to the BT model can accurately track the pulsar's rotational phase for eccentricities $e$ below some threshold value.  In Appendix~\ref{s:highorder}, we show how this threshold depends on the spin frequency $f$ and semimajor axis $x$.

	Later in the paper, in Section~\ref{s:J0523}, we design a search for \ellisrc{}, which is a gamma-ray source predicted to harbor a redback pulsar in an eccentric orbit.  For that case, the ELL1 model is insufficient and a third-order-in-$e$ model is needed.  In Appendix~\ref{s:highorder}, we derive higher-order-in-$e$ approximations to $r_{z,\text{BT}}$, and demonstrate how they improve the match (decrease the mismatch) to the BT model.

%%%%%%%%%%%%%%%%%%%%%%%%%%%%%%%%%%%%%%%%%%%
\extraspace
\subsection{Parameter space metrics} \label{ss:psm}

	In this section, we calculate the coherent and semicoherent parameter space metric for the ELL1 model.  Compared to the circular case, the parameter space has two extra dimensions.

	Since the ELL1 model differs at first order in $e$ from the circular model, the coherent metric also differs at first order.  However, for the $\{f, \dot{f}, n_x, n_y,\Omorb,x\}$ metric components, the first-order terms are of $\mathcal{O}(1/\Omorb\tobs)$ and can be neglected; the dominant difference is second order in $e$.  Thus, the coherent metric components given in Eqs.~\eqref{eq:g_ab_fs_d} and \eqref{eq:g_ab_orb} remain valid to first order in $e$.

	For the ELL1 model in Eq.~\eqref{eq:ell1}, the dominant components for the parameters $\{T_0, e, \omega\}$ are
	\begin{linenomath}
	\begin{equation}
		\begin{aligned}
			g_{T_0 T_0} =& 2 \pi^2 f^2 x^2 \Omorb^2 [1 + \mathcal{O}(1/\Omorb\tobs)] \,,\\
			g_{e e} =& \frac{1}{2} \pi^2 f^2 x^2 [1 + \mathcal{O}(1/\Omorb\tobs)] \,,\\
			g_{\omega \omega} =& 2 \pi^2 f^2 x^2 [1 + \mathcal{O}(1/\Omorb\tobs)] \,,\\
			g_{T_0 \omega} =& -2 \pi^2 f^2 x^2 \Omorb [1 + \mathcal{O}(1/\Omorb\tobs)] \,.
		\end{aligned}
	\end{equation}
	\end{linenomath}
	Note that the off-diagonal component $g_{T_0 \omega}$ does not vanish.  As described in the previous section, this complicates the form of the mismatch.

	We now change to the parameters $\{\tasc, \epsilon_1, \epsilon_2\}$, for which it is convenient to use Eq.~\eqref{eq:ELL1}.  For these, the diagonal components are
	\begin{linenomath}
	\begin{equation}
		\begin{aligned}
			g_{\tasc \tasc} =& 2 \pi^2 f^2 x^2 \Omorb^2 [1 + \mathcal{O}(1/\Omorb\tobs)] \,,\\
			g_{\epsilon_1 \epsilon_1} =& \frac{1}{2} \pi^2 f^2 x^2 [1 + \mathcal{O}(1/\Omorb\tobs)] \,,\\
			g_{\epsilon_2 \epsilon_2} =& \frac{1}{2} \pi^2 f^2 x^2 [1 + \mathcal{O}(1/\Omorb\tobs)] \,.
		\end{aligned}
	\end{equation}
	\end{linenomath}
	These diagonal metric components are of $\mathcal{O}(e^0)$.  The terms that are linear in $e$ are of $\mathcal{O}(1/\Omorb\tobs)$, and can be neglected.  Thus, the dominant diagonal $e$-dependent terms are of $\mathcal{O}(e^2)$.  However, there are off-diagonal terms of $\mathcal{O}(e^1)$.

	For small eccentricities $e$, the dominant metric components are given above. For completeness, we list the $\mathcal{O}(e^1)$-corrections, which are all off-diagonal:
	\begin{linenomath}
	\begin{equation}
		\begin{aligned}
			g_{x \epsilon_1} &= \frac{1}{2} \pi^2 f^2 x \epsilon_1 [1 + \mathcal{O}(1/\Omorb\tobs)] \,,\\
			g_{x \epsilon_2} &= \frac{1}{2} \pi^2 f^2 x \epsilon_2 [1 + \mathcal{O}(1/\Omorb\tobs)] \,,\\
			g_{\Omorb \epsilon_1} &= -\pi^2 f^2 x^2 \epsilon_2 \tobs \left[\frac{(t_0 - \tasc)}{\tobs} + \mathcal{O}(1/\Omorb\tobs) \right] \,,\\
			g_{\Omorb \epsilon_2} &= \pi^2 f^2 x^2 \epsilon_1 \tobs \left[\frac{(t_0 - \tasc)}{\tobs} + \mathcal{O}(1/\Omorb\tobs) \right] \,,\\
			g_{\tasc \epsilon_1} &= \pi^2 f^2 x^2 \Omorb \epsilon_2 [1 + \mathcal{O}(1/\Omorb\tobs)] \,,\\
			g_{\tasc \epsilon_2} &= -\pi^2 f^2 x^2 \Omorb \epsilon_1 [1 + \mathcal{O}(1/\Omorb\tobs)] \,.
		\end{aligned}
	\end{equation}
	\end{linenomath}
	The remaining off-diagonal components of the orbital metric are of $\mathcal{O}(1/\Omorb\tobs)$.

	These metric components have been found to be a good approximation even for higher eccentricities where the ELL1 model is not sufficient to track the rotational phase in a search and higher-order models need to be used.  This might be because many of the linear-in-$e$ terms vanish from the metric.

	The semicoherent metric components are very similar to the coherent ones.  The components associated with the noneccentric parameters $\{f, \dot{f}, n_x, n_y,\Omorb,x\}$, calculated in the circular case in Eqs.~\eqref{eq:semicoherentisolated} and \eqref{eq:gbar_ab_orb}, remain valid; they have only second-order corrections in $e$.  For the remaining orbital parameters $\{\tasc,\epsilon_1,\epsilon_2\}$ the semicoherent metric components are the same as in the coherent case (this follows from Eq.~\eqref{eq:con_g_gbar_orb}).  Thus, the diagonal components for $\{\tasc,\epsilon_1,\epsilon_2\}$ are
	\begin{linenomath}
	\begin{equation}
		\begin{aligned}
			\bar{g}_{\tasc \tasc} =& 2 \pi^2 f^2 x^2 \Omorb^2 \,,\\
			\bar{g}_{\epsilon_1 \epsilon_1} =& \frac{1}{2} \pi^2 f^2 x^2 \,,\\
			\bar{g}_{\epsilon_2 \epsilon_2} =& \frac{1}{2} \pi^2 f^2 x^2 \,,
		\end{aligned}
	\end{equation}
	\end{linenomath}
	where we omit terms of $\mathcal{O}(1/\Omorb\tobs)$.  Thus, the semicoherent metric for the ELL1 model simply adds the components above to the semicoherent metric for the circular model.

\begin{figure*}
		\centering
		\includegraphics[width=\textwidth]{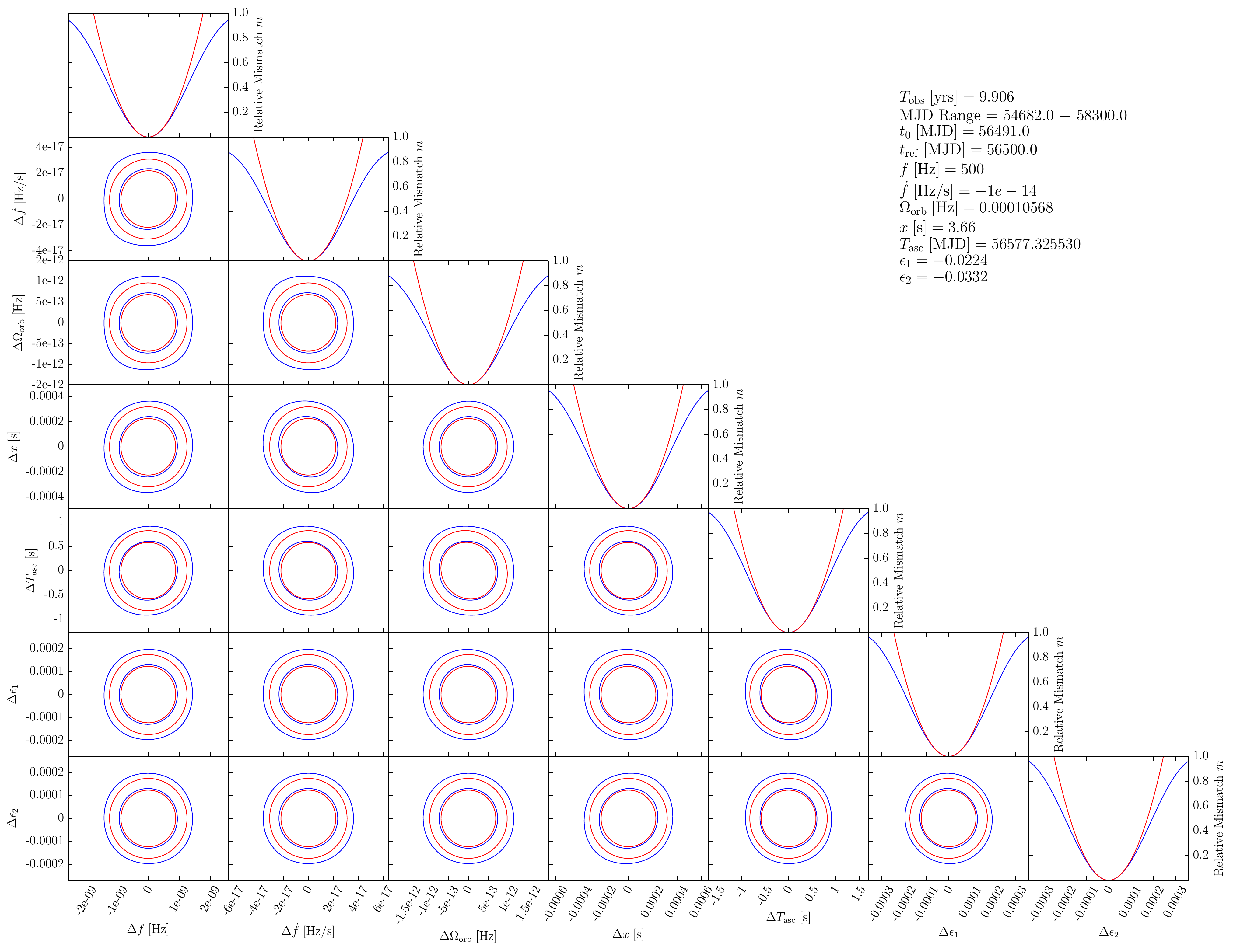}
		\caption{\label{f:triplot_elli} Comparison of the coherent metric approximation to the actual mismatch, for parameters of a simulated eccentric orbit binary pulsar in \ellisrc{}. Blue contours show the actual mismatch and red contours the metric approximation, at $m = 0.25$ and $0.5$.}
	\end{figure*}

	In Figure \ref{f:triplot_elli}, the mismatch and its coherent metric approximation are compared for small parameter offsets, for a realistic simulated pulsar.  Apart from different $f$- and $\dot{f}$-scales, the corresponding plot for the semicoherent mismatch looks very similar.  The mismatch agrees well with its metric approximation for mismatch $m \le 0.5$, which is typical: in Appendix~\ref{s:optimalmismatch}, we show that the highest sensitivity at given computing cost for an elliptical search is obtained with an average mismatch $\hat{m} = 0.471$ (see Table~\ref{t:optimalmismatch}).

	The sky position parameters $\{n_x,n_y\}$ are not shown in Figure~\ref{f:triplot_elli} because we assume that for spider pulsars they are known to high precision from optical observations.

	A \replaced{full blind}{partially informed} search for binary pulsars in elliptic orbits\added{, without exact information about the sky position and constraints on the orbital parameters,} is computationally impossible.  There are too many parameter space dimensions \mdash{} even for circular orbits with reasonable parameter ranges, the grid has too many points.  To make a search possible, one needs tight constraints derived from optical/X-ray observations of the pulsar's companion star.  In the next section, we will discuss constraints and the search design for the probable eccentric orbit binary gamma-ray pulsar in \ellisrc{} \citep{strader2014,parkinson2016}.

%%%%%%%%%%%%%%%%%%%%%%%%%%%%%%%%%%%%%%%%%%%
\extraspace
\subsection{Search design for low-eccentricity binary} \label{s:J0523}

	In this section, we discuss how to reduce the search parameter space using orbital constraints for the gamma-ray source \ellisrc{}, presumed to be a pulsar in an eccentric binary orbit.  \added{The source is named 3FGL\,J0523.3$-$2528, 2FGL\,J0523.3$-$2530, or 1FGL\,J0523.5$-$2529 in previous \ac{LAT} source catalogs.}  This is similar to the circular example of Section~\ref{s:J1653}.

	The gamma-ray source itself was investigated by \cite{parkinson2016} and ranked ninth highest in a list of most significant \ac{3FGL} unassociated sources predicted to be pulsars.  It shows typical pulsar-like properties: the photon flux is stable over time, and the spectrum is fit by an exponential cutoff power law.  The source is not in the Galactic disk, which increases the odds that it hosts an \ac{MSP}.

	Earlier optical observations identified a likely companion and indicate an orbit with small, but not negligible, eccentricity of $e=0.04$ \citep{strader2014}.  In contrast to the previous paragraph, this suggests that the pulsar is a \ac{YP}, because binary \acp{MSP} tend to have rather circular orbits \citep{phinney1992}.

	The frequency and spin-down search ranges are chosen following the logic of the previous search design (Section~\ref{s:J1653}).  For \acp{YP} we search $f \in [0,44]\,\text{Hz}$ and $\dot{f} \in [-10^{-10},0]\,\text{Hz/s}$.  For \acp{MSP} we search $f \in [44,1500]\,\text{Hz}$ and $\dot{f} \in [-10^{-13},0]\,\text{Hz/s}$.  The $f$-dimension is efficiently searched using \acp{FFT} with bandwidth $f_{\rm BW} = 8\,\text{Hz}$, and the $\dot{f}$-dimension is covered by a uniformly spaced lattice.

	The sky position search range of the probable pulsar within \ellisrc{} is tightly constrained from the X-ray and optical observations of the likely companion discussed above \citep{strader2014}.  At the time, the best estimate for the optical position was from the USNO-B1.0 Catalog \citep{monet2003}.  It is now also identified in the Gaia DR2 Catalog \citep{gaia2018}, whose pointing is so precise (see Table~\ref{t:const2}) that even at $f = 1.5\,\text{kHz}$ no search over sky position is required.

	The orbital parameter search ranges shown in Table~\ref{t:const2} come from the \cite{strader2014} analysis of the photometric and spectroscopic optical data.  The orbital period and eccentricity parameters are constrained by the periodic optical flux modulation. They assume that this arises from viewing a tidally locked and deformed (ellipsoidal) companion at different aspect angles.  Hence, the orbital period is twice the observed modulation period.  (Another possible explanation for the modulation would be irradiation, but spectroscopic data do not show the orbital-phase-dependent temperature change that would be expected.)  The orbital period is constrained to $\Porb = 0.688134 \pm 0.000028$\,d at epoch of superior conjunction $T_{0.5} = 56577.14636 \pm 0.0037$\,MJD.  The eccentric parameters $\{e,\omega\}$ fall in the ranges $e = 0.040 \pm 0.006$ and $\omega = 214 \pm 10 \, \deg$.

	The semimajor axis $x$ is constrained using Eq.~\eqref{eq:x}.  This is similar to our previous example in Section~\ref{s:J1653}, but requires fewer assumptions because the mass ratio $q=M_2/M_1$ is directly bounded from the observations.  To do this, \cite{strader2014} estimate the rotational velocity of the companion's Roche lobe from high-quality optical spectra.  Combined with the companion's radial velocity $K_2 = 190.3 \pm 1.1$\,km\,s$^{-1}$, this constrains the mass ratio $q=0.61 \pm 0.06$. Returning to Eq.~\eqref{eq:x}, this gives $x = 3.66 \pm 0.38$.

	The parameters $\{e,\omega\}$ can be converted directly to the quantities $\{\epsilon_1,\epsilon_2\}$ needed for our search, using Eqs.~\eqref{eq:eps1} and \eqref{eq:eps2}.

	For our search, we also need the epoch of ascending node $\tasc$.  However, the results of \cite{strader2014} are given in terms of the epoch of superior conjunction $T_{0.5}$.  For circular orbits, $T_{0.5}$ and $\tasc$ differ by $\Porb/4$, but for eccentric orbits the relation is more complicated.  To second order in $e$, it is
	\begin{linenomath}
	\begin{equation}
		\tasc = T_{0.5} + \Porb \left( \frac{1}{4} - \frac{\epsilon_2}{\pi} - \frac{3 \epsilon_1 \epsilon_2}{4 \pi} + \mathcal{O}(e^3) \right) \,.
	\end{equation}
	\end{linenomath}
	For \ellisrc{} with $e = 0.04$, this $\mathcal{O}(e^2)$ approximation is more accurate than the uncertainties in the measured quantities on the right-hand side.  (Higher-order approximations in $e$ would be required for pulsars in binary orbits with larger eccentricities or longer orbital periods.)  The resulting $\tasc$ is given in Table~\ref{t:const2}.

%------------------------------------------------------------------------------
	\begin{deluxetable}{ll}
		\tablewidth{\columnwidth}
		\tablecaption{\label{t:const2} Parameters and constraints for \ellisrc }
		\tablecolumns{2}
		\tablehead{
			\colhead{Parameter} &
			\colhead{Value}
		}
		\startdata
		Range of observational data (MJD) \dotfill  & $54682$ -- $58300$ \\[0.15em]
		Reference epoch (MJD)\dotfill  & $56500.0$   \\[-0.2em]
		\cutinhead{Initial companion location from \citetalias{monet2003} catalog}
		R.A., $\alpha$ (J2000.0)\dotfill & $05^{\rm h}23^{\rm m}16\fs925(4)$ \\[0.15em]
		Decl., $\delta$ (J2000.0)\dotfill    & $-25\arcdeg27\arcmin36\farcs92(6)$ \\[-0.2em]
		\cutinhead{Precise companion location from \citetalias{gaia2018} catalog}
		R.A., $\alpha$ (J2000.0)\dotfill & $05^{\rm h}23^{\rm m}16\fs931203(2)$ \\[0.15em]
		Decl., $\delta$ (J2000.0)\dotfill    & $-25\arcdeg27\arcmin37\farcs12468(4)$ \\[-0.2em]
		\cutinhead{Constraints from probable counterpart \citep{strader2014}}
		Superior conjunction epoch, $T_{0.5}$ (MJD) \dotfill & $56577.14636 \pm 0.0037$\\[0.15em]
		Companion velocity, $K_2$ ($\text{km} \,\, \text{s}^{-1}$) \dotfill & $190.3 \pm 1.1$\\[0.15em]
		Mass ratio, $q=M_2/M_1$ \dotfill & $0.61 \pm 0.06$\\[0.15em]
		Eccentricity, $e$ \dotfill & $0.040 \pm 0.006$\\[0.15em]
		Longitude of pericenter, $\omega$ (deg) \dotfill & $214 \pm 10$ \\[0.15em]
		Orbital period, $\Porb$ (d) \dotfill & $0.688134 \pm 0.000028$\\[0.1em]
		{\hfill equivalent to \hfill}\\
		Orbital frequency, $\Omorb$ (Hz) \dotfill & $0.0001056801 \pm 4.3\times10^{-9}$ \\[-0.2em]
		\cutinhead{Derived search parameters and corresponding uncertainties}
		Projected semimajor axis, $x$ (s) \dotfill &  $3.66 \pm 0.38$ \\[0.15em]
		Ascending node epoch, $\tasc$ (MJD) \dotfill & $56577.32553 \pm 0.00567$\\[0.15em]
		First Lagrange parameter, $\epsilon_1$ \dotfill & $-0.0224 \pm 0.0091$ \\[0.15em]
		Second Lagrange parameter, $\epsilon_2$ \dotfill & $-0.0332 \pm 0.0089$ \\[0.5em]
		\enddata
		\tablecomments{
			The JPL DE405 solar system ephemeris has been used, and times refer to~TDB.}
	\end{deluxetable}
%------------------------------------------------------------------------------

	A search for a pulsar in an eccentric orbit is very similar to one for a pulsar in a circular orbit.  The only differences are that a more general model for the R\o mer delay is required to track the pulsar phase, and the orbital grids need to cover five orbital dimensions.  While the latter is much more complex, it can be done with the same optimized stochastic search grid construction methods that are used in the circular case.

	To accurately track the rotational phase of the pulsar requires a higher-order-in-$e$ approximation to $r_{z,\text{BT}}$ than the ELL1 model, unless the eccentricity is very small.  Such approximations are computed in Appendix~\ref{s:highorder}.  There, we also determine which order in $e$ is sufficient.

	For the case of \ellisrc{}, a model of $\mathcal{O}(e^3)$ is sufficient.  In Figure~\ref{f:model_error}, we show that the rotational phase error is negligible for the constrained parameter ranges given above.

	Analogously to Eq.~\eqref{eq:norb}, the minimum number of grid points for the orbital parameter space can be computed from the proper $5$-volume
	\begin{linenomath}
	\begin{equation}
		N_\text{orb} \approx m^{-5/2} \int \sqrt{\det \bar{g}} \diff \pvec_\text{orb} \,.
	\end{equation}
	\end{linenomath}
	Here the metric has the five dimensions $\{x,\Omorb,\tasc,\epsilon_1,\epsilon_2\}$.  This integral is proportional to
	\begin{linenomath}
	\begin{equation} \label{eq:neorb_dep}
		N_\text{orb} \propto f^5 \tobs \left( x_\text{max}^5 - x_\text{min}^5 \right) \Omega_\text{orb}\Delta\Omega_\text{orb} \Delta\tasc \Delta\epsilon_1 \Delta\epsilon_2 \,,
	\end{equation}
	\end{linenomath}
	where $x \in [x_\text{min},x_\text{max}]$.  $\Delta\Omorb$, $\Delta\tasc$, $\Delta\epsilon_1$, and $\Delta\epsilon_2$ are the search ranges for the corresponding parameters, and we made the assumption that $\Delta\Omorb \ll \Omorb$.  The number of orbital grid points and subsequently the computing cost depend even more strongly on $f$ and $x$ in an eccentric search than in a circular one.

	The computing cost of a search for \ellisrc{} is estimated based on the number of grid points.  We assume search ranges in $f$ and $\dot{f}$ as given earlier in this section.  The remaining parameter space ranges are given in Table~\ref{t:const2}.  The required total computing cost of the search is estimated by multiplying the cost of one \ac{FFT} by the number of $\dot{f}$-grid points and the $f$-dependent number of orbital grid points and then summing over the $f$ intervals.

\begin{figure}
		\centering
		\includegraphics[width=\columnwidth]{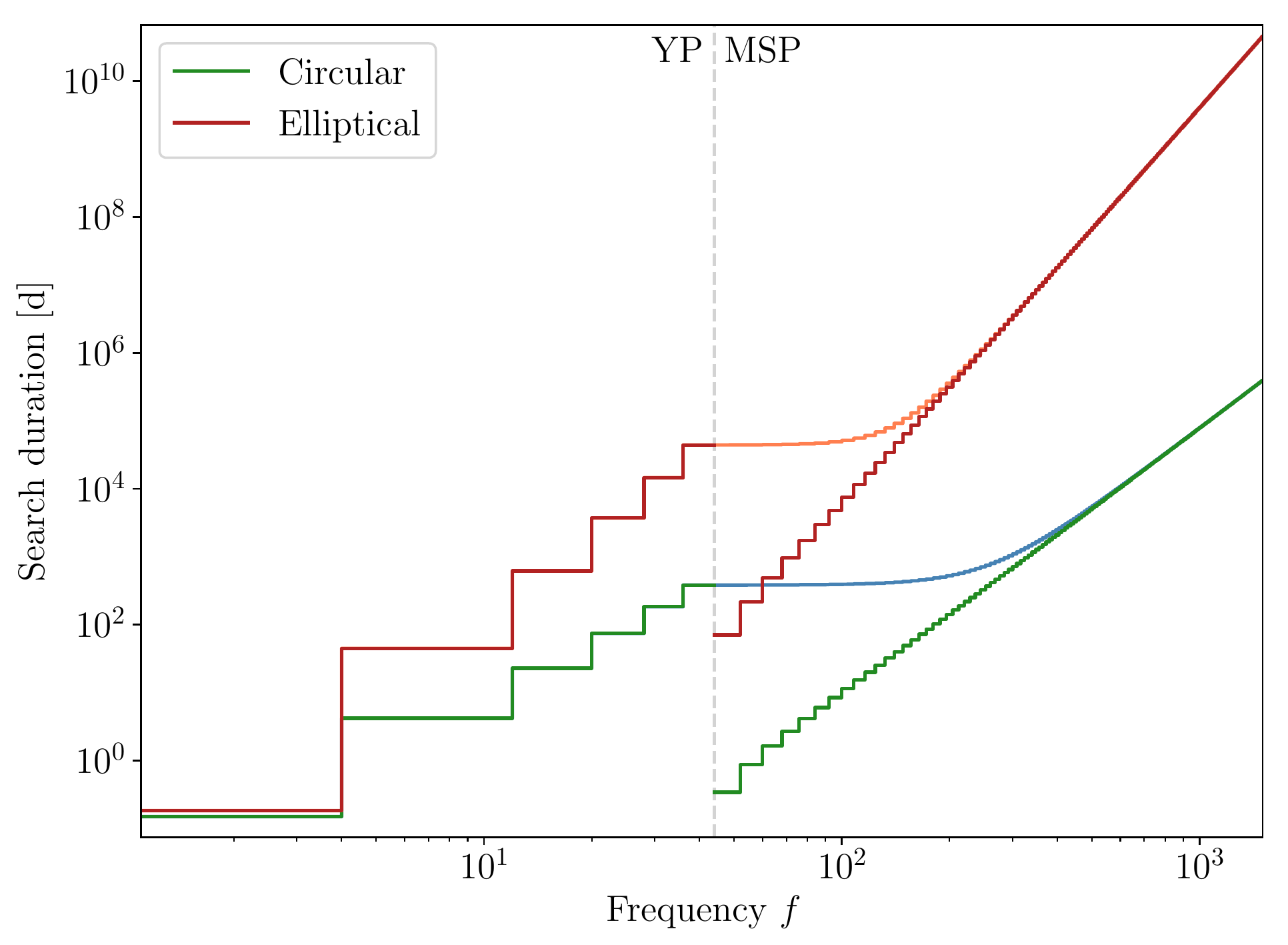}
		\caption{\label{f:fft_elli} Predicted days on \EatH{} needed to search \ellisrc{}, assuming a circular (green/blue) or elliptical (red/orange) orbit. The left curves show the cumulative duration of a \ac{YP} search from $0$\,Hz up to maximum frequency $f$. The right curves show the cumulative duration of an \ac{MSP} search from $44$\,Hz up to maximum frequency $f$. Their slopes are $\propto f^4$ and $\propto f^6$; they are integrals over the number of orbital templates. The larger slope for the elliptical search arises from the two extra dimensions of search parameter space. The blue and orange curves show the sums: the cumulative duration of a combined YP and \ac{MSP} search.}
	\end{figure}

	To exemplify the computing cost of a search for \ellisrc{}, we express it in terms of search duration on \EatH{}, assuming that the project provides $25{,}000$ GPU-hr per week.  This is shown in Figure~\ref{f:fft_elli} as a function of the maximum searched frequency.  For comparison, we also show the search duration for a circular binary search, i.e. setting $e=0$ and not searching over $\{\epsilon_1,\epsilon_2\}$.  An eccentric \ac{MSP} search up $1.5\,\text{kHz}$ would take more than $100$ million years on \EatH{}, and even a \ac{YP} search would take more than $100$ years.  Circular searches for \acp{YP} or \acp{MSP} up to $400\,\text{Hz}$ would take a few hundreds days.  Note that the search ranges are still the $1\sigma$ ranges, so searches within the $3\sigma$ range would be more computing intensive.

	In summary, this section has shown how computing intensive a search for \ellisrc{} would be.  An eccentric \ac{MSP} search even to low frequencies $~\sim 100\,\text{Hz}$ is not feasible with the current constraints, and a \ac{YP} search would be very expensive.  In the optical data, \cite{strader2014} do not see evidence for a ``false'' eccentricity, but a circular search would be much less computing intensive than an eccentric one.  With slightly tighter constraints, searches up to $800\,\text{Hz}$ could be feasible.

%%%%%%%%%%%%%%%%%%%%%%%%%%%%%%%%%%%%%%%%%%%
\extraspace
\vspace{0.9cm}
\section{Comparison with other Methods} \label{s:alternatives}

	Similar and alternative methods are used to search for binary pulsars in data from radio telescopes and gravitational-wave detectors.  In this section, we will review these, compare them to the methods presented here, and discuss their applicability to searches for binary gamma-ray pulsars.

	In addition to coming from diverse messengers and frequencies, the data have other key differences.  The gamma-ray data are similar to the gravitational-wave data: the length of the data sets is months to years, and the instruments simultaneously detect signals from a substantial fraction of the sky.  In contrast, typical radio surveys collect data in stretches of minutes from tiny fractions of the sky.  While gamma-ray data consist of discrete photon arrival times, radio and gravitational-wave data are continuous.  Therefore, it is not surprising that some pulsation search methods might work for one kind of data but not for the other.

	For these other data sources, many methods have been employed by many individuals and groups.  Here we are guided by reviews from \cite{lorimer2004} for radio search methods and \cite{messenger2015} for gravitational-wave methods. We exclude methods that require data from two detectors.

\extraspace
\subsection{Acceleration searches}

	Time-domain ``acceleration searches'' have been very successful in finding new radio pulsars in binaries with orbital periods shorter than a day \citep[see, e.g.,][]{camilo2000}.  Fourier-domain acceleration searches have also been successfully used to discover binary radio pulsars \citep[see, e.g.,][]{ransom2001,andersen2018}.  A similar approach to search for continuous gravitational waves is the ``polynomial search'' \citep{putten2010}.

	These searches do not use a model that describes periodic orbital motion.	 Instead, they assume constant acceleration along a straight line \citep[see also][]{johnston1991}.  This accurately describes an orbiting system only if the data set is much shorter than one orbital period.  Since the \ac{LAT} data set is more than a decade long, acceleration searches would only find binary gamma-ray pulsars whose orbital periods were decades or longer.

	It is straightforward to quantify the range of orbital periods an acceleration search is sensitive to.  Assume that the data set is less than $\sim 10\%$ of the orbital period and is near the superior or inferior conjunction, where the velocity is changing linearly with time \citep{johnston1991}.  An acceleration $a$ along the line of sight (``los'') toward Earth contributes an amount
	\begin{linenomath}
	\begin{equation}
		\dot{f}_\text{los} = \frac{f a}{c}
	\end{equation}
	\end{linenomath}
	to the observed spin frequency derivative.  The maximum acceleration at inferior or superior conjunction is for a circular orbit $a = c x \, \Omega_\text{orb}^2$, and for an eccentric orbit $a = c x \, \Omega_\text{orb}^2 (1+e)/(1-e)$.  Therefore, searches would be sensitive if the sum of the intrinsic pulsar spin-down and this line-of-sight contribution to the spin-down were within the search range.  Since the intrinsic spin-down is usually negative, this is most likely if the acceleration toward Earth is positive, i.e. if the pulsar is near the superior conjunction.

	Current \replaced{blind }{partially informed }search surveys for isolated gamma-ray pulsars are a form of acceleration search because they scan over spin-down \citep{clark2017}.  For \acp{YP} they search down to $\dot{f} = -10^{-9}$\,Hz\,s$^{-1}$ and for \acp{MSP} down to $\dot{f} = -10^{-13}$\,Hz\,s$^{-1}$.  In principle, these searches are sensitive to pulsars like the young ($f \approx 7\,\text{Hz}$) binary pulsar PSR\,J2032+4127, which is in a $45-50$ yr orbit around its companion \citep{ho2017}.  It was found in an isolated gamma-ray search \citep{abdo2009d}, and only afterward was it discovered to be in a binary system \citep{lyne2015}.  The orbit is highly eccentric ($e \approx 0.93-0.99$) with $x \approx 7{,}000 - 20{,}000\,\text{s}$.  The maximum spin-down contribution should therefore be of order $|\max\{\dot{f}_\text{los}\}| = 10^{-10}$\,Hz\,s$^{-1}$.  This is in the search range if the pulsar is near superior conjunction during the mission time.

	\replaced{Blind searches}{Searches} that assume linear acceleration, i.e., that search over constant $\dot{f}$, are only sensitive to binary pulsars with $\Porb \gtrsim 10 \tobs$.  To become sensitive to shorter orbital periods, higher-order frequency derivatives must be searched.  ``Jerk'' searches, which include the second-order frequency derivative $\ddot{f}$, improve the sensitivity for pulsars with orbital periods in the range $\Porb \in [7\tobs,20\tobs]$ and have been successfully used in a radio pulsar search \citep{andersen2018}. Alternatively, the full orbital motion may be taken into account, as in \cite{allen2013}.

\extraspace
\subsection{Stack/slide search}

	The ``stack/slide'' method has been used in radio pulsar searches like the Parkes Multibeam Pulsar survey to account for binary motion \citep{faulkner2004}.  This led to the discovery of the double neutron star system PSR\,J1756$-$2251 with an orbital period of $7.7\,\text{hr}$ \citep{faulkner2005}.  (The words ``stack/slide'' are used in continuous gravitational-wave searches, not to account for binary pulsar motion but rather to remove the effects of Earth rotation and motion around the \ac{SSB} \citep{brady2000,riles2017}.  That is also the case for the semicoherent searches we describe in this paper to account for the \ac{LAT}'s motion around the \ac{SSB}.)

	In a stack/slide search the data set is broken into subsets of length $\tcoh$, corresponding to frequency bins of width $\Delta f = 1/\tcoh$.  $\tcoh$ is chosen to be small enough that the Doppler modulation induced by motion of the detector around the \ac{SSB}, or of the pulsar around the binary center of mass, remains within a single bin.  For circular binary motion, provided that $\tcoh$ is a factor of a few smaller than $\Porb$, this implies
	\begin{linenomath}
	\begin{equation} \label{eq:sscondition}
		f x \Omorb^2 \tcoh < 1/\tcoh \,.
	\end{equation}
	\end{linenomath}
	Each of these subsets is then Fourier transformed.  The resulting power spectra are added (stacked) together after the Doppler modulation is compensated by shifting the frequency (slide) in each of the spectra; sources give rise to peaks in the stacked spectra.  This technique is only sensitive if the subsets are much shorter than the Doppler modulation period.

	This technique is useless for spider gamma-ray pulsars because detection statistics are constructed from the differences of photon arrival times.  Spider pulsars have typical orbital periods of $\Porb \lesssim 1\,\text{d}$, so data subsets would have to be shorter than a few hours.  Most data subsets would contain no photons.  A few would contain one photon.  Almost none would contain enough photons to compute the differences of photon arrival times.

	Stack/slide could be used for gamma-ray pulsars in orbits where $\Porb$ is too small for an acceleration search but is much larger than the $\tcoh \approx 24\,\text{d}$ used in this paper.  Using Kepler's third law, the condition of Eq.~\eqref{eq:sscondition} can be written as
	\begin{linenomath}
	\begin{equation}
		\frac{G M_1}{c^3} \frac{q^3}{(1+q)^2} f^3 \tcoh^6 \Omorb^4 < 1 \,,
	\end{equation}
	\end{linenomath}
	where $M_1$ is the pulsar mass and $q=M_2/M_1$ is the mass ratio. (In fact, this applies provided that $\tcoh \lesssim \Porb$.)  This shows that with our choice of $\tcoh$, stack/slide methods might be able to find gamma-ray pulsars with planetary companions, with orbital periods longer than $\sim 1$\,yr and masses up to $\mathcal{O}(10)$ Earth masses.

\extraspace
\subsection{Power spectrum search}

	The basic assumption of a ``power spectrum search'' is that the data set can be broken into subsets short enough that the observed spin frequency is constant in each one.  This is the same assumption as in a stack/slide search.  That technique is based on visual inspection and has been used to discover binary radio pulsars \citep[see, e.g.,][]{lyne2000}.

	To carry out the search, power spectra are computed for each subset. The spectra are binned in frequency and plotted with a frequency-versus-time color map.  The colors show the power and make it easy to visually identify peaks in the power spectrum.  A binary pulsar signal appears as a peak whose frequency varies sinusoidally with time.

	The method ``TwoSpect'' uses a similar method to perform all-sky searches for continuous gravitational waves from sources in binary systems.  The visual inspection is replaced by a second Fourier transform \citep[hence the name TwoSpect;][]{goetz2011}.  While no continuous gravitational waves have been detected, this technique has been used to put upper limits on continuous gravitational-wave emission from the low-mass X-ray binary Scorpius X-1 \citep{aasi2014}.

	The power spectrum search is not suitable for detecting gamma-ray spider pulsars for the same reasons as the stack/slide method.

\extraspace
\subsection{Sideband search}

	``Sideband searches'' have found many binary radio pulsars within globular clusters \citep{lorimer2004}.  The method has also been adapted to search for continuous gravitational waves from sources in binary systems \citep{messenger2007,sammut2014}.  One first carries out a search for isolated systems, as if there were no binary motion, and then looks for a characteristic structure in the results of that isolated search.

	If a binary is present, orbital motion produces sidebands around a central peak at the spin frequency of the pulsar \citep{ransom2003}.  Since the isolated search does not remove the effects of the binary motion, a pulsar's power is spread over many Fourier bins (also called sidebands).  This reduces the sensitivity compared to a matched-filter search.

	The method is particularly useful for tight orbit binary pulsars where the orbital period is much smaller than the observation time span, which is the case of interest for spider pulsars.  After detecting a signal, the binary parameters can be inferred from the locations and magnitudes of the sidebands and the central peak.

	To see how this works, we compute the \ac{SNR} of the coherent detection statistic $P_n$ for an isolated pulsar template, with parameters $\{\nu,\dot{f},n_x,n_y,0,0,0\}$, arising from a circular binary pulsar with parameters $\{f,\dot{f},n_x,n_y,x,\forb,\tasc\}$, where $\forb = \Omorb/2\pi$.  This \ac{SNR} is given by Eq.~\eqref{eq:SNR_Pn}, which depends on the rotational phase difference due to the parameter mismatch:
	\begin{linenomath}
		\begin{equation} \label{eq:phasemismatch}
		\Delta\Phi(t) =  2\pi (\nu - f) (t - \tref) + 2\pi f x \sin[2\pi\forb (t - \tasc)] \,.
		\end{equation}
	\end{linenomath}
	One can think of $\nu$ as denoting the pulsar frequency in the isolated search.  Our derivation closely follows \cite{ransom2003}.

	To compute the detection statistic $P_n$, we evaluate Eq.~\eqref{eq:SNR_Pn} with the phase mismatch~\eqref{eq:phasemismatch}.  We first reexpress ${\rm e}^{i n \Delta \Phi}$ using the Jacobi-Anger expansion
	\begin{linenomath}
	\begin{equation}
		{\rm e}^{i z \sin \vartheta} = \sum\limits_{m = -\infty}^{\infty} J_m (z) {\rm e}^{i m \vartheta} \,,
	\end{equation}
	\end{linenomath}
	with $z = 2 \pi n f x$ and $\vartheta = 2\pi\forb (t - \tasc)$, where $J_m$ is a Bessel function of the first kind.  Multiplying this by ${\rm e}^{i 2\pi n (\nu - f) (t - \tref)}$ gives
	\begin{linenomath}
	\begin{equation}
		{\rm e}^{i n \Delta \Phi} = \sum\limits_{m = -\infty}^{\infty} J_m (2 \pi n f x) {\rm e}^{i 2\pi [n (\nu - f) + m \forb] (t - \tasc)} \,,
	\end{equation}
	\end{linenomath}
	where, without loss of generality, we have set $\tref = \tasc$.  Since the \ac{SNR} only depends on the modulus of ${\rm e}^{i n \Delta \Phi}$, we may also set $\tasc = 0$.  We assume that there are a large number of photons from the hypothetical pulsar, which have equal weights and arrive at uniformly spaced intervals in time.  The double sum $\sum_{ j \ne k}$ in Eq.~\eqref{eq:SNR_Pn} may then be replaced by an integral over time, since
	\begin{linenomath}
	\begin{equation} \label{eq:simplification}
		\begin{aligned}
			& \sum_{j \ne k} {\rm e}^{i n (\Delta \Phi_j - \Delta \Phi_k)} \approx \Biggl| \sum_j {\rm e}^{i n \Delta \Phi_j} \Biggr|^2 \\
			& \approx \Biggl| \frac{N}{\tobs} \sum_{m = -\infty}^{\infty} J_m (2 \pi n f x) \int\limits_{-\tobs/2}^{\tobs/2} \! \! \! {\rm e}^{i 2\pi [n (\nu - f) + m \forb] t} \diff t \Biggr|^2 \,.
		\end{aligned}
	\end{equation}
	\end{linenomath}
	On the right-hand side we have included the diagonal $j=k$ term, which is absent on the left-hand side, but is negligible in the limit where the number of photons $N$ is large.  The integral over time is
	\begin{linenomath}
	\begin{equation} \label{eq:sincfunc}            
		\frac{1}{\tobs} \int\limits_{-\tobs/2}^{\tobs/2} {\rm e}^{i 2\pi F t} \diff t = \frac{\sin( \pi F \tobs)}{ \pi F \tobs},
	\end{equation}
	\end{linenomath}
	with  $F = n (\nu - f) + m \forb$.  For observation times that include many orbits, the right-hand side of Eq.~\eqref{eq:sincfunc} is unity for $F = 0$ and is negligible otherwise.  Thus, the only terms in Eq.~\eqref{eq:simplification} that survive are those for which $ \nu = f - m \forb / n$.  When that is satisfied, we have
	\begin{linenomath}
	\begin{equation} \label{eq:doublesum}
		\sum_{j \ne k} {\rm e}^{i n \bigl(\Delta \Phi_j - \Delta \Phi_k \bigr)} \approx N^2 J_m^2 (2 \pi n f x) \,,
	\end{equation}
	\end{linenomath}
	where $m$ is constrained by $F = 0$.  Thus, the double sum in Eq.~\eqref{eq:simplification} vanishes at all frequencies $\nu$ except for the ``sideband frequencies'' $\nu = \nu_m = f - m \forb / n$, where $m$ takes on all integer values.

	We now evaluate the \ac{SNR} $\theta_{P_n}^2 (\nu)$ from Eq.~\eqref{eq:SNR_Pn} by substituting in Eq.~\eqref{eq:doublesum}, assuming that the weights $w_j$ are constant.  For the reasons just given, $\theta_{P_n}^2 (\nu)$ vanishes except at the discrete sideband frequencies $\nu_m = f - m \forb / n$.  We obtain
	\begin{linenomath}
	\begin{equation} \label{eq:snr_onesideband}
		\theta_{P_n}^2(\nu) =
		\begin{cases}
			J_m^2( 2 \pi n f x ) \, \theta_{P_n}^2 & \,\, \text{for $\nu = \nu_m$, $m \in \mathbb{Z}$\,,} \\
			0 & \,\, \text{otherwise\,.}
		\end{cases}
	\end{equation}
	\end{linenomath}
	The quantity $\theta_{P_n}^2$ that appears on the right-hand side is given by Eq.~\eqref{eq:coh_snr_p}.  It is the \ac{SNR} that the pulsar would have in an isolated search if the binary motion were absent.  It is also the \ac{SNR} that the pulsar would have in a binary pulsar search at the true signal parameter values.

	The structure in frequency space $\nu$ is evident from Eq.\eqref{eq:snr_onesideband}.  As described by \cite{ransom2003}, the \ac{SNR} is spread over equally spaced sidebands around the pulsar frequency $f$, whose spacing is commensurate with the orbital frequency.  The sideband width is $\sim 1/\tobs$, as can be seen from Eq.~\eqref{eq:sincfunc}.

	In comparison with a binary pulsar search, the isolated pulsar search has lost some \ac{SNR}, since $J_m^2 \le 1$.  To recover some of the lost \ac{SNR} within the isolated pulsar search, we introduce a new test statistic that sums over the first $m_\text{orb}$ sidebands around the central pulsar frequency.  This cumulative sideband power may be written as
	\begin{linenomath}
	\begin{equation} \label{eq:def_bn}
		B_n (\nu) = \sum\limits_{m = -m_\text{orb}}^{m_\text{orb}} P_n \left(\nu - \frac{m f_\text{orb}}{n}\right) \,, \\
	\end{equation}
	\end{linenomath}
	with the detection statistic $P_n (\nu)$ appropriate to an isolated pulsar search with parameters $\{\nu,\dot{f},n_x,n_y,0,0,0\}$.  (A test statistic weighing the $m$th sideband in Eq.~\eqref{eq:def_bn} by $J_m^2(2 \pi n f x)$ would be more sensitive, but for simplicity it is not considered here.)

	The \ac{SNR} for the cumulative sideband power $B_n$ is easily calculated.  It is defined as
	\begin{linenomath}
	\begin{equation} \label{eq:defsnr}
		\theta_{B_n}^2  = \frac{\operatorname{E}_p[B_n]-\operatorname{E}_0[B_n]}{\sqrt{\operatorname{E}_0[B_n^2]-\operatorname{E}_0^2[B_n]}} \,,
	\end{equation}
	\end{linenomath}
	where $p$ is the pulsed fraction defined in Eq.~\eqref{eq:probdist1}.  The numerator of Eq.~\eqref{eq:defsnr} can be found from Eq.~\eqref{eq:snr_onesideband}, which implies that $\operatorname{E}_p[P_n]-\operatorname{E}_0[P_n] = \operatorname{Var}_0[P_n] \theta_{P_n}^2 J_m^2( 2 \pi n f x ) = 4 \theta_{P_n}^2 J_m^2( 2 \pi n f x )$.  Summing this over $m$ gives the numerator:
	\begin{linenomath}
	\begin{equation}
	\operatorname{E}_p[B_n]-\operatorname{E}_0[B_n] = 2  \theta_{P_n}^2 \sum_{m=-m_\text{orb}}^{m_\text{orb}}  J_m^2( 2 \pi n f x ) \,.
	\end{equation}
	\end{linenomath}
	The denominator of Eq.~\eqref{eq:defsnr} is defined in the absence of a signal, with $p=0$.  It is easily calculated if the noise at the different frequencies that contribute to the sum is independent.  Since Poisson noise is stationary, these contributing terms will be independent if they are spaced more than one frequency bin apart, where the bins have width $1/n\tobs$.  Since the sideband frequencies are separated by $f_\text{orb}/n$, these different terms will be independent if there are many orbits in the observation time: $f_\text{orb} \tobs \gg 1$.  Each term in the denominator then has variance 4, so the sum yields $\operatorname{E}_0[B_n^2] - \operatorname{E}_0^2[B_n] = 4(2 m_\text{orb} + 1)$.  Thus, the \ac{SNR} for $B_n$ is
	\begin{linenomath}
	\begin{equation} \label{eq:snr_sb}
		\theta_{B_n}^2 = \frac{\theta_{P_n}^2}{\sqrt{ 2 m_\text{orb} + 1 }} \sum\limits_{m = -m_\text{orb}}^{m_\text{orb}} J_m^2( 2 \pi n f x ) \,.
	\end{equation}
	\end{linenomath}
	To maximize this \ac{SNR}, what is the optimal number of sidebands $m_\text{orb}$ to include?

	As shown by \cite{ransom2003}, the optimal number of sidebands to include depends on
	\begin{linenomath}
	\begin{equation}
		M_\text{orb} = [ 2 \pi n f x ] \,,
	\end{equation}
	\end{linenomath}
	where square brackets denote ``integer part''.  To see this, consider the sum that appears in Eq.~\eqref{eq:snr_sb}:
	\begin{linenomath}
	\begin{equation} \label{eq:besselsum}
		\sum\limits_{m = -m_\text{orb}}^{m_\text{orb}} J_m^2( 2 \pi n f x ) \,.
	\end{equation}
	\end{linenomath}
	For $m_\text{orb} < M_\text{orb}$ this sum grows (approximately linearly) with increasing $m_\text{orb}$.  But the addition theorem for Bessel functions ensures that Eq.~\eqref{eq:besselsum} stops growing and approaches unity as soon as $m_\text{orb}$ exceeds $M_\text{orb} $.  Since the denominator of the \ac{SNR} in Eq.~\eqref{eq:snr_sb} has a term that grows like $\sqrt{ 2 m_\text{orb} + 1 }$, the \ac{SNR} is maximized for $m_\text{orb} = M_\text{orb}$.  For this number of sidebands, one thus obtains
	\begin{linenomath}
	\begin{equation} \label{eq:snr_sb_Morb}
		\theta_{B_n}^2 \approx \frac{\theta_{P_n}^2}{\sqrt{ 2 M_\text{orb} + 1 }}
	\end{equation}
	\end{linenomath}
	for the expected \ac{SNR} of the cumulative sideband power.

	The behavior we have just described, considered alongside the definition \eqref{eq:defsnr} of the \ac{SNR}, shows the main weakness of sideband searches.  The numerator grows (approximately) linearly as we include more sidebands, meaning that we can recover all of the signal power.  But, in the absence of a signal, $B_n$ undergoes a random walk as sidebands are included, and so the denominator of Eq.~\eqref{eq:defsnr} (the root-mean-squared of $B_n$ in the absence of a signal) increases as $\sqrt{2M_{\text{orb}} + 1}$.  Thus, in comparison with an optimal matched filter, the incoherent summation over sidebands loses a factor of $\sqrt{ 2 M_\text{orb} + 1 }$ in the \ac{SNR}.  This is explicit in Eq.~\eqref{eq:snr_sb_Morb} and makes sideband searches ineffective if there are many sidebands, as is often the case. For example, consider the potential circular binary pulsar in \circsrc{} and the potential eccentric binary pulsar in \ellisrc{} discussed earlier in this paper.  Their estimated parameter ranges in $f$ and $x$ give rise to large numbers of sidebands.

	This means that sideband searches work best if only a few sidebands are expected, meaning that $2 \pi x f$, the total rotational phase arising from the orbital modulation, is small.  This is the case for black widow systems, which have very light companions.  The small companion mass means that the pulsar orbits very close to the center of mass, so the projected semimajor axis $x$ is extremely small.  Note that the modulation can be small even for the high frequencies $f$ typically found for black widows.

\begin{figure}
		\centering
		\includegraphics[width=\columnwidth]{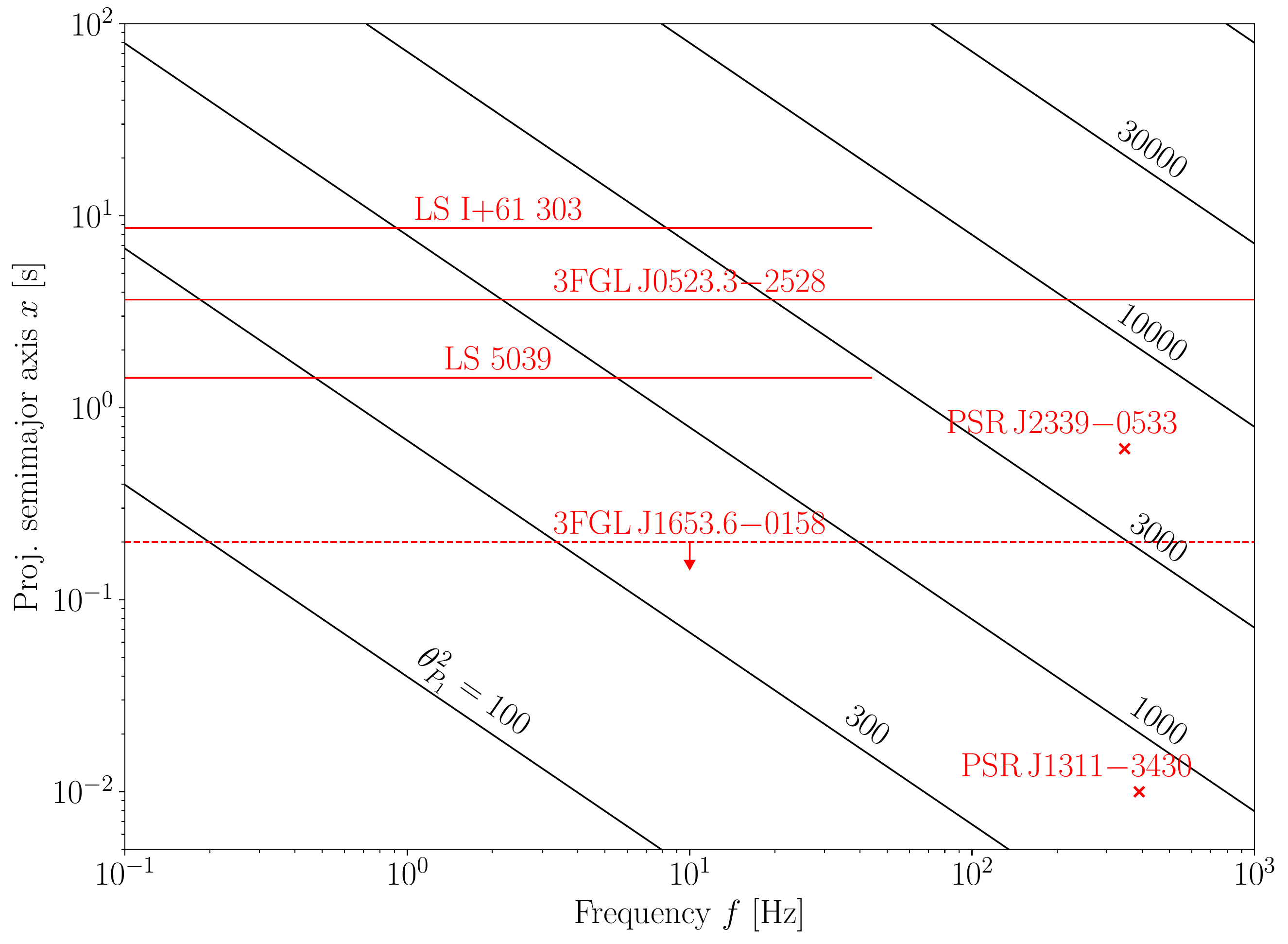}
		\caption{\label{f:snr_sideband} Comparison between the expected cumulative sideband \ac{SNR} $\theta_{B_1}^2$ and the expected optimal matched-filter \ac{SNR} $\theta_{P_1}^2$.  For given frequency $f$ and semimajor axis $x$, the black contours show the $\theta_{P_1}^2$ required to exceed a threshold $\theta_{B_1}^2 > 100$.  The crosses are at the locations of two known pulsars: PSR\,J1311$-$3430 and PSR\,J2339$-$0533.  The red lines show four potential sideband search candidates.  For the \ac{YP} candidates LS 5039 and LS I+61 303, and the spider candidate \ellisrc{}, the approximate values for the semimajor axes are known.  The dashed line shows the maximum semimajor axis value for  \circsrc{}.}
	\end{figure}

	Figure~\ref{f:snr_sideband} illustrates this, for example, for the black widow pulsar \psr{}, which would have been a candidate for a sideband search.  The figure shows the expected optimal matched-filter \ac{SNR} $\theta_{P_1}^2$ required to exceed a threshold in the expected cumulative sideband \ac{SNR} $\theta_{B_1}^2 > 100$, which is a reasonable threshold for confident detection.  From Eq.~\eqref{eq:snr_sb_Morb}, this requires $\theta_{P_1}^2$ to exceed $100\sqrt{ 2 M_\text{orb} + 1 }$.  Hence, $M_\text{orb}$ is constant on the contour lines, which therefore denote boundaries of constant $f\,x$.  Since the largest observed $\theta_{P_1}^2$ values for known pulsars are $\sim 1000$, the region below and to the left of the contour line corresponding to $\theta_{P_1}^2 = 1000$ might be considered for sideband searches.

	Sideband searches within gamma-ray binaries like LS\,$5039$ and LS\,I\,$+61\,303$ would also be justified.  These systems contain a compact object: a black hole or neutron star.  Since both binaries are highly eccentric \citep[0.3 < e < 0.6; ][]{aragona2009}, the compact objects could be \acp{YP}.  These two candidate pulsars are both displayed in Figure~\ref{f:snr_sideband}.  This is purely illustrative, since the sideband power $B_n$ defined here is only suitable for circular binary pulsars.  Eccentric pulsars will have additional sidebands \citep{ransom2003} and thus must have an even higher pulsed fraction to be detectable in a sideband search.

	This section has not discussed the implementation of a practical sideband search.  We would need some constraints on the parameters $\forb$ and $x$ to hunt for the sidebands.  If those are available from optical observations, then the sky position will be known precisely.  This in turn would make a fully coherent isolated pulsar search computationally feasible.  The resulting test statistics could then be used to construct the sideband search statistic $B_n$ of Eq.~\eqref{eq:def_bn}.

\extraspace
\subsection{Discussion}

	The methods discussed in this section have little applicability to searches for gamma-ray pulsars in spider systems, which are the main focus of this paper.  But they are of interest for other types of binary systems.

	Acceleration searches could discover binary pulsars with orbital periods comparable to or longer than our observation time $\tobs \sim 10\,\text{yr}$.  These binaries have pulsars whose companions are very low mass stars or planets, in wide orbits.  These pulsars might have been missed by isolated pulsar searches.

	Stack/slide and power spectrum methods do not appear suitable for spider gamma-ray pulsar searches.  They might potentially detect systems with orbital periods longer than our typical coherence time $\tcoh \sim 24\,\text{d}$ and shorter than $\tobs \sim 10\,\text{yr}$.  However, these searches are very expensive computationally.

	Sideband searches could be used to hunt for binary pulsars with low spin frequencies or in very close orbits.  While these are computationally less expensive than the search methods discussed earlier in this paper, they are also considerably less sensitive.

	All of these methods have a domain of applicability.  Given prior knowledge and constraints on a specific target, one can investigate these different methods to determine which are feasible and to estimate which one is potentially the most sensitive.

%%%%%%%%%%%%%%%%%%%%%%%%%%%%%%%%%%%%%%%%%%%
\extraspace
\section{Conclusions} \label{s:conclusions}

	This work presents computationally efficient methods to detect circular and eccentric orbit binary gamma-ray pulsars\deleted{ in blind searches}.  These generalize techniques that have been previously developed to search for isolated pulsars \citep{pletsch2014}.

	We have presented all of the elements of this generalization.  Physically, the central element is a model that accurately describes the rotational phase of a pulsar over time as would be observed at the \acl{SSB}.  In comparison with the isolated model, this must also account for the R\o mer delay caused by the binary motion.  A second key element are semicoherent and coherent test statistics, along with their expected \aclp{SNR}.  The last key element are the metrics for these statistics, which measure the ``distance'' in parameter space between two different rotational phase models.  This metric quantifies the expected fractional loss in \acl{SNR}, and enables the construction of efficient parameter space grids for a search.

	We have shown how these different elements can be used together to search for gamma-ray pulsars.  This is analogous to the isolated pulsar case \citep{pletsch2014}: the most computationally efficient approach is a multistage search with several semicoherent and coherent stages.  The computing cost is proportional to the number of points in the parameter space grid.  We compute this from the metric and show how the computing cost depends on the search parameters.  This in turn allows the grid spacing to be optimized, achieving the highest possible sensitivity at fixed computing cost.  These methods have been very successful in discovering isolated gamma-ray pulsars \citep{pletsch2012a,clark2015,clark2016,clark2017,clark2018}.

	\replaced{A truly blind}{Currently, a} search for binary pulsars \added{without partial information about the sky position and constraints on the orbital parameters} is computationally impossible.  Because the parameter space has at least seven of the nine possible dimensions $\{f, \dot{f},\alpha,\delta,x,\Porb,\tasc,\epsilon_1,\epsilon_2\}$, too many grid points are needed to cover it.  However, in some cases, the number of dimensions can be reduced and/or the corresponding search ranges can be tightly constrained by multiwavelength observations.  \deleted{Such searches may be characterized as ``vision impaired'' rather than ``blind''.}

	This paper considers two illustrative examples of this type, drawn from potential spider pulsars.  Here, analysis of optical observations constrains the orbital parameters, and we show that searches of reasonable sensitivity (in some cases limited to \aclp{YP}) are feasible.  This enables \replaced{``blind''}{partially informed} searches for binary gamma-ray pulsars that were previously not feasible.  This is important because these pulsars might be impossible to detect in other wave bands.

	The methods of this paper, particularly the metric in parameter space, have applications beyond \replaced{blind}{partially informed} searches.  There are binary pulsars that are visible in radio, optical, or X-ray, for which gamma-ray pulsations have not yet been found.  For recent discoveries, precise determination of their orbital and other parameters is often not possible, since it requires observations spanning several years.  The methods here are useful in such cases, to carry out efficient follow-up searches to discover gamma-ray pulsations.  This way, within days or weeks after radio pulsations are discovered, the pulsar's parameters can be precisely measured over the $>10\,\text{yr}$ of elapsed \ac{LAT} mission time.  This approach led to the discovery of gamma-ray pulsations soon after the radio detection of the $707\,\text{Hz}$ black widow pulsar PSR\,J0952$-$0607 \citep{bassa2017b,nieder2019}.

	A significant shortcoming of this paper's methods is that the number of grid points and hence the required computing resources grow quickly with increasing frequency $f$ and semimajor axis $x$.  To make searches feasible, it might be necessary to balance a reduced search range (smaller maximum $f$ and/or $x$) versus a reduced search sensitivity (wider grid spacing and/or shorter coherence time).  Even with large computing resources like \EatH{}, \acl{MSP} searches for binaries with $x \sim $ seconds are only feasible if the orbital parameters are precisely constrained.

	The second significant shortcoming is that search sensitivity is lost if the  pulsar's rotational phase does not match our model.  This can happen for several types of pulsars and binary systems.  This paper assumes that the intrinsic spin frequency $f$ varies linearly with time.  It does not include the time-dependent variations or the unpredictable frequency glitches often seen in \aclp{YP}.  This means that pulsars could be ``detected'' in the semicoherent stages of a search but are then discarded after the coherent stage, because they did not match the phase model well enough to produce a significant detection statistic \citep[see, e.g.,][]{clark2017}.  Phase model mismatch can also arise from time-dependent variations of the orbital period $\Porb$, which seems to be common in redback systems \citep[see, e.g.,][]{pletsch2015}.  For pulsars in short orbital period binaries with heavy companions, post-Keplerian gravitational corrections also have to be taken into account \citep[see, e.g.,][]{damour1986,edwards2006}.

	Because of these limitations, this paper also evaluates alternative search methods, which have previously been used in radio and gravitational-wave searches.  While these may be applied to search for binary gamma-ray pulsars, only the sideband search methods appear to have some chance to detect tight-orbit spider pulsars, which are the main focus of this paper.

	A more detailed study is necessary to make a fair sensitivity comparison between the sideband search and this paper's methods.  Indeed, while the cumulative sideband power loses a lot of \acl{SNR} compared to this paper's methods, it might be improved.  Since the sideband structure follows a known form, one could obtain a larger \ac{SNR} by assigning weights to the sidebands before summing them, rather than using equal weights as done here.

	We have implemented the new methods developed in this paper in a mixture of C and Python codes.  These have been tested using simulated pulsar signals, both with our own code and with the widely used TEMPO2 package \citep{hobbs2006}.  We are confident that these codes work correctly, in part because they have discovered new spider pulsars, soon to be published.

	We are currently using these codes and methods to hunt for spider pulsars in the unassociated sources of the \acl{4FGL}.  These \replaced{``blind''}{partially-informed} searches are guided by orbital constraints from optical observations.  The orbital grids are constructed on the computing cluster \atlas{} at the Albert Einstein Institute in Hannover.  The first two (semicoherent) stages and the third (coherent) stage are all done on \EatH{}, whose volunteers provide a massive computing pool.  The final, less computation-demanding ($H$ statistic) follow-up stage is done on \atlas{}.  To increase the computing power available in the initial stages of the search, we ported the search codes to work on \EatH{} volunteer's GPUs.  The \atlas{} cluster is also used to carry out follow-up gamma-ray searches of newly discovered radio pulsars, to refine the parameters as discussed above and in \cite{nieder2019}.

	This paper has used the two gamma-ray sources \circsrc{} and \ellisrc{} as examples, to show how a realistic search might be structured.  Both of these searches are being or have been carried out, and the results will be discussed in upcoming papers.

	\added{The reader might wonder if these methods work in practice.  They do, and they have already detected three spider pulsars.}  \replaced{A similar search for a pulsar within 3FGL\,J2039.6$-$5618 \citep{romani2015b,salvetti2015} successfully detected pulsations using the methods presented here (C.J. Clark et al. 2020, in prep.).}{A preliminary version detected \psr{} \citep{pletsch2012}.  The current version successfully detected pulsations within 4FGL\,J2039.5$-$5617, by exploiting partial information \citep{romani2015b,salvetti2015}.  This confirmed that it is a redback and provides an $11$ yr phase-connected rotational ephemeris \citep{clark2020}.  The search for \circsrc{}, described in Section~\ref{s:J1653}, also resulted in a black widow \ac{MSP} discovery \citep{nieder2020b}.}

	\added{That these methods work is not surprising: the different parts have been tested and demonstrated.  \textit{The metric approximation for the orbital parameters} was demonstrated to be a good fit to the actual mismatch for typical parameters as presented in Figures~\ref{f:triplot_circ} and \ref{f:triplot_elli}.  The metric was used in a successful follow-up search shortly after the radio discovery of the fastest-spinning pulsar known in the Galactic field \citep{nieder2019}.  \textit{The approximate phase model for elliptical orbits} was verified on simulated data with the results up to fifth order in eccentricity shown in Fig.~\ref{f:model_error_ex}.  \textit{The test statistics and the multistage search} approach have already detected more than $30$ isolated pulsars \citep{clark2017,clark2018}.}

	\added{A topic we have not addressed is timing analysis.  Following detection, this ``pins down'' the parameters as precisely as possible.  An interesting and useful feature is that, regardless of the path to detection, if the pulsar is bright enough in gamma rays, the \Fermi{}-\ac{LAT} all-sky data immediately allow one to extend the ephemeris back to the launch of the \Fermi{} satellite in August 2008 \citep{ray2011,kerr2015b}.  This determines many of the pulsar's parameters with much higher precision than is typical soon after radio/X-ray discoveries.  For those, an additional campaign of timing observations is required to infer astrophysical properties.}

	\added{Gamma-ray timing analyses of \ac{LAT}-discovered pulsars, which often remain undetected in radio, have led to several interesting discoveries \citep[see, e.g.,][]{allafort2013,lyne2015,schinzel2019}.  \ac{LAT} data were used to resolve the variations in the orbital period of a binary pulsar, which was difficult to observe in radio owing to large eclipses \citep{pletsch2015}.  The previously mentioned study of PSR\,J0952$-$0607 is another example.}

	The outlook for future searches is promising.  The \citetalias{gaia2018} Catalog provides sky locations for the spider companions, which are precise enough so that no search in $\{\alpha,\delta\}$ is required.  In addition, since the \acl{LAT} mission is ongoing, data sets are getting longer.  Current searches use $\tobs \sim 11\,\text{yr}$ of data, compared with initial searches with $\tobs\sim 4\,\text{yr}$.  Furthermore, our available computing power is also increasing with time.  This means that current searches employ $\tcoh\sim 24\,\text{d}$ in the first stage, compared with initial searches with $\tcoh\sim 12\,\text{d}$.  Since search sensitivity scales with $(\tcoh\tobs)^{1/4}$ \citep{pletsch2014}, our current sensitivity has increased by more than $50\%$.  We believe that $\mathcal{O} (10-30)$ of the unassociated sources in the \ac{4FGL} Catalog are undiscovered spider pulsars and that we can find some of them.

	There are systems that are very likely to be spider gamma-ray pulsars for which the orbital constraints are not yet good enough to perform searches.  These include the five redback pulsar candidates: \replaced{3FGL\,J0212.1$+$5320}{4FGL\,J0212.1$+$5321} \citep{li2016,linares2017b}, \replaced{3FGL\,J0744.1$-$2523}{4FGL\,J0744.0$-$2525} \citep{salvetti2017}, \replaced{3FGL\,J0838.8$-$2829}{4FGL\,J0838.7$-$2827} \citep{halpern2017}, \replaced{3FGL\,J0954.8$-$3948}{4FGL\,J0955.3$-$3949} \citep{li2018}, and the recent 4FGL\,J2333.1$-$5527 \citep{swihart2020}.  We hope that this work helps motivate additional optical observations to improve these constraints and enable new gamma-ray pulsar discoveries.

%%%%%%%%%%%%%%%%%%%%%%%%%%%%%%%%%%%%%%%%%%%
\acknowledgments{}
	We thank Anne Archibald and Andrea Belfiore for encouraging us to look into sideband search methods.  This work was supported by the Max-Planck-Gesellschaft~(MPG), by the Deutsche Forschungsgemeinschaft~(DFG) through Emmy Noether research grant No. PL~710/1-1 (PI: Holger~J.~Pletsch), and by National Science Foundation grants 1104902 and 1816904.  This work was supported by an STSM Grant from COST Action CA16214.  C.J.C. acknowledges support from the ERC under the European Union's Horizon 2020 research and innovation program (grant agreement No. 715051; Spiders).

\software{\textsc{tempo2} \citep{hobbs2006,edwards2006}, \texttt{matplotlib} \citep{matplotlib2007}, \texttt{NumPy} \citep{numpy2006,numpy2011}}

%%%%%%%%%%%%%%%%%%%%%%%%%%%%%%%%%%%%%%%%%%%
\appendix

\section{Expectation values of signal statistics} \label{s:expectedvalues}

	Here we show how to calculate the expectation values of signal statistics.  The statistics depend on the $j=1, \dots, N$ modeled pulsar rotation phases at the photon arrival times $t_j$.  To simplify the language and notation, we suppose that the vector of parameters $\pvec = \{f,\dot{f},\alpha,\delta\}$ is fixed and denote the modeled rotation phases by $\Phi_j = \Phi(t_j, \pvec) = \Phi(t_{\text{psr}}(t_j,\alpha, \delta),f,\dot{f})$.  Sums and products over $j,k,\ell$\\ run from $1, \dots ,N$ unless otherwise specified.  Finally, we write ``the phase of the $j$th photon'', rather than ``the modeled pulsar rotational phase associated with the $j$th photon''.

	Our key assumption is that the phase of each photon is an independent (hence uncorrelated) random variable.  This is justified because the number of photons detected is much less than one per pulsar revolution.  The phase $\Phi_j$ of the $j$th photon is drawn from the distribution $F_j(\Phi_j)$ as given in Eq.~(\ref{eq:probdist1}).  Thus, using Eq.~(\ref{eq:probdist2}), the probability distribution function of $\Phi_j$ is
	\begin{linenomath}
	\begin{equation} \label{eq:probdist3}
		F_j(\Phi_j) = \frac{1}{2 \pi} + \frac{p w_j}{2 \pi} \sum_{n=-\infty}^\infty \gamma_n {\rm e}^{i n \Phi_j} \,,
	\end{equation}
	\end{linenomath}
	where the Fourier coefficients $\gamma_n$ are defined by Eq.~(\ref{eq:gammadef}) for $n>0$, by $\gamma_n = \gamma^*_{-n}$ for $n<0$, and by $\gamma_0 = 0$ for $n=0$.

	The expectation value of any quantity $Q(\Phi_1, \dots, \Phi_N)$ is now given by
	\begin{linenomath}
	\begin{equation}
		E[Q] = \int_0^{2 \pi} \! \! \! \! \! \! \! \diff\Phi_1 F_1(\Phi_1) \cdots \int_0^{2 \pi} \! \! \! \! \! \! \! \diff\Phi_N F_N(\Phi_N) Q(\Phi_1, \cdots, \Phi_N) \,,
	\end{equation}
	\end{linenomath}
	where the statistical independence of the rotation phases allows the probability density to be written as a product.  For example, the expected value of $\exp(-i n \Phi_j)$ is
	\begin{linenomath}
	\begin{equation} \label{eq:onephase}
		\begin{aligned}
			E[  {\rm e}^{-i n \Phi_j}] &  =    \int_0^{2 \pi} \! \! \! \! \! \! \! \diff\Phi_j F_j(\Phi_j) {\rm e}^{-i n \Phi_j} \\
			& = \delta_{n0} + \frac{p w_j}{2 \pi} \sum_{m=-\infty}^\infty \int_0^{2 \pi} \! \! \! \! \! \! \! \diff\Phi \gamma_m {\rm e}^{i(m-n)\Phi} \\
			& = \delta_{n0} + p\, w_j\, \gamma_n \,,
		\end{aligned}
	\end{equation}
	\end{linenomath}
	where $\delta_{nm}$ is the Kronecker delta, giving unity for $n=0$.

	The expected value of the coherent power signal statistic Eq.~(\ref{eq:cohpower}) is
	\begin{linenomath}
	\begin{equation}
		E[P_n] = \kappa^{-2} \sum_{j,k} w_j w_k \prod_{\ell} \int_0^{2 \pi} \diff\Phi_\ell F_\ell(\Phi_\ell) {\rm e}^{i n(\Phi_k - \Phi_j)}.
	\end{equation}
	\end{linenomath}
	In the product above, only two terms are nontrivial, for which either $\ell = k$ or $\ell = j$.  The integrand does not depend on the other $N-2$ integration variables, whose corresponding integrals give unity, since the probability density is normalized.  One obtains
	\begin{linenomath}
	\begin{equation} \label{eq:finalexpec}
		\begin{aligned}
			E[P_n] &= \kappa^{-2} \sum_j w_j^2 + \kappa^{-2} p^2 \sum_{\substack{j,k \\j \ne k}} w^2_j w^2_k | \gamma_n |^2  \\
			&= 2 + 2 p^2 |\gamma_n|^2 \biggl[ \sum_j w^2_j - \frac{\sum_j w_j^4}{\sum_j w_j^2} \biggr] \,.
		\end{aligned}
	\end{equation}
	\end{linenomath}
	On the first line, the first sum comes from terms with $j = k$ and the second sum from terms where $j \ne k$, and we have used Eq.~\ref{eq:onephase} to simplify both terms.

\section{Maximal Sensitivity at Fixed Computing Cost} \label{s:optimalmismatch}

	The sensitivity of a search can be quantified via the pulsed fraction $p$ defined in Eq.~\eqref{eq:probdist1}.  More sensitive searches can detect sources with smaller values of $p$.

	If infinite computing power were available, we would employ the fully coherent detection statistics $H$ or $P_1$, and the sensitivity of a search would only be limited by the data.  To determine that ultimate sensitivity, consider the expected \ac{SNR} $\theta_{P_1}^2$ given in Eq.~\eqref{eq:coh_snr_p}.  A point in parameter space where $\theta_{P_1}^2$ exceeded some threshold $\theta_{\rm threshold}^2$ (established by the desired false-alarm and false-dismissal probabilities) would be counted as a detection.  A reasonable detection threshold might be $\theta_{\rm threshold}^2 = 50$, corresponding to pulsed fraction sensitivity $p^2 > \theta_{\rm threshold}^2 /|\gamma_1|^2 \mu \tobs$.  For typical values of $\mu \tobs = 500$ effective photons and $|\gamma_1|^2 =0.8$, this gives an ultimate, data-limited sensitivity of $p^2 > 0.13$.

	In practice, with limited computing power, we adopt the multistage hierarchical approach described in Section~\ref{s:multistage}.  A sensible choice is to use most of the computing power in the first, semicoherent stage.  Roughly speaking, this is because a signal will only be found if it rises above the detection threshold in the first stage of the search\footnote{Of course, this depends on the choice of threshold and the region of parameter space around a candidate that is searched in the subsequent stages.  If the full parameter space is searched for each candidate, then the statement is false!}. Hence, we will assume that our sensitivity is limited by the first semicoherent search stage.

	The maximum possible sensitivity of the semicoherent stage is determined by the threshold on the semicoherent \ac{SNR}, whose expected value is given in Eq.~\eqref{eq:scoh_snr_p}.  The threshold is lower than before, typically $\theta_{S_1}^2 > \theta_{\rm threshold}^2 = 10$.  Using search parameters from Eq.~\eqref{eq:assumptions} and later in that section gives a minimum detectable pulsed fraction of $p^2 > \theta_{\rm threshold}^2/|\gamma_1|^2 \mu \sqrt{\tobs \tcoh} = 0.31$.  As before, this is the theoretical sensitivity that could be achieved with unlimited computing power, but employing the semicoherent statistic.

	In practice, we must take the computing cost into account.  This cost is proportional to the number of grid points in parameter space at which the detection statistic is calculated.  Reducing the number of grid points (corresponding to a larger average mismatch) loses some \ac{SNR} but the additional computing power may be used to increase the coherence time $\tcoh$, which increases the \ac{SNR}.  What compromise maximizes the search sensitivity for a given computing cost?

	To find the optimal balance between the worst-case grid mismatch $m$ and the coherent integration time $\tcoh$, we maximize the sensitivity with the constraint that the computing power is fixed, as described in \cite{prix2012} and \cite{pletsch2014}.  What is important is the rate at which the number of grid points grows with increasing $\tcoh$, which in turn depends on the dimension of the parameter space.

	The number of dimensions $d$ in the search parameter space is determined by our prior knowledge.  To quantify that, we use $\norb$ (possible values 3 or 5) for the number of orbital parameters searched and $\nsky$ (possible values 0 or 2) for the number of sky dimensions searched, so $d=2+\norb+\nsky$.  In the case of an eccentric binary with poorly known position, we have the full parameter space discussed in the main text, $\{f, \dot{f}, n_x, n_y, \Omorb, x, \tasc, \epsilon_1, \epsilon_2\}$, so $\norb=5$, $\nsky=2$, and $d=9$.  For an eccentric binary whose position is precisely known (for example, from optical observations), $\{n_x, n_y\}$ are omitted from the search, $\norb=5$, $\nsky=0$, and $d=7$.  For a circular binary whose position is precisely known, $\{\epsilon_1, \epsilon_2\}$ are also omitted, so $\norb=3$, $\nsky=0$, and $d=5$.

	The smallest detectable pulsed fraction (averaged over signal location in parameter space) may be written as
	\begin{linenomath}
	\begin{equation} \label{eq:minpulsedfraction}
		p_{S_1}^{2} = \frac{\theta_{\rm threshold}^2}{ (1- \hat{m}) |\gamma_1|^2 \mu \sqrt{\tobs \tcoh}} \,.
	\end{equation}
	\end{linenomath}
	Here $\hat{m}$ represents the average (over parameter space) mismatch of the grid \citep{prix2012}.

	The construction of our parameter space grid is described following Eq.~\eqref{eq:gbar_ab_orb}; its average mismatch may be estimated as follows.  Within a given $8$\,Hz frequency interval, the grid is the direct product of an equally spaced grid in the frequency direction, an equally spaced grid in the $\dot f$ direction, a two-dimensional hexagonal lattice in sky position $\{n_x, n_y\}$, and an optimized stochastic grid in the orbital parameters.  Below, we call these ``subgrids''.  To determine the computing cost, we need to count the number of grid points in these subgrids and multiply them together.

	Because the metric has no off-diagonal terms that couple the different subgrids, the average parameter space mismatch $\hat m$ can be written as
	\begin{linenomath}
	\begin{equation} \label{eq:mbar}
		\hat{m} = \hmf + \hmfd + \hms + \hmo \,,
	\end{equation}
	\end{linenomath}
	where $\hmf$ is the average mismatch in the frequency dimension (if all other parameters are exactly matched to the signal) and $\hmfd$, $\hms$, $\hmo$ are the corresponding average mismatches in the $\dot f$, sky, and orbital subgrids (if all other parameters are exactly matched to the signal).

	The frequency dimension is searched with an \ac{FFT} whose frequency spacing $\diff f = 1/\tcoh$.  For the worst case, which is two points separated by $\diff f/2$, the quadratic metric approximation predicts a mismatch $\bar{g}_{ff}/(2 \tcoh)^2 = \pi^2/24 = 0.411$, and hence an average mismatch $\hat m_f = 0.14$.  As is often the case, the quadratic approximation slightly overestimates the mismatch; the spherical ansatz of \cite{allen2019} predicts a worst-case $m=\sin^2(\sqrt{\pi^2/24}) \approx 0.36$ which agrees well with the numerically measured value given in Section~5.2 of \cite{pletsch2014}.  In fact, as described before Eq.~(42) of that paper, we can reduce the average mismatch to $\hat m_f = 0.075$ at almost no extra computational cost, by interpolating the frequency spectrum.

	The $\dot f$ subgrid has uniform spacing $\diff \dot f$, and is an example of a regular lattice.  For regular lattices, the average mismatch $\hat m$ is related to the worst-case mismatch $m$ via $\hat m = \xi m$, where $\xi \in [0,1]$ is a lattice-dependent dimensionless geometrical factor called ``thickness' \citep{prix2012}.  Here we have a (one-dimensional) hypercubic grid, for which $\xi = 1/3$, so the average mismatch $\hmfd = m_{\dot f}/3$, where $m_{\dot f} = {\bar g}_{\dot{f} \dot{f}} (\diff\dot f/2)^2 = \pi^2 \tcoh^2 \tobs^2 \diff\dot f^2/288$ is the maximal mismatch in the $\dot f$ dimension.  (Since the differences are small, for simplicity we do not employ the spherical ansatz further.)

	The sky subgrid is a hexagonal lattice with thickness $\xi=5/12 \approx 0.416$.  Hence, $ \hms = 0.416 m_\text{sky}$, where $m_\text{sky}$ is the worst-case sky mismatch.

	The orbital parameter grid has an average mismatch that is well estimated during the process of its construction and can be easily controlled via the parameter that determines when new points are added to the stochastic bank.

	The computing cost is the product of the number of grid points in the nonfrequency dimensions with the cost of a single \ac{FFT}.  The number of grid points can be estimated using arguments like those given in deriving Eq.~\eqref{eq:norb}.  In each of the different subgrids, the number of grid points is proportional to ${\hat m}^{-D/2}$, where $\hat m$ is the average mismatch in that subgrid and $D$ is the dimension of that subgrid.  Hence, the number of grid points in the $\dot f$ subgrid is proportional to $\tcoh \hmfd^{-1/2}$, and the number of grid points in the sky subgrid is proportional to $\tcoh^{\nsky} \hms^{-\nsky/2}$.  The number of grid points in the orbital subgrid is proportional to $\hmo^{-\norb/2}$ and is independent of $\tcoh$.  Since the cost of an \ac{FFT} is proportional to $\tcoh \log \tcoh$, this gives a total computing cost $C$,
	\begin{linenomath}
	\begin{equation}
		C = C_0 \, \hmfd^{-1/2} \, \hms^{-\nsky/2} \, \hmo^{-\norb/2}\, \tcoh^{2+\nsky} \,.
	\end{equation}
	\end{linenomath}
	Here $C_0$ is a constant, and following \cite{pletsch2014}, we have omitted the slowly varying logarithmic factor from the cost of the \ac{FFT}.

	The method of Lagrange multipliers can be used to maximize sensitivity $p_{S_1}^{-2}$ at fixed computing cost\footnote{One obtains the same result by maximizing any negative power of $p_{S_1}$.}.  The quantity we extremize is
	\begin{linenomath}
	\begin{align}
		\mathcal{L} & = p_{S_1}^{-2} + \lambda C \\ \nonumber
		& = c_1 (1-\hat m) \tcoh^{1/2} + \lambda c_2 \hmfd^{-1/2} \hms^{-\nsky/2}  \hmo^{-\norb/2}  \tcoh^s \,,
	\end{align}
	\end{linenomath}
	where $\lambda$ is the Lagrange multiplier, $s=2+\nsky$, and $c_1$ and $c_2$ are constants (independent of the average mismatches and $\tcoh$).  Extremizing $\mathcal{L}$ with respect to the coherence time and the three different average mismatches gives
	\begin{linenomath}
	\begin{align}
		\nonumber
		\frac{\partial \mathcal{L} }{\partial \tcoh}    & = \frac{c_1}{2} (1-\hat m) \tcoh^{\, -1/2} + s \lambda c_2  \hmfd^{-1/2} \hms^{-\nsky/2}  \hmo^{-\norb/2} \tcoh^{s-1} =0 \nonumber \\
		\nonumber
		\frac{\partial \mathcal{L}  }{\partial \hmfd}   & = -  c_1 \tcoh^{1/2} - \tfrac{1}{2}\lambda c_2 \hmfd^{-3/2} \hms^{-\nsky/2}  \hmo^{-\norb/2}  \tcoh^s  =0 \\
		\nonumber
		\frac{\partial \mathcal{L}  }{\partial \hms}    & = -  c_1 \tcoh^{1/2} - \tfrac{{\nsky}}{2}\lambda c_2 \hmfd^{-1/2} \hms^{-\nsky/2-1}  \hmo^{-\norb/2}  \tcoh^s  =0 \\
		\nonumber
		\frac{\partial \mathcal{L}  }{\partial \hmo}    & = -  c_1 \tcoh^{1/2} - \tfrac{{\norb}}{2}\lambda c_2 \hmfd^{-1/2} \hms^{-\nsky/2}  \hmo^{-\norb/2-1}  \tcoh^s  =0 \,,
	\end{align}
	\end{linenomath}
	where we have made use of Eq.~\eqref{eq:mbar} to evaluate the derivatives of $\hat m$.

	\begin{deluxetable*}{lccccccc}[t]
		\tablecaption{\label{t:optimalmismatch} Comparison of computationally unlimited and optimal computationally limited semicoherent searches, showing mismatches and sensitivity}
		\tablecolumns{8}
		\tablehead{ Search & $\hat m$ & $\hmfd$ & $\hms$ & $\hmo$ & $m_{\dot f}$ & $m_{sky}$ & $p_{S_1}^{2}$ }
		\startdata
		Infinite computing cost (zero mismatch) grid      & 0     & 0     & 0     & 0     & 0     & 0     & 0.307 \\
		All parameters unknown ($\nsky=2$, $\norb=5$)     & 0.383 & 0.039 & 0.077 & 0.193 & 0.116 & 0.093 & 0.497 \\
		Elliptical, known position ($\nsky=0$, $\norb=5$) & 0.471 & 0.066 & 0     & 0.330 & 0.198 & 0.159 & 0.580 \\
		Circular, known position ($\nsky=0$, $\norb=3$)   & 0.383 & 0.077 & 0     & 0.231 & 0.231 & 0.185 & 0.497 \\
		Isolated ($\nsky=2$, $\norb=0$)                   & 0.221 & 0.049 & 0.097 & 0     & 0.146 & 0.117 & 0.394 \\
		\enddata
		\tablecomments{The columns show the average template bank mismatch $\hat m$, and the average mismatches in the $\dot f$, sky and orbital subgrids. (Note that the average per-dimension mismatch is constant.)  Then the corresponding maximum $\dot f$ and sky mismatch are listed with the (square of the) minimum detectable pulsed fraction $p$.  The first row shows the ideal semicoherent case where the grid points are infinitesimally spaced and the computing cost is infinite.  The next three rows illustrate smaller and smaller binary system parameter spaces.  The final row is for an isolated pulsar with unknown sky position.}
	\end{deluxetable*}

	To find the average mismatches that maximize the sensitivity at fixed computing cost, combine the first equation in turn with the second or third or fourth: $\tcoh$ drops out, and one obtains a closed form for the corresponding average mismatch.  The independence from coherence time $\tcoh$ in the binary pulsar case was previously shown for the isolated pulsar case by \cite{pletsch2014}.  For example, to solve for $\hmfd$, multiply the first equation by $\tcoh^{1/2}$, multiply the second equation by $2s \hmfd \tcoh^{-1/2}$, and add them.  One obtains $(1-\hat m)/2 - 2 s \hmfd =0$, whose solution is $\hmfd = (1-\hat m)/4s$.  Doing this for all three combinations yields
	\begin{linenomath}
	\begin{align}
		\nonumber
		\hmfd &= \frac{1-\hat m}{4(2+\nsky)} \,, \\ 
		\hmo  &= \frac{1-\hat m}{4(2+\nsky)} \norb \,, \text{ and} \\
		\nonumber
		\hms  &= \frac{1-\hat m}{4(2+\nsky)} \nsky \,.
	\end{align}
	\end{linenomath}
	Note that the optimal solution has equal average ``per-dimension'' mismatch in the non-frequency subgrids.  From Eq.~\eqref{eq:mbar} it follows that $ \hat m -\hmf$ is the sum of the three terms above, and since $d-1 = 1+\norb+\nsky$, we have $\hat m -\hmf = (1-\hat m)(d-1)/4(2+\nsky)$.  The solution is
	\begin{linenomath}
	\begin{equation}
		\hat m = \frac{1+\norb+\nsky + 4(2+\nsky)\hmf}{9+\norb+5\nsky} \,.
	\end{equation}
	\end{linenomath}
	Thus we have
	\begin{linenomath}
	\begin{align}
		\nonumber
		\hmfd &= \frac{1-\hmf}{9+\norb+5\nsky} \,, \\ 
		\hmo  &=  \frac{1-\hmf}{9+\norb+5\nsky} \norb \,, \text{ and} \\
		\nonumber
		\hms  &= \frac{1-\hmf}{9+\norb+5\nsky} \nsky \,,
	\end{align}
	\end{linenomath}
	which in turn allows us to determine the average and maximum mismatch in each of the subgrids, and the corresponding search sensitivity compared with an extremely finely spaced (but computationally very expensive) semicoherent search.

	In practice, after setting the mismatch as given by this optimal point, one adjusts the coherence time $\tcoh$ to be as long as allowed by the available computing resources. What does this imply about the sensitivity?  Above, we showed that with reasonable assumptions a semicoherent search can detect a pulsed fraction $p^2>0.31$ if there are infinite computing resources.  With finite computing resources, this is increased by a factor of $1/(1-\hat m) = (9+\norb+5 \nsky)/4(2+\nsky)(1-\hmf)$, as can be seen from Eq.~\eqref{eq:minpulsedfraction}.  The corresponding loss of sensitivity is shown in Table~\ref{t:optimalmismatch}.  The achievable pulsed fraction sensitivity is not far from the ideal case.

	This analysis extends previous work \citep{pletsch2014}, which assumed a grid with fixed thickness $\xi = 1/3$ in all dimensions.  However, this is not the case for current searches.  Here we have considered a grid that is a product of subgrids, each of which can have different geometrical properties, as used in existing searches.  If we assume fixed thickness, then our results and in particular the final line of Table~\ref{t:optimalmismatch} agree with Eq.~(H2) from \cite{pletsch2014}.

\section{High-order phase model for elliptical binaries}
\label{s:highorder}

	The main text uses a linear-in-$e$ ``ELL1'' approximation to the correct ``BT'' line-of-sight motion in eccentric orbits.  Here we consider higher orders in the eccentricity $e$.  The BT model is given in Eq.~\eqref{ecc_los_motion}:
	\begin{linenomath}
		\begin{gather}
			r_{z, \text{BT}}(t) = x \left[ \sin \omega (\cos E - e) + \cos \omega \sqrt{1 - e^2} \sin E \right]\,,\\
			E - e \sin E = M\,, \label{eq:app_kepler} \\ 
			M = \Omega_\text{orb} (t - T_0) \,.
		\end{gather}
	\end{linenomath}
	We express this as
	\begin{linenomath}
		\begin{equation}
			r_{z, \text{BT}}(t) = x \left[ \sin \omega \sum_{n = 0}^{\infty} \alpha_n(e) \cos(n\,M) + \cos \omega \sum_{n = 1}^{\infty} \beta_n(e) \sin(n\,M) \right] \,,
		\end{equation}
	\end{linenomath}
	where $\alpha_n(e)$ and $\beta_n(e)$ are power series in $e$. The goal here is to find these functions, and to determine the appropriate order needed for our searches.  (\cite{taff1985} gives an expansion of $\sin E$ and $\cos E$ in powers of $e$, but does not give a similar expansion for the line-of-sight motion.)

	For the derivation of the power series we introduce the Bessel functions and some of their properties. For positive integers $n$ the Bessel function can be expressed as the power series
	\begin{linenomath}
		\begin{equation}
			J_n(x) = \sum_{m = 0}^{\infty} \frac{(-1)^m \left(\frac{x}{2}\right)^{2m+n}}{m! (n+m)!}
		\end{equation}
	\end{linenomath}
	or in integral form as
	\begin{linenomath}
		\begin{equation}
			J_n(x) = \frac{1}{2 \pi} \int_{0}^{2 \pi} \cos( n \theta - x \sin \theta ) \diff \theta \,.
		\end{equation}
	\end{linenomath}
	The relation
	\begin{linenomath}
		\begin{equation}
			J_{n-1}(x) + J_{n+1}(x) = \frac{2 n}{x} J_n(x) \label{eq:propbessel}
		\end{equation}
	\end{linenomath}
	is also needed.

	Following \cite{taff1985}, we start with the Fourier expansion of $\cos E$:
	\begin{linenomath}
		\begin{equation}
			\cos E = \frac{\hat{\alpha}_0}{2} + \sum_{n = 1}^{\infty} \hat{\alpha}_n \cos(n \, M)\,.
		\end{equation}
	\end{linenomath}
	It has Fourier coefficients
	\begin{linenomath}
		\begin{equation}
			\hat{\alpha}_n = \frac{1}{\pi} \int_{0}^{2 \pi} \cos E \cos (n \, M) \diff M \,.
		\end{equation}
	\end{linenomath}
	Using Kepler's equation \eqref{eq:app_kepler} to write $M$ and $\diff M$ as functions of $E$, along with the integral form above, one obtains
	\begin{linenomath}
		\begin{equation}
			\hat{\alpha}_n = \frac{2}{n} \frac{\diff J_n (n\,e)}{\diff (n\,e)} \,.
		\end{equation}
	\end{linenomath}
	Using the power series above, this may be written as
	\begin{linenomath}
		\begin{equation}
			\hat{\alpha}_n(e) =
			\begin{cases}
				\sum\limits_{m = 0}^{\infty} \frac{(-1)^m \, (2m+n)}{n \, m! \, (m+n)!} \left( \frac{n}{2} \right)^{2m+n-1} e^{2m+n-1} &, \,\, n \ge 1 \\
			-\frac{1}{2}e &, \,\, n = 0 \,.
			\end{cases}
		\end{equation}
	\end{linenomath}
	The analogous calculation for $\sin E$ gives
	\begin{linenomath}
		\begin{equation}
			\sin E = \sum_{n=1}^{\infty} \hat{\beta}_n \sin(n \, M)
		\end{equation}
	\end{linenomath}
	with coefficients
	\begin{linenomath}
		\begin{equation}
			\hat{\beta}_n = \frac{1}{\pi} \int_{0}^{2 \pi} \sin E \sin(n\,M) \diff M \,,
		\end{equation}
	\end{linenomath}
	where
	\begin{linenomath}
		\begin{equation}
		\begin{aligned}
			\hat{\beta}_n =& \frac{2}{n\,e} J_n(n\,e) \\
			=& \sum_{m=0}^{\infty} \frac{(-1)^m \left(\frac{n\,e}{2}\right)^{2m+n-1}}{m! \, (n+m)!} e^{2m + n -1} \label{eq:coeff_b}
		\end{aligned}
		\end{equation}
	\end{linenomath}
	is obtained using the recursion relation above.

	To obtain $\beta$ from $\sqrt{1-e^2}\hat{\beta}$, we first express
	\begin{linenomath}
		\begin{equation}
		\sqrt{1 - e^2} = \sum_{k = 0}^{\infty} e^{2k} \left( \sum_{l = 0}^{2k} (-1)^l \binom{1/2}{l} \binom{1/2}{2k - l} \right) \,, \label{eq:sqrt}
		\end{equation}
	\end{linenomath}
	where we have introduced the generalized binomial coefficient
	\begin{linenomath}
		\begin{equation}
			\binom{r}{k} = \frac{r \cdot (r - 1) \cdots (r - (k-1))}{k!} \,.
		\end{equation}
	\end{linenomath}
	The Cauchy product of $\sqrt{1-e^2}$ and $\hat{\beta}$ gives
	\begin{linenomath}
		\begin{equation}
			\begin{aligned}
			\beta_n(e) = \sum_{m = 0}^{\infty}& e^{2m + n - 1} \sum_{k = 0}^{m} \frac{\left(\frac{n}{2}\right)^{(2m - 2k + n -1)}}{(m-k)! (n + m - k)!} \\ &\times \sum_{l = 0}^{2k} (-1)^{m - k + l} \binom{1/2}{l} \binom{1/2}{2k - l} \,.
			\end{aligned}
		\end{equation}
	\end{linenomath}
	The $\alpha_n$ follow directly from $\hat{\alpha}_n$, and differ only for $n=0$.

	\begin{figure*}
		\centering
		\includegraphics[width=\textwidth]{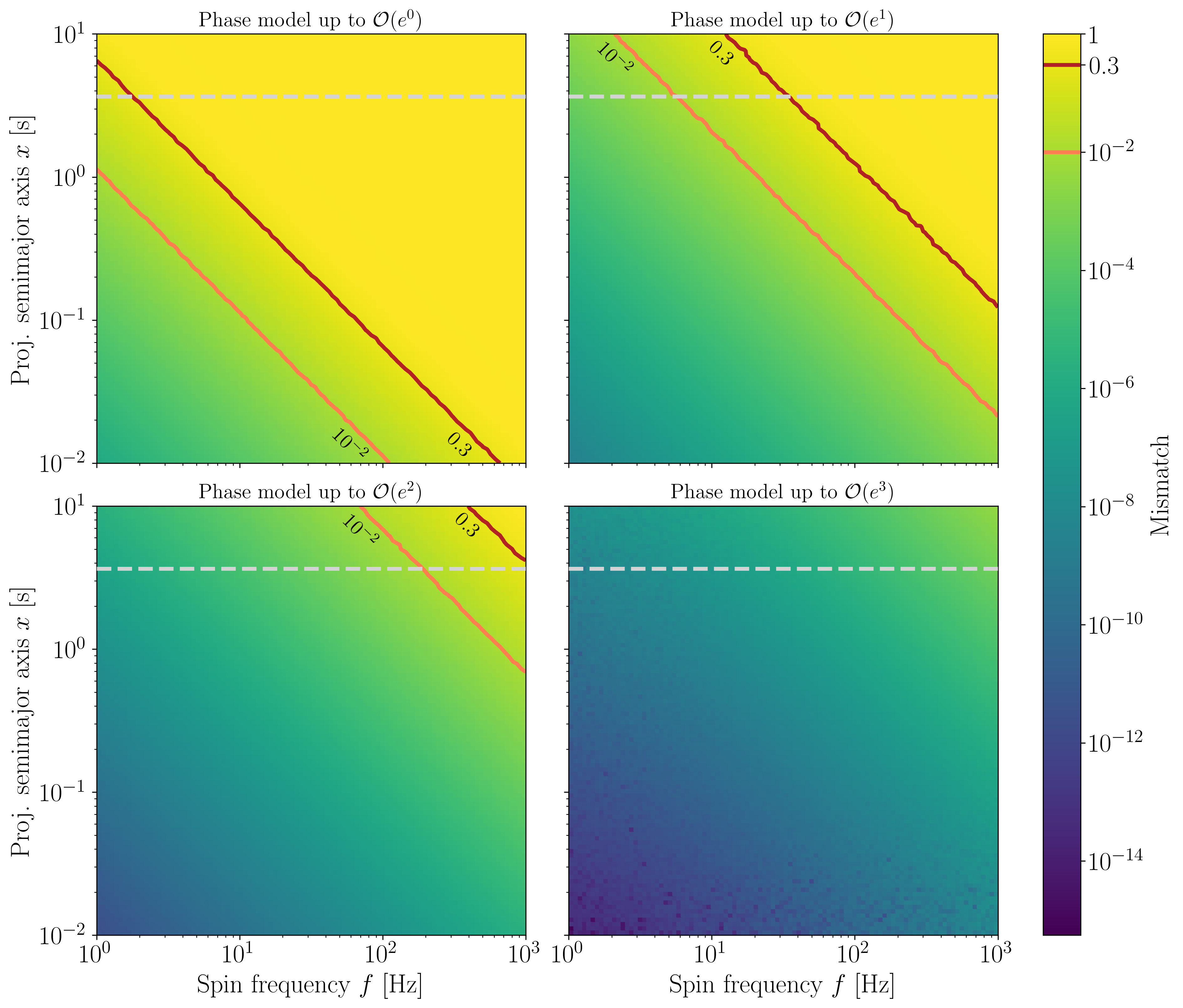}
		\caption{Mismatch between the BT model and models truncated at orders $e^0, e^1, e^2,$ and $e^3$, for the source \ellisrc{} with $e=0.04$.  This is computed on a grid of $100 \times 100$ simulated pulsar signals, with equally spaced $\log_{10} f/\text{Hz} \in [0,3]$, and $\log_{10} x/\text{s} \in [-2,1]$. The gray dashed line indicates the semimajor axis $x = 3.66$ of the likely pulsar in \ellisrc{}. The slopes of the constant-mismatch contours are the same for different models because $e$ is fixed.}
		\label{f:model_error}
	\end{figure*}

	\begin{figure*}
		\centering
		\includegraphics[width=0.95\textwidth]{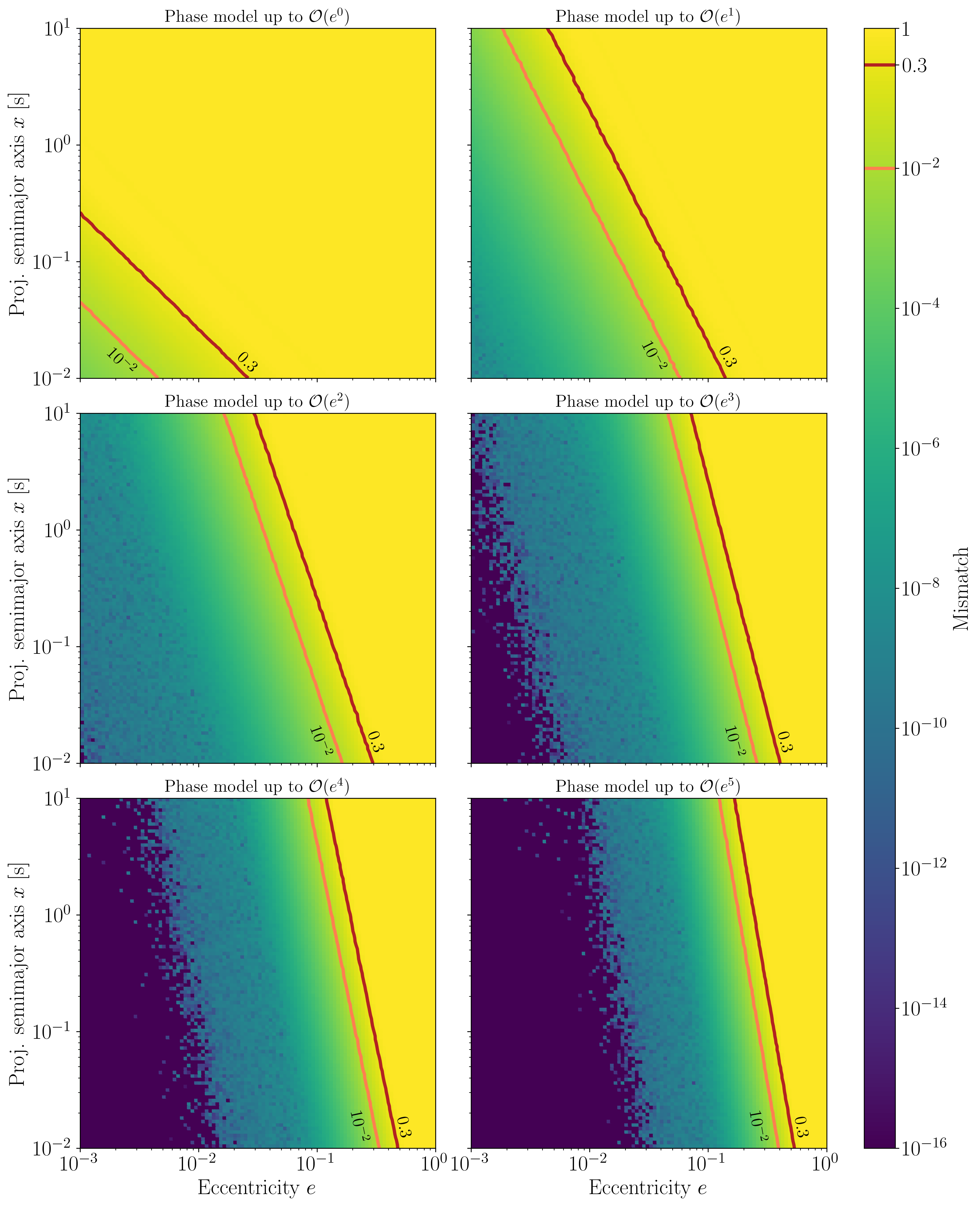}
		\caption{Same as Figure~\ref{f:model_error}, but varying the eccentricity $e$ with fixed frequency $f = 1\,\text{kHz}$, and going up to $e^5$.  The mismatch is computed on a grid of $100 \times 100$ simulated pulsar signals, with equally spaced $\log_{10} e \in [-3,0]$, and $\log_{10} x/\text{s} \in [-2,1]$.}
		\label{f:model_error_ex}
	\end{figure*}

	We list the results to $11$th order.  (A similar calculation \citep{dhurandhar2001} gives the coefficients to seventh order, but without a general formula.) The $\alpha$ values are given by
	\begin{linenomath}
		\begin{subequations}
			\begin{align}
				\alpha_0 =& -\frac{3}{2} e \,,\\
				\alpha_1 =& 1 -\frac{3}{8} e^2 +\frac{5}{192} e^4 -\frac{7}{9216} e^6 +\frac{1}{81920} e^8\\
				& -\frac{11}{88473600} e^{10} \,, \nonumber\\
				\alpha_2 =& \frac{1}{2} e -\frac{1}{3} e^3 +\frac{1}{16} e^5 -\frac{1}{180} e^7 +\frac{1}{3456} e^9 \,,\\
				\alpha_3 =& \frac{3}{8} e^2 -\frac{45}{128} e^4 +\frac{576}{5120} e^6 -\frac{729}{40960} e^8 \\
				& +\frac{8019}{4587520} e^{10} \,, \nonumber\\
				\alpha_4 =& \frac{1}{3} e^3 -\frac{2}{5} e^5 +\frac{8}{45} e^7 -\frac{8}{189} e^9 \,,\\
				\alpha_5 =& \frac{125}{384} e^4 -\frac{4375}{9216} e^6 +\frac{15625}{57344} e^8 -\frac{4296875}{49545216} e^{10} \,,\\
				\alpha_6 =& \frac{27}{80} e^5 -\frac{81}{140} e^7 +\frac{729}{1792} e^9 \,,\\
				\alpha_7 =& \frac{16807}{46080} e^6 -\frac{117649}{163840} e^8 + \frac{63412811}{106168320} e^{10} \,,\\
				\alpha_8 =& \frac{128}{315} e^7 -\frac{512}{567} e^9 \,,\\
				\alpha_9 =& \frac{531441}{1146880} e^8 -\frac{52612659}{45875200} e^{10} \,,\\
				\alpha_{10} =& \frac{78125}{145152} e^9 \,,\\
				\alpha_{11} =& \frac{2357947691}{3715891200} e^{10} \,.
			\end{align}
		\end{subequations}
	\end{linenomath}
	The $\beta$ values are given by
	\begin{linenomath}
		\begin{subequations}
			\begin{align}
				\beta_1 =& 1 -\frac{5}{8} e^2 -\frac{11}{192} e^4 -\frac{457}{9216} e^6 -\frac{23479}{737280} e^8 \\
				& -\frac{2014861}{88473600} e^{10} \,, \nonumber\\
				\beta_2 =& \frac{1}{2} e -\frac{5}{12} e^3 +\frac{1}{24} e^5 -\frac{1}{45} e^7 -\frac{379}{34560} e^9 \,,\\
				\beta_3 =& \frac{3}{8} e^2 -\frac{51}{128} e^4 +\frac{543}{5120} e^6 -\frac{219}{8192} e^8 \\
				& -\frac{18141}{4587520} e^{10} \,, \nonumber\\
				\beta_4 =& \frac{1}{3} e^3  -\frac{13}{30} e^5 +\frac{13}{72} e^7 -\frac{739}{15120} e^9\,,\\
				\beta_5 =& \frac{125}{384} e^4 -\frac{4625}{9216} e^6 +\frac{144625}{516096} e^8 -\frac{4611125}{49545216} e^{10} \,,\\
				\beta_6 =& \frac{27}{80} e^5 -\frac{135}{224} e^7 +\frac{3753}{8960} e^9 \,,\\
				\beta_7 =& \frac{16807}{46080} e^6 -\frac{218491}{294912} e^8 +\frac{65160739}{106168320} e^{10} \,,\\
				\beta_8 =& \frac{128}{315} e^7 -\frac{2624}{2835} e^9 \,,\\
				\beta_9 =& \frac{531441}{1146880} e^8 -\frac{53675541}{45875200} e^{10} \,,\\
				\beta_{10} =& \frac{78125}{145152} e^9 \,,\\
				\beta_{11} =& \frac{2357947691}{3715891200} e^{10} \,.
			\end{align}
		\end{subequations}
	\end{linenomath}
	A line-of-sight model accurate to $\mathcal{O}(e^k)$ requires retaining terms up to and including $\alpha_{k+1}$ and $\beta_{k+1}$.  Depending on the search parameters $\{f,x,e\}$, different orders of these Taylor series will be required.

	Consider the source \ellisrc{}. The expected eccentricity is $e\sim 0.04$.  To find the appropriate order in $e$, we simulated $10{,}000$ realizations of a pulsar in \ellisrc{}, with different spin frequencies $f$ and semimajor axes $x$.  Figure~\ref{f:model_error} shows the mismatches that arise from using approximations of different orders in $e$, compared to the full BT model.  For high frequencies the  mismatch $m$ is significant, $m \sim 0.3$ (\ac{SNR} loss of up to $30\%$), for the $\mathcal{O} (e^2)$-model. A sensible choice is the $\mathcal{O} (e^3)$-model, for which the mismatch is below $1\%$ for frequencies $f < 1\,\text{kHz}$.

	For systems with different eccentricities, we can also provide guidance.	  Since most of the known spider pulsars are \acp{MSP}, we simulated $10{,}000$ realizations of a $1\,\text{kHz}$ pulsar with different semimajor axes $x$ and eccentricities $e$.  Figure~\ref{f:model_error_ex} shows the mismatches that arise up to sixth order in $e$.

%%%%%%%%%%%%%%%%%%%%%%%%%%%%%%%%%%%%%%%%%%%
\clearpage
\bibliographystyle{aasjournal}
	
\bibliography{library}

\begin{thebibliography}{}
\expandafter\ifx\csname natexlab\endcsname\relax\def\natexlab#1{#1}\fi
\providecommand{\url}[1]{\href{#1}{#1}}
\providecommand{\dodoi}[1]{doi:~\href{http://doi.org/#1}{\nolinkurl{#1}}}
\providecommand{\doeprint}[1]{\href{http://ascl.net/#1}{\nolinkurl{http://ascl.net/#1}}}
\providecommand{\doarXiv}[1]{\href{https://arxiv.org/abs/#1}{\nolinkurl{https://arxiv.org/abs/#1}}}

\bibitem[{{Aasi} {et~al.}(2014){Aasi}, {Abbott}, {Abbott}, {Abbott},
  {Abernathy}, {Accadia}, {Acernese}, {Ackley}, {Adams}, {Adams}, \&
  et~al.}]{aasi2014}
{Aasi}, J., {Abbott}, B.~P., {Abbott}, R., {et~al.} 2014, \prd, 90, 062010,
  \dodoi{10.1103/PhysRevD.90.062010}

\bibitem[{{Abdo} {et~al.}(2009{\natexlab{a}}){Abdo}, {Ackermann}, {Atwood},
  {Baldini}, {Ballet}, {Barbiellini}, {Baring}, {Bastieri}, {Baughman},
  {Bechtol}, {Bellazzini}, {Berenji}, {Bloom}, {Bonamente}, {Borgland},
  {Bregeon}, {Brez}, {Brigida}, {Bruel}, {Burnett}, {Caliandro}, {Cameron},
  {Caraveo}, {Casandjian}, {Cecchi}, {Charles}, {Chekhtman}, {Cheung},
  {Chiang}, {Ciprini}, {Claus}, {Cohen-Tanugi}, {Cominsky}, {Conrad}, {Dermer},
  {de Angelis}, {de Palma}, {Digel}, {Donato}, {Dormody}, {do Couto e Silva},
  {Drell}, {Dubois}, {Dumora}, {Edmonds}, {Farnier}, {Favuzzi}, {Fleury},
  {Focke}, {Frailis}, {Fukazawa}, {Funk}, {Fusco}, {Gargano}, {Gasparrini},
  {Gehrels}, {Germani}, {Giebels}, {Giglietto}, {Giordano}, {Glanzman},
  {Godfrey}, {Grenier}, {Grondin}, {Grove}, {Guillemot}, {Guiriec}, {Harding},
  {Hayashida}, {Hays}, {Hughes}, {J{\'o}hannesson}, {Johnson}, {Johnson},
  {Johnson}, {Johnson}, {Johnston}, {Kamae}, {Katagiri}, {Kataoka}, {Kawai},
  {Kerr}, {Kn{\"o}dlseder}, {Komin}, {Kramer}, {Kuehn}, {Kuss}, {Latronico},
  {Lee}, {Lemoine-Goumard}, {Longo}, {Loparco}, {Lott}, {Lovellette},
  {Lubrano}, {Makeev}, {Marelli}, {Mazziotta}, {McConville}, {McEnery},
  {Meurer}, {Michelson}, {Mitthumsiri}, {Mizuno}, {Moiseev}, {Monte},
  {Monzani}, {Morselli}, {Moskalenko}, {Murgia}, {Nolan}, {Nuss}, {Ohsugi},
  {Omodei}, {Orlando}, {Ormes}, {Paneque}, {Panetta}, {Parent}, {Pepe},
  {Pesce-Rollins}, {Piron}, {Porter}, {Rain{\`o}}, {Rando}, {Razzano},
  {Reimer}, {Reimer}, {Reposeur}, {Ritz}, {Rochester}, {Rodriguez}, {Romani},
  {Roth}, {Ryde}, {Sadrozinski}, {Sanchez}, {Sander}, {Parkinson}, {Sgr{\`o}},
  {Siskind}, {Smith}, {Smith}, {Spandre}, {Spinelli}, {Starck}, {Strickman},
  {Suson}, {Tajima}, {Takahashi}, {Tanaka}, {Thayer}, {Thayer}, {Thompson},
  {Thorsett}, {Tibaldo}, {Torres}, {Tosti}, {Tramacere}, {Uchiyama}, {Usher},
  {Van Etten}, {Vilchez}, {Vitale}, {Waite}, {Watters}, {Wood}, {Ylinen},
  {Ziegler}, {Hobbs}, {Keith}, {Manchester}, \& {Weltevrede}}]{abdo2009a}
{Abdo}, A.~A., {Ackermann}, M., {Atwood}, W.~B., {et~al.} 2009{\natexlab{a}},
  \apjl, 695, L72, \dodoi{10.1088/0004-637X/695/1/L72}

\bibitem[{{Abdo} {et~al.}(2009{\natexlab{b}}){Abdo}, {Ackermann}, {Atwood},
  {Axelsson}, {Baldini}, {Ballet}, {Barbiellini}, {Bastieri}, {Battelino},
  {Baughman}, {Bechtol}, {Bellazzini}, {Berenji}, {Bloom}, {Bonamente},
  {Borgland}, {Bregeon}, {Brez}, {Brigida}, {Bruel}, {Burnett}, {Caliandro},
  {Cameron}, {Caraveo}, {Casandjian}, {Cecchi}, {Charles}, {Chekhtman},
  {Cheung}, {Chiang}, {Ciprini}, {Claus}, {Cognard}, {Cohen-Tanugi},
  {Cominsky}, {Conrad}, {Cutini}, {Dermer}, {de Angelis}, {de Palma}, {Digel},
  {Dormody}, {Silva}, {Drell}, {Dubois}, {Dumora}, {Farnier}, {Favuzzi},
  {Focke}, {Frailis}, {Fukazawa}, {Funk}, {Fusco}, {Gargano}, {Gasparrini},
  {Gehrels}, {Germani}, {Giebels}, {Giglietto}, {Giordano}, {Glanzman},
  {Godfrey}, {Grenier}, {Grondin}, {Grove}, {Guillemot}, {Guiriec}, {Hanabata},
  {Harding}, {Hayashida}, {Hays}, {Hughes}, {J{\'o}hannesson}, {Johnson},
  {Johnson}, {Johnson}, {Johnson}, {Kamae}, {Katagiri}, {Kataoka}, {Kawai},
  {Kerr}, {Kn{\"o}dlseder}, {Kocian}, {Komin}, {Kuehn}, {Kuss}, {Lande},
  {Latronico}, {Lee}, {Lemoine-Goumard}, {Longo}, {Loparco}, {Lott},
  {Lovellette}, {Lubrano}, {Madejski}, {Makeev}, {Marelli}, {Mazziotta},
  {McConville}, {McEnery}, {Meurer}, {Michelson}, {Mitthumsiri}, {Mizuno},
  {Moiseev}, {Monte}, {Monzani}, {Morselli}, {Moskalenko}, {Murgia}, {Nolan},
  {Nuss}, {Ohsugi}, {Omodei}, {Orlando}, {Ormes}, {Pancrazi}, {Paneque},
  {Panetta}, {Parent}, {Pepe}, {Pesce-Rollins}, {Piron}, {Porter}, {Rain{\`o}},
  {Rando}, {Razzano}, {Reimer}, {Reimer}, {Reposeur}, {Ritz}, {Rochester},
  {Rodriguez}, {Romani}, {Ryde}, {Sadrozinski}, {Sanchez}, {Sander},
  {Parkinson}, {Sgr{\`o}}, {Siskind}, {Smith}, {Smith}, {Spandre}, {Spinelli},
  {Starck}, {Strickman}, {Suson}, {Tajima}, {Takahashi}, {Tanaka}, {Thayer},
  {Thayer}, {Theureau}, {Thompson}, {Tibaldo}, {Torres}, {Tosti}, {Tramacere},
  {Uchiyama}, {Usher}, {Van Etten}, {Vilchez}, {Vitale}, {Waite}, {Watters},
  {Webb}, {Wood}, {Ylinen}, \& {Ziegler}}]{abdo2009c}
---. 2009{\natexlab{b}}, \apj, 699, 1171, \dodoi{10.1088/0004-637X/699/2/1171}

\bibitem[{{Abdo} {et~al.}(2009{\natexlab{c}}){Abdo}, {Ackermann}, {Ajello},
  {Anderson}, {Atwood}, {Axelsson}, {Baldini}, {Ballet}, {Barbiellini},
  {Baring}, {Bastieri}, {Baughman}, {Bechtol}, {Bellazzini}, {Berenji},
  {Bignami}, {Blandford}, {Bloom}, {Bonamente}, {Borgland}, {Bregeon}, {Brez},
  {Brigida}, {Bruel}, {Burnett}, {Caliandro}, {Cameron}, {Caraveo},
  {Casandjian}, {Cecchi}, {{\c C}elik}, {Chekhtman}, {Cheung}, {Chiang},
  {Ciprini}, {Claus}, {Cohen-Tanugi}, {Conrad}, {Cutini}, {Dermer}, {de
  Angelis}, {de Luca}, {de Palma}, {Digel}, {Dormody}, {do Couto e Silva},
  {Drell}, {Dubois}, {Dumora}, {Farnier}, {Favuzzi}, {Fegan}, {Fukazawa},
  {Funk}, {Fusco}, {Gargano}, {Gasparrini}, {Gehrels}, {Germani}, {Giebels},
  {Giglietto}, {Giommi}, {Giordano}, {Glanzman}, {Godfrey}, {Grenier},
  {Grondin}, {Grove}, {Guillemot}, {Guiriec}, {Gwon}, {Hanabata}, {Harding},
  {Hayashida}, {Hays}, {Hughes}, {J{\'o}hannesson}, {Johnson}, {Johnson},
  {Johnson}, {Kamae}, {Katagiri}, {Kataoka}, {Kawai}, {Kerr}, {Kn{\"o}dlseder},
  {Kocian}, {Kuss}, {Lande}, {Latronico}, {Lemoine-Goumard}, {Longo},
  {Loparco}, {Lott}, {Lovellette}, {Lubrano}, {Madejski}, {Makeev}, {Marelli},
  {Mazziotta}, {McConville}, {McEnery}, {Meurer}, {Michelson}, {Mitthumsiri},
  {Mizuno}, {Monte}, {Monzani}, {Morselli}, {Moskalenko}, {Murgia}, {Nolan},
  {Norris}, {Nuss}, {Ohsugi}, {Omodei}, {Orlando}, {Ormes}, {Paneque},
  {Parent}, {Pelassa}, {Pepe}, {Pesce-Rollins}, {Pierbattista}, {Piron},
  {Porter}, {Primack}, {Rain{\`o}}, {Rando}, {Ray}, {Razzano}, {Rea}, {Reimer},
  {Reimer}, {Reposeur}, {Ritz}, {Rochester}, {Rodriguez}, {Romani}, {Ryde},
  {Sadrozinski}, {Sanchez}, {Sander}, {Parkinson}, {Scargle}, {Sgr{\`o}},
  {Siskind}, {Smith}, {Smith}, {Spandre}, {Spinelli}, {Starck}, {Strickman},
  {Suson}, {Tajima}, {Takahashi}, {Takahashi}, {Tanaka}, {Thayer}, {Thompson},
  {Tibaldo}, {Tibolla}, {Torres}, {Tosti}, {Tramacere}, {Uchiyama}, {Usher},
  {Van Etten}, {Vasileiou}, {Vilchez}, {Vitale}, {Waite}, {Wang}, {Watters},
  {Winer}, {Wolff}, {Wood}, {Ylinen}, {Ziegler}, \& {Fermi LAT
  Collaboration}}]{abdo2009d}
{Abdo}, A.~A., {Ackermann}, M., {Ajello}, M., {et~al.} 2009{\natexlab{c}},
  Science, 325, 840, \dodoi{10.1126/science.1175558}

\bibitem[{{Abdollahi} {et~al.}(2020){Abdollahi}, {Acero}, {Ackermann},
  {Ajello}, {Atwood}, {Axelsson}, {Baldini}, {Ballet}, {Barbiellini},
  {Bastieri}, {Becerra Gonzalez}, {Bellazzini}, {Berretta}, {Bissaldi}, {Bland
  ford}, {Bloom}, {Bonino}, {Bottacini}, {Brandt}, {Bregeon}, {Bruel},
  {Buehler}, {Burnett}, {Buson}, {Cameron}, {Caputo}, {Caraveo}, {Casandjian},
  {Castro}, {Cavazzuti}, {Charles}, {Chaty}, {Chen}, {Cheung}, {Chiaro},
  {Ciprini}, {Cohen-Tanugi}, {Cominsky}, {Coronado-Bl{\'a}zquez}, {Costantin},
  {Cuoco}, {Cutini}, {D'Ammando}, {DeKlotz}, {Torre Luque}, {de Palma},
  {Desai}, {Digel}, {Lalla}, {Mauro}, {Venere}, {Dom{\'\i}nguez}, {Dumora},
  {Dirirsa}, {Fegan}, {Ferrara}, {Franckowiak}, {Fukazawa}, {Funk}, {Fusco},
  {Gargano}, {Gasparrini}, {Giglietto}, {Giommi}, {Giordano}, {Giroletti},
  {Glanzman}, {Green}, {Grenier}, {Griffin}, {Grondin}, {Grove}, {Guiriec},
  {Harding}, {Hayashi}, {Hays}, {Hewitt}, {Horan}, {J{\'o}hannesson},
  {Johnson}, {Kamae}, {Kerr}, {Kocevski}, {Kovac'evic'}, {Kuss}, {Landriu},
  {Larsson}, {Latronico}, {Lemoine-Goumard}, {Li}, {Liodakis}, {Longo},
  {Loparco}, {Lott}, {Lovellette}, {Lubrano}, {Madejski}, {Maldera},
  {Malyshev}, {Manfreda}, {Marchesini}, {Marcotulli}, {Mart{\'\i}-Devesa},
  {Martin}, {Massaro}, {Mazziotta}, {McEnery}, {Mereu}, {Meyer}, {Michelson},
  {Mirabal}, {Mizuno}, {Monzani}, {Morselli}, {Moskalenko}, {Negro}, {Nuss},
  {Ojha}, {Omodei}, {Orienti}, {Orlando}, {Ormes}, {Palatiello}, {Paliya},
  {Paneque}, {Pei}, {Pe{\~n}a-Herazo}, {Perkins}, {Persic}, {Pesce-Rollins},
  {Petrosian}, {Petrov}, {Piron}, {Poon}, {Porter}, {Principe}, {Rain{\`o}},
  {Rando}, {Razzano}, {Razzaque}, {Reimer}, {Reimer}, {Remy}, {Reposeur},
  {Romani}, {Parkinson}, {Schinzel}, {Serini}, {Sgr{\`o}}, {Siskind}, {Smith},
  {Spandre}, {Spinelli}, {Strong}, {Suson}, {Tajima}, {Takahashi}, {Tak},
  {Thayer}, {Thompson}, {Tibaldo}, {Torres}, {Torresi}, {Valverde}, {Klaveren},
  {Zyl}, {Wood}, {Yassine}, \& {Zaharijas}}]{4fgl}
{Abdollahi}, S., {Acero}, F., {Ackermann}, M., {et~al.} 2020, \apjs, 247, 33,
  \dodoi{10.3847/1538-4365/ab6bcb}

\bibitem[{{Allafort} {et~al.}(2013){Allafort}, {Baldini}, {Ballet},
  {Barbiellini}, {Baring}, {Bastieri}, {Bellazzini}, {Bonamente}, {Bottacini},
  {Brandt}, {Bregeon}, {Bruel}, {Buehler}, {Buson}, {Caliandro}, {Cameron},
  {Caraveo}, {Cecchi}, {Chaves}, {Chekhtman}, {Chiang}, {Chiaro}, {Ciprini},
  {Claus}, {D'Ammando}, {de Palma}, {Digel}, {Di Venere}, {Drell}, {Favuzzi},
  {Ferrara}, {Franckowiak}, {Fusco}, {Gargano}, {Gasparrini}, {Giglietto},
  {Giroletti}, {Glanzman}, {Godfrey}, {Grenier}, {Guiriec}, {Hadasch},
  {Harding}, {Hayashida}, {Hayashi}, {Hays}, {Hewitt}, {Hill}, {Horan}, {Hou},
  {Jogler}, {Johnson}, {Johnson}, {Kerr}, {Kn{\"o}dlseder}, {Kuss}, {Lande},
  {Larsson}, {Latronico}, {Lemoine-Goumard}, {Longo}, {Loparco}, {Lubrano},
  {Malyshev}, {Marelli}, {Mayer}, {Mazziotta}, {Mehault}, {Mizuno}, {Monzani},
  {Morselli}, {Murgia}, {Nemmen}, {Nuss}, {Ohsugi}, {Omodei}, {Orienti},
  {Orlando}, {Paneque}, {Pesce-Rollins}, {Pierbattista}, {Piron}, {Pivato},
  {Porter}, {Rain{\`o}}, {Rando}, {Ray}, {Razzano}, {Reimer}, {Reposeur},
  {Romani}, {Sartori}, {Saz Parkinson}, {Sgr{\`o}}, {Siskind}, {Smith},
  {Spinelli}, {Strong}, {Takahashi}, {Thayer}, {Thompson}, {Tibaldo},
  {Tinivella}, {Torres}, {Tosti}, {Uchiyama}, {Usher}, {Vandenbroucke},
  {Vasileiou}, {Venter}, {Vianello}, {Vitale}, {Winer}, \&
  {Wood}}]{allafort2013}
{Allafort}, A., {Baldini}, L., {Ballet}, J., {et~al.} 2013, \apjl, 777, L2,
  \dodoi{10.1088/2041-8205/777/1/L2}

\bibitem[{{Allen}(2019)}]{allen2019}
{Allen}, B. 2019, \prd, 100, 124004, \dodoi{10.1103/PhysRevD.100.124004}

\bibitem[{{Allen} {et~al.}(2013){Allen}, {Knispel}, {Cordes}, {Deneva},
  {Hessels}, {Anderson}, {Aulbert}, {Bock}, {Brazier}, {Chatterjee},
  {Demorest}, {Eggenstein}, {Fehrmann}, {Gotthelf}, {Hammer}, {Kaspi},
  {Kramer}, {Lyne}, {Machenschalk}, {McLaughlin}, {Messenger}, {Pletsch},
  {Ransom}, {Stairs}, {Stappers}, {Bhat}, {Bogdanov}, {Camilo}, {Champion},
  {Crawford}, {Desvignes}, {Freire}, {Heald}, {Jenet}, {Lazarus}, {Lee}, {van
  Leeuwen}, {Lynch}, {Papa}, {Prix}, {Rosen}, {Scholz}, {Siemens}, {Stovall},
  {Venkataraman}, \& {Zhu}}]{allen2013}
{Allen}, B., {Knispel}, B., {Cordes}, J.~M., {et~al.} 2013, \apj, 773, 91,
  \dodoi{10.1088/0004-637X/773/2/91}

\bibitem[{{Andersen} \& {Ransom}(2018)}]{andersen2018}
{Andersen}, B.~C., \& {Ransom}, S.~M. 2018, \apjl, 863, L13,
  \dodoi{10.3847/2041-8213/aad59f}

\bibitem[{{Aragona} {et~al.}(2009){Aragona}, {McSwain}, {Grundstrom}, {Marsh},
  {Roettenbacher}, {Hessler}, {Boyajian}, \& {Ray}}]{aragona2009}
{Aragona}, C., {McSwain}, M.~V., {Grundstrom}, E.~D., {et~al.} 2009, \apj, 698,
  514, \dodoi{10.1088/0004-637X/698/1/514}

\bibitem[{{Atwood} {et~al.}(2009){Atwood}, {Abdo}, {Ackermann}, {Althouse},
  {Anderson}, {Axelsson}, {Baldini}, {Ballet}, {Band}, {Barbiellini}, \&
  et~al.}]{atwood2009}
{Atwood}, W.~B., {Abdo}, A.~A., {Ackermann}, M., {et~al.} 2009, \apj, 697,
  1071, \dodoi{10.1088/0004-637X/697/2/1071}

\bibitem[{{Babak}(2008)}]{babak2008}
{Babak}, S. 2008, Classical and Quantum Gravity, 25, 195011,
  \dodoi{10.1088/0264-9381/25/19/195011}

\bibitem[{{Balasubramanian} {et~al.}(1996){Balasubramanian}, {Sathyaprakash},
  \& {Dhurandhar}}]{balasubramanian1996}
{Balasubramanian}, R., {Sathyaprakash}, B.~S., \& {Dhurandhar}, S.~V. 1996,
  \prd, 53, 3033, \dodoi{10.1103/PhysRevD.53.3033}

\bibitem[{{Bassa} {et~al.}(2017){Bassa}, {Pleunis}, {Hessels}, {Ferrara},
  {Breton}, {Gusinskaia}, {Kondratiev}, {Sanidas}, {Nieder}, {Clark}, {Li},
  {van Amesfoort}, {Burnett}, {Camilo}, {Michelson}, {Ransom}, {Ray}, \&
  {Wood}}]{bassa2017b}
{Bassa}, C.~G., {Pleunis}, Z., {Hessels}, J.~W.~T., {et~al.} 2017, \apjl, 846,
  L20, \dodoi{10.3847/2041-8213/aa8400}

\bibitem[{{Bickel} {et~al.}(2008){Bickel}, {Kleijn}, \& {Rice}}]{bickel2008}
{Bickel}, P., {Kleijn}, B., \& {Rice}, J. 2008, \apj, 685, 384,
  \dodoi{10.1086/590399}

\bibitem[{{Blandford} \& {Teukolsky}(1976)}]{blandford1976}
{Blandford}, R., \& {Teukolsky}, S.~A. 1976, \apj, 205, 580,
  \dodoi{10.1086/154315}

\bibitem[{{Brady} \& {Creighton}(2000)}]{brady2000}
{Brady}, P.~R., \& {Creighton}, T. 2000, \prd, 61, 082001,
  \dodoi{10.1103/PhysRevD.61.082001}

\bibitem[{{Bruel}(2019)}]{bruel2019}
{Bruel}, P. 2019, \aap, 622, A108, \dodoi{10.1051/0004-6361/201834555}

\bibitem[{{Camilo} {et~al.}(2000){Camilo}, {Lorimer}, {Freire}, {Lyne}, \&
  {Manchester}}]{camilo2000}
{Camilo}, F., {Lorimer}, D.~R., {Freire}, P., {Lyne}, A.~G., \& {Manchester},
  R.~N. 2000, \apj, 535, 975, \dodoi{10.1086/308859}

\bibitem[{{Caraveo}(2014)}]{caraveo2014}
{Caraveo}, P.~A. 2014, \araa, 52, 211,
  \dodoi{10.1146/annurev-astro-081913-035948}

\bibitem[{{Chen} {et~al.}(2013){Chen}, {Chen}, {Tauris}, \& {Han}}]{chen2013}
{Chen}, H.-L., {Chen}, X., {Tauris}, T.~M., \& {Han}, Z. 2013, \apj, 775, 27,
  \dodoi{10.1088/0004-637X/775/1/27}

\bibitem[{{Clark} {et~al.}(2015){Clark}, {Pletsch}, {Wu}, {Guillemot},
  {Ackermann}, {Allen}, {de Angelis}, {Aulbert}, {Baldini}, {Ballet},
  {Barbiellini}, {Bastieri}, {Bellazzini}, {Bissaldi}, {Bock}, {Bonino},
  {Bottacini}, {Brandt}, {Bregeon}, {Bruel}, {Buson}, {Caliandro}, {Cameron},
  {Caragiulo}, {Caraveo}, {Cecchi}, {Champion}, {Charles}, {Chekhtman},
  {Chiang}, {Chiaro}, {Ciprini}, {Claus}, {Cohen-Tanugi}, {Cu{\'e}llar},
  {Cutini}, {D'Ammando}, {Desiante}, {Drell}, {Eggenstein}, {Favuzzi},
  {Fehrmann}, {Ferrara}, {Focke}, {Franckowiak}, {Fusco}, {Gargano},
  {Gasparrini}, {Giglietto}, {Giordano}, {Glanzman}, {Godfrey}, {Grenier},
  {Grove}, {Guiriec}, {Harding}, {Hays}, {Hewitt}, {Hill}, {Horan}, {Hou},
  {Jogler}, {Johnson}, {J{\'o}hannesson}, {Kramer}, {Krauss}, {Kuss}, {Laffon},
  {Larsson}, {Latronico}, {Li}, {Li}, {Longo}, {Loparco}, {Lovellette},
  {Lubrano}, {Machenschalk}, {Manfreda}, {Marelli}, {Mayer}, {Mazziotta},
  {Michelson}, {Mizuno}, {Monzani}, {Morselli}, {Moskalenko}, {Murgia}, {Nuss},
  {Ohsugi}, {Orienti}, {Orlando}, {de Palma}, {Paneque}, {Pesce-Rollins},
  {Piron}, {Pivato}, {Rain{\`o}}, {Rando}, {Razzano}, {Reimer}, {Saz
  Parkinson}, {Schaal}, {Schulz}, {Sgr{\`o}}, {Siskind}, {Spada}, {Spandre},
  {Spinelli}, {Suson}, {Takahashi}, {Thayer}, {Tibaldo}, {Torne}, {Torres},
  {Tosti}, {Troja}, {Vianello}, {Wood}, {Wood}, \& {Yassine}}]{clark2015}
{Clark}, C.~J., {Pletsch}, H.~J., {Wu}, J., {et~al.} 2015, \apjl, 809, L2,
  \dodoi{10.1088/2041-8205/809/1/L2}

\bibitem[{{Clark} {et~al.}(2016){Clark}, {Pletsch}, {Wu}, {Guillemot},
  {Camilo}, {Johnson}, {Kerr}, {Allen}, {Aulbert}, {Beer}, {Bock},
  {Cu{\'e}llar}, {Eggenstein}, {Fehrmann}, {Kramer}, {Machenschalk}, \&
  {Nieder}}]{clark2016}
---. 2016, \apjl, 832, L15, \dodoi{10.3847/2041-8205/832/1/L15}

\bibitem[{{Clark} {et~al.}(2017){Clark}, {Wu}, {Pletsch}, {Guillemot}, {Allen},
  {Aulbert}, {Beer}, {Bock}, {Cu{\'e}llar}, {Eggenstein}, {Fehrmann}, {Kramer},
  {Machenschalk}, \& {Nieder}}]{clark2017}
{Clark}, C.~J., {Wu}, J., {Pletsch}, H.~J., {et~al.} 2017, \apj, 834, 106,
  \dodoi{10.3847/1538-4357/834/2/106}

\bibitem[{{Clark} {et~al.}(2018){Clark}, {Pletsch}, {Wu}, {Guillemot}, {Kerr},
  {Johnson}, {Camilo}, {Salvetti}, {Allen}, {Anderson}, {Aulbert}, {Beer},
  {Bock}, {Cu{\'e}llar}, {Eggenstein}, {Fehrmann}, {Kramer}, {Kwang},
  {Machenschalk}, {Nieder}, {Ackermann}, {Ajello}, {Baldini}, {Ballet},
  {Barbiellini}, {Bastieri}, {Bellazzini}, {Bissaldi}, {Blandford}, {Bloom},
  {Bonino}, {Bottacini}, {Brandt}, {Bregeon}, {Bruel}, {Buehler}, {Burnett},
  {Buson}, {Cameron}, {Caputo}, {Caraveo}, {Cavazzuti}, {Cecchi}, {Charles},
  {Chekhtman}, {Ciprini}, {Cominsky}, {Costantin}, {Cutini}, {D'Ammando}, {De
  Luca}, {Desiante}, {Di Venere}, {Di Mauro}, {Di Lalla}, {Digel}, {Favuzzi},
  {Ferrara}, {Franckowiak}, {Fukazawa}, {Funk}, {Fusco}, {Gargano},
  {Gasparrini}, {Giglietto}, {Giordano}, {Giroletti}, {Gomez-Vargas}, {Green},
  {Grenier}, {Guiriec}, {Harding}, {Hewitt}, {Horan}, {J{\'o}hannesson},
  {Kensei}, {Kuss}, {La Mura}, {Larsson}, {Latronico}, {Li}, {Longo},
  {Loparco}, {Lovellette}, {Lubrano}, {Magill}, {Maldera}, {Manfreda},
  {Mazziotta}, {McEnery}, {Michelson}, {Mirabal}, {Mitthumsiri}, {Mizuno},
  {Monzani}, {Morselli}, {Moskalenko}, {Nuss}, {Ohsugi}, {Omodei}, {Orienti},
  {Orlando}, {Palatiello}, {Paliya}, {de Palma}, {Paneque}, {Perkins},
  {Persic}, {Pesce-Rollins}, {Porter}, {Principe}, {Rain{\`o}}, {Rando}, {Ray},
  {Razzano}, {Reimer}, {Reimer}, {Romani}, {Saz Parkinson}, {Sgr{\`o}},
  {Siskind}, {Smith}, {Spada}, {Spandre}, {Spinelli}, {Thayer}, {Thompson},
  {Torres}, {Troja}, {Vianello}, {Wood}, \& {Wood}}]{clark2018}
{Clark}, C.~J., {Pletsch}, H.~J., {Wu}, J., {et~al.} 2018, Science Advances, 4,
  eaao7228, \dodoi{10.1126/sciadv.aao7228}

\bibitem[{{Clark} {et~al.}(2020){Clark}, {Nieder}, {Voisin}, {Allen},
  {Aulbert}, {Behnke}, {Breton}, {Choquet}, {Corongiu}, {Dhillon},
  {Eggenstein}, {Fehrmann}, {Guillemot}, {Harding}, {Kennedy}, {Machenschalk},
  {Marsh}, {Mata S{\'a}nchez}, {Mignani}, {Stringer}, {Wadiasingh}, \&
  {Wu}}]{clark2020}
{Clark}, C.~J., {Nieder}, L., {Voisin}, G., {et~al.} 2020, arXiv e-prints.
\newblock \doarXiv{2007.14849}

\bibitem[{{Cromartie} {et~al.}(2020){Cromartie}, {Fonseca}, {Ransom},
  {Demorest}, {Arzoumanian}, {Blumer}, {Brook}, {DeCesar}, {Dolch}, {Ellis},
  {Ferdman}, {Ferrara}, {Garver-Daniels}, {Gentile}, {Jones}, {Lam}, {Lorimer},
  {Lynch}, {McLaughlin}, {Ng}, {Nice}, {Pennucci}, {Spiewak}, {Stairs},
  {Stovall}, {Swiggum}, \& {Zhu}}]{cromartie2020}
{Cromartie}, H.~T., {Fonseca}, E., {Ransom}, S.~M., {et~al.} 2020, Nature
  Astronomy, 4, 72, \dodoi{10.1038/s41550-019-0880-2}

\bibitem[{{Damour} \& {Deruelle}(1986)}]{damour1986}
{Damour}, T., \& {Deruelle}, N. 1986, Ann. Inst. Henri Poincar{\'e} Phys.
  Th{\'e}or, 44, 263

\bibitem[{{de Jager} {et~al.}(1989){de Jager}, {Raubenheimer}, \&
  {Swanepoel}}]{dejager1989}
{de Jager}, O.~C., {Raubenheimer}, B.~C., \& {Swanepoel}, J.~W.~H. 1989, \aap,
  221, 180

\bibitem[{{Dhurandhar} \& {Vecchio}(2001)}]{dhurandhar2001}
{Dhurandhar}, S.~V., \& {Vecchio}, A. 2001, \prd, 63, 122001,
  \dodoi{10.1103/PhysRevD.63.122001}

\bibitem[{{Edwards} {et~al.}(2006){Edwards}, {Hobbs}, \&
  {Manchester}}]{edwards2006}
{Edwards}, R.~T., {Hobbs}, G.~B., \& {Manchester}, R.~N. 2006, \mnras, 372,
  1549, \dodoi{10.1111/j.1365-2966.2006.10870.x}

\bibitem[{{Faulkner} {et~al.}(2004){Faulkner}, {Stairs}, {Kramer}, {Lyne},
  {Hobbs}, {Possenti}, {Lorimer}, {Manchester}, {McLaughlin}, {D'Amico},
  {Camilo}, \& {Burgay}}]{faulkner2004}
{Faulkner}, A.~J., {Stairs}, I.~H., {Kramer}, M., {et~al.} 2004, \mnras, 355,
  147, \dodoi{10.1111/j.1365-2966.2004.08310.x}

\bibitem[{{Faulkner} {et~al.}(2005){Faulkner}, {Kramer}, {Lyne}, {Manchester},
  {McLaughlin}, {Stairs}, {Hobbs}, {Possenti}, {Lorimer}, {D'Amico}, {Camilo},
  \& {Burgay}}]{faulkner2005}
{Faulkner}, A.~J., {Kramer}, M., {Lyne}, A.~G., {et~al.} 2005, \apjl, 618,
  L119, \dodoi{10.1086/427776}

\bibitem[{{Fehrmann} \& {Pletsch}(2014)}]{fehrmann2014}
{Fehrmann}, H., \& {Pletsch}, H.~J. 2014, \prd, 90, 124049,
  \dodoi{10.1103/PhysRevD.90.124049}

\bibitem[{{Frigo} \& {Johnson}(2005)}]{frigo2005}
{Frigo}, M., \& {Johnson}, S.~G. 2005, Proceedings of the IEEE, 93, 216,
  \dodoi{10.1109/JPROC.2004.840301}

\bibitem[{{Fruchter} {et~al.}(1988){Fruchter}, {Stinebring}, \&
  {Taylor}}]{fruchter1988}
{Fruchter}, A.~S., {Stinebring}, D.~R., \& {Taylor}, J.~H. 1988, \nat, 333,
  237, \dodoi{10.1038/333237a0}

\bibitem[{{Gaia Collaboration} {et~al.}(2018){Gaia Collaboration}, {Brown},
  {Vallenari}, {Prusti}, {de Bruijne}, {Babusiaux}, {Bailer-Jones}, {Biermann},
  {Evans}, {Eyer}, \& et~al.}]{gaia2018}
{Gaia Collaboration}, {Brown}, A.~G.~A., {Vallenari}, A., {et~al.} 2018, \aap,
  616, A1, \dodoi{10.1051/0004-6361/201833051}

\bibitem[{{Goetz} \& {Riles}(2011)}]{goetz2011}
{Goetz}, E., \& {Riles}, K. 2011, Classical and Quantum Gravity, 28, 215006,
  \dodoi{10.1088/0264-9381/28/21/215006}

\bibitem[{{Guillemot} {et~al.}(2012){Guillemot}, {Johnson}, {Venter}, {Kerr},
  {Pancrazi}, {Livingstone}, {Janssen}, {Jaroenjittichai}, {Kramer}, {Cognard},
  {Stappers}, {Harding}, {Camilo}, {Espinoza}, {Freire}, {Gargano}, {Grove},
  {Johnston}, {Michelson}, {Noutsos}, {Parent}, {Ransom}, {Ray}, {Shannon},
  {Smith}, {Theureau}, {Thorsett}, \& {Webb}}]{guillemot2012}
{Guillemot}, L., {Johnson}, T.~J., {Venter}, C., {et~al.} 2012, \apj, 744, 33,
  \dodoi{10.1088/0004-637X/744/1/33}

\bibitem[{{Halpern} {et~al.}(2017){Halpern}, {Strader}, \& {Li}}]{halpern2017}
{Halpern}, J.~P., {Strader}, J., \& {Li}, M. 2017, \apj, 844, 150,
  \dodoi{10.3847/1538-4357/aa7cff}

\bibitem[{{Harry} {et~al.}(2009){Harry}, {Allen}, \&
  {Sathyaprakash}}]{harry2009}
{Harry}, I.~W., {Allen}, B., \& {Sathyaprakash}, B.~S. 2009, \prd, 80, 104014,
  \dodoi{10.1103/PhysRevD.80.104014}

\bibitem[{{Ho} {et~al.}(2017){Ho}, {Ng}, {Lyne}, {Stappers}, {Coe}, {Halpern},
  {Johnson}, \& {Steele}}]{ho2017}
{Ho}, W.~C.~G., {Ng}, C.-Y., {Lyne}, A.~G., {et~al.} 2017, \mnras, 464, 1211,
  \dodoi{10.1093/mnras/stw2420}

\bibitem[{{Hobbs} {et~al.}(2006){Hobbs}, {Edwards}, \&
  {Manchester}}]{hobbs2006}
{Hobbs}, G.~B., {Edwards}, R.~T., \& {Manchester}, R.~N. 2006, \mnras, 369,
  655, \dodoi{10.1111/j.1365-2966.2006.10302.x}

\bibitem[{{Hui} {et~al.}(2015){Hui}, {Park}, {Hu}, {Lin}, {Li}, {Kong}, {Tam},
  {Takata}, {Cheng}, {Jin}, {Yen}, \& {Kim}}]{hui2015}
{Hui}, C.~Y., {Park}, S.~M., {Hu}, C.~P., {et~al.} 2015, \apj, 809, 68,
  \dodoi{10.1088/0004-637X/809/1/68}

\bibitem[{{Hunter}(2007)}]{matplotlib2007}
{Hunter}, J.~D. 2007, Computing in Science and Engineering, 9, 90,
  \dodoi{10.1109/MCSE.2007.55}

\bibitem[{{Johnston} \& {Kulkarni}(1991)}]{johnston1991}
{Johnston}, H.~M., \& {Kulkarni}, S.~R. 1991, \apj, 368, 504,
  \dodoi{10.1086/169715}

\bibitem[{{Kerr}(2011)}]{kerr2011}
{Kerr}, M. 2011, \apj, 732, 38, \dodoi{10.1088/0004-637X/732/1/38}

\bibitem[{{Kerr} {et~al.}(2015){Kerr}, {Ray}, {Johnston}, {Shannon}, \&
  {Camilo}}]{kerr2015b}
{Kerr}, M., {Ray}, P.~S., {Johnston}, S., {Shannon}, R.~M., \& {Camilo}, F.
  2015, \apj, 814, 128, \dodoi{10.1088/0004-637X/814/2/128}

\bibitem[{{Knispel} {et~al.}(2015){Knispel}, {Lyne}, {Stappers}, {Freire},
  {Lazarus}, {Allen}, {Aulbert}, {Bock}, {Bogdanov}, {Brazier}, {Camilo},
  {Cardoso}, {Chatterjee}, {Cordes}, {Crawford}, {Deneva}, {Eggenstein},
  {Fehrmann}, {Ferdman}, {Hessels}, {Jenet}, {Karako-Argaman}, {Kaspi}, {van
  Leeuwen}, {Lorimer}, {Lynch}, {Machenschalk}, {Madsen}, {McLaughlin},
  {Patel}, {Ransom}, {Scholz}, {Siemens}, {Spitler}, {Stairs}, {Stovall},
  {Swiggum}, {Venkataraman}, {Wharton}, \& {Zhu}}]{knispel2015}
{Knispel}, B., {Lyne}, A.~G., {Stappers}, B.~W., {et~al.} 2015, \apj, 806, 140,
  \dodoi{10.1088/0004-637X/806/1/140}

\bibitem[{{Kong} {et~al.}(2014){Kong}, {Jin}, {Yen}, {Hu}, {Hui}, {Tam},
  {Takata}, {Lin}, {Cheng}, {Park}, \& {Kim}}]{kong2014}
{Kong}, A.~K.~H., {Jin}, R., {Yen}, T.-C., {et~al.} 2014, \apjl, 794, L22,
  \dodoi{10.1088/2041-8205/794/2/L22}

\bibitem[{{Lange} {et~al.}(2001){Lange}, {Camilo}, {Wex}, {Kramer}, {Backer},
  {Lyne}, \& {Doroshenko}}]{lange2001}
{Lange}, C., {Camilo}, F., {Wex}, N., {et~al.} 2001, \mnras, 326, 274,
  \dodoi{10.1046/j.1365-8711.2001.04606.x}

\bibitem[{{Li} {et~al.}(2016){Li}, {Kong}, {Hou}, {Mao}, {Strader}, {Chomiuk},
  \& {Tremou}}]{li2016}
{Li}, K.-L., {Kong}, A.~K.~H., {Hou}, X., {et~al.} 2016, \apj, 833, 143,
  \dodoi{10.3847/1538-4357/833/2/143}

\bibitem[{{Li} {et~al.}(2018){Li}, {Hou}, {Strader}, {Takata}, {Kong},
  {Chomiuk}, {Swihart}, {Hui}, \& {Cheng}}]{li2018}
{Li}, K.-L., {Hou}, X., {Strader}, J., {et~al.} 2018, \apj, 863, 194,
  \dodoi{10.3847/1538-4357/aad243}

\bibitem[{{Linares} {et~al.}(2017){Linares}, {Miles-P{\'a}ez},
  {Rodr{\'\i}guez-Gil}, {Shahbaz}, {Casares}, {Fari{\~n}a}, \&
  {Karjalainen}}]{linares2017b}
{Linares}, M., {Miles-P{\'a}ez}, P., {Rodr{\'\i}guez-Gil}, P., {et~al.} 2017,
  \mnras, 465, 4602, \dodoi{10.1093/mnras/stw3057}

\bibitem[{{Lorimer} \& {Kramer}(2004)}]{lorimer2004}
{Lorimer}, D.~R., \& {Kramer}, M. 2004, {Handbook of Pulsar Astronomy}

\bibitem[{{Lyne} {et~al.}(2000){Lyne}, {Mankelow}, {Bell}, \&
  {Manchester}}]{lyne2000}
{Lyne}, A.~G., {Mankelow}, S.~H., {Bell}, J.~F., \& {Manchester}, R.~N. 2000,
  \mnras, 316, 491, \dodoi{10.1046/j.1365-8711.2000.03517.x}

\bibitem[{{Lyne} {et~al.}(2015){Lyne}, {Stappers}, {Keith}, {Ray}, {Kerr},
  {Camilo}, \& {Johnson}}]{lyne2015}
{Lyne}, A.~G., {Stappers}, B.~W., {Keith}, M.~J., {et~al.} 2015, \mnras, 451,
  581, \dodoi{10.1093/mnras/stv236}

\bibitem[{{Manchester} {et~al.}(2005){Manchester}, {Hobbs}, {Teoh}, \&
  {Hobbs}}]{manchester2005}
{Manchester}, R.~N., {Hobbs}, G.~B., {Teoh}, A., \& {Hobbs}, M. 2005, \aj, 129,
  1993, \dodoi{10.1086/428488}

\bibitem[{{Meinshausen} {et~al.}(2009){Meinshausen}, {Bickel}, \&
  {Rice}}]{meinshausen2009}
{Meinshausen}, N., {Bickel}, P., \& {Rice}, J. 2009, Ann. Appl. Stat., 3, 38,
  \dodoi{10.1214/08-AOAS180}

\bibitem[{{Messenger} \& {Woan}(2007)}]{messenger2007}
{Messenger}, C., \& {Woan}, G. 2007, Classical and Quantum Gravity, 24, S469,
  \dodoi{10.1088/0264-9381/24/19/S10}

\bibitem[{{Messenger} {et~al.}(2015){Messenger}, {Bulten}, {Crowder},
  {Dergachev}, {Galloway}, {Goetz}, {Jonker}, {Lasky}, {Meadors}, {Melatos},
  {Premachandra}, {Riles}, {Sammut}, {Thrane}, {Whelan}, \&
  {Zhang}}]{messenger2015}
{Messenger}, C., {Bulten}, H.~J., {Crowder}, S.~G., {et~al.} 2015, \prd, 92,
  023006, \dodoi{10.1103/PhysRevD.92.023006}

\bibitem[{{Monet} {et~al.}(2003){Monet}, {Levine}, {Canzian}, {Ables}, {Bird},
  {Dahn}, {Guetter}, {Harris}, {Henden}, {Leggett}, {Levison}, {Luginbuhl},
  {Martini}, {Monet}, {Munn}, {Pier}, {Rhodes}, {Riepe}, {Sell}, {Stone},
  {Vrba}, {Walker}, {Westerhout}, {Brucato}, {Reid}, {Schoening}, {Hartley},
  {Read}, \& {Tritton}}]{monet2003}
{Monet}, D.~G., {Levine}, S.~E., {Canzian}, B., {et~al.} 2003, \aj, 125, 984,
  \dodoi{10.1086/345888}

\bibitem[{{Nieder} {et~al.}(2019){Nieder}, {Clark}, {Bassa}, {Wu}, {Singh},
  {Donner}, {Allen}, {Breton}, {Dhillon}, {Eggenstein}, {Hessels}, {Kennedy},
  {Kerr}, {Littlefair}, {Marsh}, {Mata S{\'a}nchez}, {Papa}, {Ray}, {Steltner},
  \& {Verbiest}}]{nieder2019}
{Nieder}, L., {Clark}, C.~J., {Bassa}, C.~G., {et~al.} 2019, \apj, 883, 42,
  \dodoi{10.3847/1538-4357/ab357e}

\bibitem[{{Nieder} {et~al.}(2020){Nieder}, {Clark}, {Kandel}, {Romani},
  {Bassa}, {Allen}, {Ashok}, {Cognard}, {Fehrmann}, {Freire}, {Karuppusamy},
  {Kramer}, {Li}, {Machenschalk}, {Pan}, {Papa}, {Ransom}, {Ray}, {Roy},
  {Wang}, {Wu}, {Aulbert}, {Barr}, {Beheshtipour}, {Behnke}, {Bhattacharyya},
  {Breton}, {Camilo}, {Choquet}, {Dhillon}, {Ferrara}, {Guillemot}, {Hessels},
  {Kerr}, {Kwang}, {Marsh}, {Mickaliger}, {Pleunis}, {Pletsch}, {Roberts},
  {Sanpa-arsa}, \& {Steltner}}]{nieder2020b}
{Nieder}, L., {Clark}, C.~J., {Kandel}, D., {et~al.} 2020, arXiv e-prints.
\newblock \doarXiv{2009.01513}

\bibitem[{Oliphant(2006)}]{numpy2006}
Oliphant, T.~E. 2006, A guide to NumPy, Vol.~1 (Trelgol Publishing USA)

\bibitem[{{Owen}(1996)}]{owen1996}
{Owen}, B.~J. 1996, \prd, 53, 6749, \dodoi{10.1103/PhysRevD.53.6749}

\bibitem[{{Phinney}(1992)}]{phinney1992}
{Phinney}, E.~S. 1992, Philosophical Transactions of the Royal Society of
  London Series A, 341, 39, \dodoi{10.1098/rsta.1992.0084}

\bibitem[{{Pletsch} \& {Clark}(2014)}]{pletsch2014}
{Pletsch}, H.~J., \& {Clark}, C.~J. 2014, \apj, 795, 75,
  \dodoi{10.1088/0004-637X/795/1/75}

\bibitem[{{Pletsch} \& {Clark}(2015)}]{pletsch2015}
---. 2015, \apj, 807, 18, \dodoi{10.1088/0004-637X/807/1/18}

\bibitem[{{Pletsch} {et~al.}(2012{\natexlab{a}}){Pletsch}, {Guillemot},
  {Allen}, {Kramer}, {Aulbert}, {Fehrmann}, {Ray}, {Barr}, {Belfiore},
  {Camilo}, {Caraveo}, {{\c C}elik}, {Champion}, {Dormody}, {Eatough},
  {Ferrara}, {Freire}, {Hessels}, {Keith}, {Kerr}, {de Luca}, {Lyne},
  {Marelli}, {McLaughlin}, {Parent}, {Ransom}, {Razzano}, {Reich}, {Saz
  Parkinson}, {Stappers}, \& {Wolff}}]{pletsch2012a}
{Pletsch}, H.~J., {Guillemot}, L., {Allen}, B., {et~al.} 2012{\natexlab{a}},
  \apj, 744, 105, \dodoi{10.1088/0004-637X/744/2/105}

\bibitem[{{Pletsch} {et~al.}(2012{\natexlab{b}}){Pletsch}, {Guillemot},
  {Fehrmann}, {Allen}, {Kramer}, {Aulbert}, {Ackermann}, {Ajello}, {de
  Angelis}, {Atwood}, {Baldini}, {Ballet}, {Barbiellini}, {Bastieri},
  {Bechtol}, {Bellazzini}, {Borgland}, {Bottacini}, {Brandt}, {Bregeon},
  {Brigida}, {Bruel}, {Buehler}, {Buson}, {Caliandro}, {Cameron}, {Caraveo},
  {Casandjian}, {Cecchi}, {{\c C}elik}, {Charles}, {Chaves}, {Cheung},
  {Chiang}, {Ciprini}, {Claus}, {Cohen-Tanugi}, {Conrad}, {Cutini},
  {D'Ammando}, {Dermer}, {Digel}, {Drell}, {Drlica-Wagner}, {Dubois}, {Dumora},
  {Favuzzi}, {Ferrara}, {Franckowiak}, {Fukazawa}, {Fusco}, {Gargano},
  {Gehrels}, {Germani}, {Giglietto}, {Giordano}, {Giroletti}, {Godfrey},
  {Grenier}, {Grondin}, {Grove}, {Guiriec}, {Hadasch}, {Hanabata}, {Harding},
  {den Hartog}, {Hayashida}, {Hays}, {Hill}, {Hou}, {Hughes},
  {J{\'o}hannesson}, {Jackson}, {Jogler}, {Johnson}, {Johnson}, {Kataoka},
  {Kerr}, {Kn{\"o}dlseder}, {Kuss}, {Lande}, {Larsson}, {Latronico},
  {Lemoine-Goumard}, {Longo}, {Loparco}, {Lovellette}, {Lubrano}, {Massaro},
  {Mayer}, {Mazziotta}, {McEnery}, {Mehault}, {Michelson}, {Mitthumsiri},
  {Mizuno}, {Monzani}, {Morselli}, {Moskalenko}, {Murgia}, {Nakamori},
  {Nemmen}, {Nuss}, {Ohno}, {Ohsugi}, {Omodei}, {Orienti}, {Orlando}, {de
  Palma}, {Paneque}, {Perkins}, {Piron}, {Pivato}, {Porter}, {Rain{\`o}},
  {Rando}, {Ray}, {Razzano}, {Reimer}, {Reimer}, {Reposeur}, {Ritz}, {Romani},
  {Romoli}, {Sanchez}, {Parkinson}, {Schulz}, {Sgr{\`o}}, {do Couto e Silva},
  {Siskind}, {Smith}, {Spandre}, {Spinelli}, {Suson}, {Takahashi}, {Tanaka},
  {Thayer}, {Thayer}, {Thompson}, {Tibaldo}, {Tinivella}, {Troja}, {Usher},
  {Vandenbroucke}, {Vasileiou}, {Vianello}, {Vitale}, {Waite}, {Winer}, {Wood},
  {Wood}, {Yang}, \& {Zimmer}}]{pletsch2012}
{Pletsch}, H.~J., {Guillemot}, L., {Fehrmann}, H., {et~al.} 2012{\natexlab{b}},
  Science, 338, 1314, \dodoi{10.1126/science.1229054}

\bibitem[{{Prix} \& {Shaltev}(2012)}]{prix2012}
{Prix}, R., \& {Shaltev}, M. 2012, \prd, 85, 084010,
  \dodoi{10.1103/PhysRevD.85.084010}

\bibitem[{{Ransom} {et~al.}(2003){Ransom}, {Cordes}, \&
  {Eikenberry}}]{ransom2003}
{Ransom}, S.~M., {Cordes}, J.~M., \& {Eikenberry}, S.~S. 2003, \apj, 589, 911,
  \dodoi{10.1086/374806}

\bibitem[{{Ransom} {et~al.}(2001){Ransom}, {Greenhill}, {Herrnstein},
  {Manchester}, {Camilo}, {Eikenberry}, \& {Lyne}}]{ransom2001}
{Ransom}, S.~M., {Greenhill}, L.~J., {Herrnstein}, J.~R., {et~al.} 2001, \apjl,
  546, L25, \dodoi{10.1086/318062}

\bibitem[{{Ray} {et~al.}(2011){Ray}, {Kerr}, {Parent}, {Abdo}, {Guillemot},
  {Ransom}, {Rea}, {Wolff}, {Makeev}, {Roberts}, {Camilo}, {Dormody}, {Freire},
  {Grove}, {Gwon}, {Harding}, {Johnston}, {Keith}, {Kramer}, {Michelson},
  {Romani}, {Saz Parkinson}, {Thompson}, {Weltevrede}, {Wood}, \&
  {Ziegler}}]{ray2011}
{Ray}, P.~S., {Kerr}, M., {Parent}, D., {et~al.} 2011, \apjs, 194, 17,
  \dodoi{10.1088/0067-0049/194/2/17}

\bibitem[{{Riles}(2017)}]{riles2017}
{Riles}, K. 2017, Modern Physics Letters A, 32, 1730035,
  \dodoi{10.1142/S021773231730035X}

\bibitem[{{Roberts}(2013)}]{roberts2013}
{Roberts}, M.~S.~E. 2013, in IAU Symposium, Vol. 291, Neutron Stars and
  Pulsars: Challenges and Opportunities after 80 years, ed. J.~{van Leeuwen},
  127--132, \dodoi{10.1017/S174392131202337X}

\bibitem[{{Romani}(2015)}]{romani2015b}
{Romani}, R.~W. 2015, \apjl, 812, L24, \dodoi{10.1088/2041-8205/812/2/L24}

\bibitem[{{Romani} {et~al.}(2014){Romani}, {Filippenko}, \&
  {Cenko}}]{romani2014}
{Romani}, R.~W., {Filippenko}, A.~V., \& {Cenko}, S.~B. 2014, \apjl, 793, L20,
  \dodoi{10.1088/2041-8205/793/1/L20}

\bibitem[{{Salvetti} {et~al.}(2015){Salvetti}, {Mignani}, {De Luca}, {Delvaux},
  {Pallanca}, {Belfiore}, {Marelli}, {Breeveld}, {Greiner}, {Becker}, \&
  {Pizzocaro}}]{salvetti2015}
{Salvetti}, D., {Mignani}, R.~P., {De Luca}, A., {et~al.} 2015, \apj, 814, 88,
  \dodoi{10.1088/0004-637X/814/2/88}

\bibitem[{{Salvetti} {et~al.}(2017){Salvetti}, {Mignani}, {De Luca}, {Marelli},
  {Pallanca}, {Breeveld}, {H{\"u}semann}, {Belfiore}, {Becker}, \&
  {Greiner}}]{salvetti2017}
---. 2017, \mnras, 470, 466, \dodoi{10.1093/mnras/stx1247}

\bibitem[{{Sammut} {et~al.}(2014){Sammut}, {Messenger}, {Melatos}, \&
  {Owen}}]{sammut2014}
{Sammut}, L., {Messenger}, C., {Melatos}, A., \& {Owen}, B.~J. 2014, \prd, 89,
  043001, \dodoi{10.1103/PhysRevD.89.043001}

\bibitem[{{Saz Parkinson} {et~al.}(2016){Saz Parkinson}, {Xu}, {Yu},
  {Salvetti}, {Marelli}, \& {Falcone}}]{parkinson2016}
{Saz Parkinson}, P.~M., {Xu}, H., {Yu}, P.~L.~H., {et~al.} 2016, \apj, 820, 8,
  \dodoi{10.3847/0004-637X/820/1/8}

\bibitem[{{Saz Parkinson} {et~al.}(2010){Saz Parkinson}, {Dormody}, {Ziegler},
  {Ray}, {Abdo}, {Ballet}, {Baring}, {Belfiore}, {Burnett}, {Caliandro},
  {Camilo}, {Caraveo}, {de Luca}, {Ferrara}, {Freire}, {Grove}, {Gwon},
  {Harding}, {Johnson}, {Johnson}, {Johnston}, {Keith}, {Kerr},
  {Kn{\"o}dlseder}, {Makeev}, {Marelli}, {Michelson}, {Parent}, {Ransom},
  {Reimer}, {Romani}, {Smith}, {Thompson}, {Watters}, {Weltevrede}, {Wolff}, \&
  {Wood}}]{parkinson2010}
{Saz Parkinson}, P.~M., {Dormody}, M., {Ziegler}, M., {et~al.} 2010, \apj, 725,
  571, \dodoi{10.1088/0004-637X/725/1/571}

\bibitem[{{Schinzel} {et~al.}(2019){Schinzel}, {Kerr}, {Rau}, {Bhatnagar}, \&
  {Frail}}]{schinzel2019}
{Schinzel}, F.~K., {Kerr}, M., {Rau}, U., {Bhatnagar}, S., \& {Frail}, D.~A.
  2019, \apjl, 876, L17, \dodoi{10.3847/2041-8213/ab18f7}

\bibitem[{{Smith} {et~al.}(2017){Smith}, {Guillemot}, {Kerr}, {Ng}, \&
  {Barr}}]{smith2017}
{Smith}, D.~A., {Guillemot}, L., {Kerr}, M., {Ng}, C., \& {Barr}, E. 2017,
  ArXiv e-prints.
\newblock \doarXiv{1706.03592}

\bibitem[{{Strader} {et~al.}(2014){Strader}, {Chomiuk}, {Sonbas}, {Sokolovsky},
  {Sand}, {Moskvitin}, \& {Cheung}}]{strader2014}
{Strader}, J., {Chomiuk}, L., {Sonbas}, E., {et~al.} 2014, \apjl, 788, L27,
  \dodoi{10.1088/2041-8205/788/2/L27}

\bibitem[{{Strader} {et~al.}(2019){Strader}, {Swihart}, {Chomiuk}, {Bahramian},
  {Britt}, {Cheung}, {Dage}, {Halpern}, {Li}, {Mignani}, {Orosz}, {Peacock},
  {Salinas}, {Shishkovsky}, \& {Tremou}}]{strader2019}
{Strader}, J., {Swihart}, S., {Chomiuk}, L., {et~al.} 2019, \apj, 872, 42,
  \dodoi{10.3847/1538-4357/aafbaa}

\bibitem[{{Swihart} {et~al.}(2020){Swihart}, {Strader}, {Urquhart}, {Orosz},
  {Shishkovsky}, {Chomiuk}, {Salinas}, {Aydi}, {Dage}, \&
  {Kawash}}]{swihart2020}
{Swihart}, S.~J., {Strader}, J., {Urquhart}, R., {et~al.} 2020, \apj, 892, 21,
  \dodoi{10.3847/1538-4357/ab77ba}

\bibitem[{{Taff}(1985)}]{taff1985}
{Taff}, L.~G. 1985, {Celestial mechanics: A computational guide for the
  practitioner}

\bibitem[{{van der Putten} {et~al.}(2010){van der Putten}, {Bulten}, {van den
  Brand}, \& {Holtrop}}]{putten2010}
{van der Putten}, S., {Bulten}, H.~J., {van den Brand}, J.~F.~J., \& {Holtrop},
  M. 2010, in Journal of Physics Conference Series, Vol. 228, Journal of
  Physics Conference Series, 012005, \dodoi{10.1088/1742-6596/228/1/012005}

\bibitem[{{van der Walt} {et~al.}(2011){van der Walt}, {Colbert}, \&
  {Varoquaux}}]{numpy2011}
{van der Walt}, S., {Colbert}, S.~C., \& {Varoquaux}, G. 2011, Computing in
  Science and Engineering, 13, 22, \dodoi{10.1109/MCSE.2011.37}

\end{thebibliography}
	
\end{document}